\documentclass[useAMS,usenatbib,usegraphicx]{mn2e}
\usepackage[draft]{hyperref}
\usepackage{scrextend}
\usepackage[british]{babel}
\usepackage{times}
\usepackage{latexsym,amssymb}
\usepackage[fleqn]{amsmath}
\usepackage{graphics,graphicx}
\usepackage[absolute]{textpos}

\usepackage{natbib}
\usepackage{journalnames}
\usepackage{color}
\usepackage{graphicx}
\usepackage{epstopdf}
\usepackage[T1]{fontenc}
\usepackage{aecompl}

\usepackage{caption}
\usepackage{comment}
\usepackage{lipsum}
\usepackage{array}
\usepackage{dcolumn}
\newcounter{subfigure}
\usepackage{amsmath} 

\newcommand{\kms}{km\,s$^{-1}$}
\newcommand{\dgr}{$^\circ$}

\newcommand{\angstrom}{\text{\normalfont\AA}}

\newcommand{\sauron}{\texttt{SAURON}}      

\newcommand{\du}{\mathrm{d}}               
\newcommand{\HI}{\textrm{H}\textsc{I}}
\newcommand{\Vrms}{\ensuremath{V_\mathrm{rms}}}

\newcommand{\MLstar}{\ensuremath{\Upsilon_\star}}
\newcommand{\MLtot}{\ensuremath{\Upsilon_\mathrm{tot}}}

\newcommand{\vcjam}{\ensuremath{v_{c,\mathrm{JAM}}}}
\newcommand{\vcadc}{\ensuremath{v_{c,\mathrm{ADC}}}}

\title[The inner mass distribution]
{The inner mass distribution of late-type spiral galaxies from \sauron\ stellar kinematic maps}
\author[Kalinova et al.]{%
Veselina Kalinova$^{1,2}$\thanks{E-mail: kalinova@mpifr.de},
Glenn van de Ven$^{1}$,
Mariya Lyubenova$^{1,3}$, 
Jes{\'u}s Falc{\'o}n-Barroso$^{4,5}$,
\newauthor{Dario Colombo$^{2,6}$,
Erik Rosolowsky$^{2}$}\\
\\
$^{1}$Max Planck Institute for Astronomy, K\"onigstuhl 17, 69117 Heidelberg, Germany\\
$^{2}$Department of Physics 4-181 CCIS, University of Alberta, Edmonton AB T6G 2E1, Canada\\
$^{3}$Kapteyn Astronomical Institute, University of Groningen, Landleven 12, 9747 AD Groningen, the Netherlands\\
$^{4}$Instituto de Astrof\'isica de Canarias, V\'ia L\'actea s/n, La Laguna, Tenerife, Spain\\
$^{5}$Departamento de Astrof\'isica, Universidad de La Laguna, E-38205 La Laguna, Tenerife, Spain\\
$^{6}$Max Planck Institute for Radio Astronomy, Auf dem H\"ugel 69, 53121 Bonn, Germany\\ 
}

\pagerange{\pageref{firstpage}--\pageref{lastpage}} \pubyear{2016}

\begin{document}



\label{firstpage}

\maketitle

\begin{abstract}
We infer the central mass distributions within 0.4--1.2 disc scale lengths of 18 late-type spiral galaxies using two different dynamical modelling approaches -- the Asymmetric Drift Correction (ADC) and 
axisymmetric Jeans Anisotropic Multi-gaussian expansion (JAM) model. ADC adopts a thin disc assumption, 
whereas JAM does a full line-of-sight velocity integration. We use stellar kinematics maps obtained with the integral-field spectrograph $\sauron$ 
to derive the corresponding circular velocity curves from the two models. To find their best-fit values, we apply Markov Chain Monte Carlo (MCMC) method.  
ADC and JAM modelling approaches are consistent within 5\% uncertainty when the ordered motions are significant comparable to the random motions, 
i.e, $\overline{v_{\phi}}/\sigma_R$ is locally greater than 1.5.  Below this value, the ratio $v_\mathrm{c,JAM}/v_\mathrm{c,ADC}$ gradually increases with decreasing $\overline{v_{\phi}}/\sigma_R$, reaching
$v_\mathrm{c,JAM}\approx 2 \times v_\mathrm{c,ADC}$.
Such conditions indicate that the stellar masses of the galaxies in our sample are not confined to their disk planes and likely 
have a non-negligible contribution from their bulges and thick disks.
\end{abstract}


\section{Introduction}
\label{S:intro}
The non-Keplerian rotation curves of spiral galaxies provided the first observational evidence that galaxies are embedded in extensive dark matter haloes 
(\citealt{Bosma1978, Rubin1983, vanAlbada1985, Begeman1987}).    
Historically, the 21-cm emission from atomic neutral hydrogen gas (HI) has been the main tool to derive galaxy rotation curves, 
because of its capability to trace the gravitational field beyond the optical stellar disc. However, HI is less useful in constraining the rotation curve in the central 
parts of discs usually due to insufficient spatial resolution and the lack of HI gas in the inner parts of galaxies 
(e.g., \citealt{Noordermeer2007b}). The interstellar medium in the centre is instead dominated by gas in the molecular and ionised phases (e.g., \citealt{Leroy2008}). 
Unfortunately, the rotation curves derived through CO emission, the common tracer for the molecular gas distribution, often show non-axisymmetric 
signatures such as wiggles (\citealt{Wada2004,Colombo2014b}). Hot ionised gas has the additional disadvantage that the observed rotation alone is often 
insufficient to trace the total mass distribution and that its velocity dispersion needs to be included. This velocity dispersion is generally influenced 
by a typically unknown contribution from non-gravitational effects such as stellar winds and shocks (e.g., \citealt{weijmans2008}). 

Some of these shortcomings of gas tracers can be overcome through careful correction and analysis.  However, a broader shortcoming of gas tracers becomes 
apparent because of their dissipative nature.  The gas is easily disturbed by perturbations in the plane from, for example, a bar or spiral arm (\citealt{Englmaier1999, Weiner2001,Kranz2003}).  
Gas also settles in the galaxy disc plane (or polar plane) and is thus less sensitive to the mass distribution perpendicular to it.  
The typical velocity dispersions of the neutral and ionized gas, respectively, are $\sim$10 \kms (\citealt{Casertano1983,Caldu-Primo2013}) and $\sim$20 \kms (e.g. \citealt{Fathi2007}).  

Instead of gas, stars could be used as a tracer of the underlying gravitational potential. Stars are present in all galaxy types, are distributed in all three dimensions, 
and are collisionless making them less susceptible to perturbations. The random motion of the collisionless stars tipycally ranges 
from 20 \kms to above 300 \kms (\citealt{Dehnen1998,Martinsson2013b,Cappellari2013,Rys2013,Ganda2006}).  This implies a scale height difference of over an order 
of magnitude between the gaseous and stellar components (e.g., \citealt{Koyama2009,Bottema1993}).  

However, we need to measure both the ordered and random motions of stars before they can be used as a dynamical tracer.  
Their random motion can be different in all three directions, an effect known as the velocity anisotropy.  
This anisotropy requires more challenging observational and modelling techniques to uncover the total mass distribution.  
Nonetheless, these techniques are becoming available.  Integral-field spectrographs like \sauron\ (\citealt{Bacon2001}), 
which is used in this study, allow us to extract high-quality stellar kinematic maps
and enable us to perform dynamical models.  


A common approach for spiral galaxies is to apply the asymmetric drift correction (ADC; \citealt{Binney2008}) to infer 
the circular velocity curve from the measured stellar mean velocity and velocity dispersion profiles. 
This approach is straightforward since the velocity and dispersion profiles can  be obtained from long-slit spectroscopy, 
and no line-of-sight integration is required due to an underlying thin-disc assumption.  
Using this method requires assuming both the disc inclination and the magnitude of the velocity anisotropy.  
As such, the ADC approach is widely adopted in studies that use the stellar kinematics to infer the circular velocity curve. 
For example, when investigating the Tully-Fisher relation for earlier-type spirals (e.g., \citealt{Bottema1993}, \citealt{Neistein1999, Williams2010}), 
the speed of bars (e.g.,  \citealt{Aguerri1998, ButaZhang2009}), as well as the inner distribution of dark matter (e.g., \citealt{Kregel2005b,weijmans2008}).

In this paper, we adopt an alternative approach, namely fitting a solution of the axisymmetric Jeans equations to stellar mean velocity and velocity dispersion fields to infer mass distributions.  
We apply this approach to the data acquired with the integral-field spectrograph SAURON from inner parts of 18 late-type Sb--Sd spiral galaxies.  
These Jeans models are less general than orbit-based models (e.g., Schwarzschild's method; \citealt{vandenBoschetal2008}) but are far less computationally expensive while 
still providing a good description of galaxies dominated by stars on disc-like orbits (e.g., \citealt{Cappellari2008}).  
The applicability of these fitted Jeans solutions even applies to dynamically hot systems such as lenticular galaxies.  
The Jeans models take into account the two-dimensional information in the stellar kinematic maps as well as integration 
along the line-of-sight. We also compare the resulting circular velocity curves with those obtained through ADC to investigate the validity of the assumptions underlying the simpler ADC approach.

In Section~\ref{S:observations}, we summarise the sample including the \sauron\ observations and near-infrared imaging.  
The Jeans and ADC modelling approaches are described in Section~\ref{S:method} and applied via Markov Chain Monte Carlo technique in Section~\ref{S:analysis}. 
We then compare the circular velocity curves from both modelling approaches in Section~\ref{S:vcirc} and discuss the possible reasons for the significant differences we find in Section~\ref{S:discussion}. 
We draw our conclusions in Section~\ref{S:concl}. 
\section{Observations}
\label{S:observations} 


\begin{table*}
 \caption{Properties of our sample of 18 galaxies (see Sec.~\ref{SS:sample}): (1) Galaxy
  identifier; (2) Hubble type (G06 from NED); (3) Galactocentric distance in Mpc (G06 from NED); 
  (4) Photometric position angle in degrees from DSS images with a typical uncertainty from 5\dgr to 10\dgr\ (G09; Sec.~\ref{SS:sample} ) ;
 (5)  Ellipticity from DSS images with a typical uncertainty of 15 \dgr\ (G09; Sec.~\ref{SS:sample} );
 (6)  Adopted inclination in ADC and JAM approaches, where the values marked with star correspond to the minimum allowed inclinations in JAM model  (see Sec.~\ref{SS:sample});
(7) Systemic velocity in km s$^{-1}$, measured using the enforced point-symmetry
method with typical uncertainty of 1 km s$^{-1}$ (Sec.~\ref{SS:stellarkinematics});
 (8)  Galaxy MGE effective radius in arcseconds with a typical uncertainty of 10\% (Sec.~\ref{SS:sample}); 
 (9) Radial extent of the stellar kinematic data in arcseconds with a typical uncertainty of 1\%  (Sec.~\ref{SS:sample});
  (10) Disk scale length in arcseconds with a typical uncertainty of 5\% (G09); 
  (11) Effective radius of the bulge in arcseconds with a typical uncertainty of 15\% (G09); 
  (12) Radial extent of the stellar kinematic data in 
  terms of disk scale lengths with a typical uncertainty of 5\% (Sec.~\ref{SS:sample})}
    \tabcolsep=2.0mm 
    \begin{tabular}{*{12}{|l}|}
      \hline
NGC      & Type &D & PA &$\epsilon$  & \emph{i} & $V_{\mathrm{sys}}$ & $R_{e}$ & $R_{\mathrm{max}}$& $h_d$  & $r_b$  & 
$(\frac{R_{\mathrm{max}}}{h_d})$\\
 \hline
 (1)      &  (2)  & (3) &  (4)&  (5) &(6) &  (7) &  (8) &   (9) & (10) &  (11) & (12)  \\
\hline
 %
%
0488  &SA(r)b    & 32.1 &  5 & 0.230  & 42 & 2299 & 53.6 & 19.2 & 42.7 &  9.9 &  0.4 \\
0772 &SA(s)b    & 35.6 & 126 & 0.340  & 51 & 2506 & 54.0 & 20.6 & 47.1 & 19.0 &  0.4 \\
4102 &SAB(s)b   & 15.5 &  42 & 0.445  & 59 & 838  & 15.0 & 22.1 & 19.5 &  1.5 &  1.1 \\
5678 &SAB(rs)b  & 31.3 &   5 & 0.475  & 61 & 1896 & 24.4 & 24.0 & 20.9 &  3.1 &  1.1 \\
3949 &SA(s)bc   & 14.6 & 122 & 0.360  & 53 & 808  & 20.4 & 22.1 & 16.8 &  4.6 &  1.3 \\
4030 &SA(s)bc   & 21.1 &  37 & 0.240  & 43 & 1443 & 30.2 & 19.2 & 26.5 & 15.3 &  0.7 \\
2964 &SAB(r)bc  & 20.6 &  96 & 0.450  & 59 & 1324 & 20.8 & 21.0 & 16.9 &  1.0 &  1.2 \\
0628 &SA(s)c    &  9.8 &  25 & 0.190  & 38 & 703  & 90.7 & 22.1 & 70.7 & 12.3 &  0.3 \\
0864 &SAB(rs)c  & 21.8 &  26 & 0.320  & 50 & 1606 & 38.3 & 15.1 & 28.1 &  1.6 &  0.5 \\
4254 &SA(s)c    & 19.4 &  50 & 0.270  & 46 & 2384 & 52.1 & 19.2 & 40.7 & 15.8 &  0.5 \\
1042 &SAB(rs)cd & 18.1 & 174 & 0.290  & 47 & 1404 & 73.4 & 15.1 & 52.5 &  4.4 &  0.3 \\
3346 &SB(rs)cd  & 18.9 & 100 & 0.160  & 42$^{\star}$ & 1257 & 55.3 & 17.8 & 35.4 &  3.2 &  0.5 \\
3423 &SA(s)cd   & 14.7 &  41 & 0.230  & 42 & 1001 & 52.6 & 20.1 & 40.1 &  9.7 &  0.5 \\
4487 &SAB(rs)cd & 14.7 &  77 & 0.370  & 53 & 1016 & 49.8 & 19.2 & 36.8 & 10.6 &  0.5 \\
2805 &SAB(rs)d  & 28.2 & 125 & 0.240  & 43 & 1742 & 65.8 & 12.6 & 50.8 & 13.0 &  0.3 \\
4775 &SA(s)d    & 22.5 &  96 & 0.135  & 34$^{\star}$ & 1547 & 29.0 & 17.8 & 19.3 &  8.3 &  0.9 \\
5585 &SAB(s)d   &  8.2 &  38 & 0.360  & 53 & 312  & 69.7 & 17.8 & 53.2 & 15.7 &  0.3 \\
5668 &SA(s)d    & 23.9 & 120 & 0.155  & 37$^{\star}$ & 1569 & 38.3 & 16.5 & 30.3 & 13.8 &  0.5 \\
\hline
 \end{tabular}
  \label{tab:prop}
\end{table*}

\subsection{Sample selection}
\label{SS:sample}
Our sample consists of 18 nearby, late-type spiral galaxies with Hubble types ranging from Sb to Sd. The sample selection, observations, and data reduction are presented in detail in \citet[hereafter G06 and G09]{Ganda2006, Ganda2009}. The galaxies all have imaging data from the Hubble Space Telescope (HST) including either WFPC2 and/or NICMOS data \citep{Carollo1997,Carollo1998,Carollo2002,Laine2002,Boeker2002}. All targets were selected to be brighter than $B_T = 12.5$ according to the values listed in the \citet{deV1991} catalogue, where interacting and Seyfert galaxies were discarded.  The sample also excludes targets that are inaccessible to the 4.2-m William Herschel Telescope (\S\ref{SS:sauronifs}), so only galaxies with  0 $<$ RA $<15 h$ and $ \delta>-20$\dgr\ have been selected. 

In Table~\ref{tab:prop} we list the main properties of the 18 galaxies, combined from the literature and our own measurements. 
Their morphological type ranging between Sb and Sd, together with the galactocentric distance (D), are taken from G06 using NASA/IPAC Extragalactic Database (NED)\footnote{http://nedwww.ipac.caltech.edu}. 
The photometric position angle (PA) is measured by G09 using Digital Sky Survey (DSS)\footnote{http://archive.eso.org/dss/dss} images, 
but the typical uncertainty of 5\dgr-10\dgr\ comes from comparing the measured values of the PA in G06 to literature values compiled from the 
Third Reference Catalog of Bright Galaxies (RC3)\footnote{http://heasarc.nasa.gov/W3Browse/all/rc3.html}. 
In the same way, we estimated the typical uncertainty of 15\% for the ellipticity $\epsilon$, which we also take from G06 to compare to the RC3 catalogue.
From the axis ratio of the galaxies ($q=1-\epsilon$), we calculate the photometric inclination (\emph{i}) of the galaxies using $\epsilon$ from Table \ref{tab:prop}
in the following way (\citealt{Hubble1926}):

\begin{equation}
  \label{eq:incl}
  \mathrm{cos(i)^2} = \frac{q^2-q_o^2}{1-q_o^2},
\end{equation}
Here $q_o$ is the intrinsic axis ratio of the galaxy (e.g. \citealt{Rodriguez2013}) and we adopt $q_o$ = 0.2 as commonly used (e.g., \citealt{Tully2000}). 
The change in the value of $q_o$ slightly affects our estimation of the inclination (1\dgr--3\dgr). The total uncertainty of the inclination, considering the errors of $q$ and $q_o$ can reach 5\dgr. 

In column 6 of Table \ref{tab:prop}, we present the inclinations we adopted throughout our entire analysis. Note here that the listed inclinations of the galaxies 
NGC3346, NGC4775 and NGC5668 are slightly larger than the corresponding photometric values calculated in eq. (\ref{eq:incl}) by 7\dgr, 1\dgr and 2\dgr, respectively. 
For these three galaxies, the estimated photometric inclination is below the minimum allowed MGE inclination ($\emph{i}^{MGE}$) of the JAM approach (see Sec. \ref{SS:jam}; eq.\ref{eq:ilim}), and 
thus we adopt $i=i^{MGE}$.
Nevertheless, the difference between the adopted and photometric inclination for these three galaxies is generally within the uncertainty of the photometric 
inclination, and does not affect our results. 
We wish to keep a constant inclination through this analysis to focus on the differences between the Jeans and ADC approaches.

Further, we measure the systemic velocity ($V_{\mathrm{sys}}$) using the enforced point-symmetry method with typical uncertainty of 1 km s$^{-1}$ 
(see Sec.~\ref{SS:stellarkinematics}) and determine the galaxy effective radius $R_e$ (half-light radius) after applying a Multi-Gaussian 
Expansion method (Sec.~\ref{SS:surfacebrightness}) to the surface brightness profiles of the galaxies. Using chi-squared values of the fit, 
we estimate that the typical $1\sigma$ uncertainty of the effective radius does not exceed 10\%.  The radial extent $R_{\mathrm{max}}$ of 
the stellar kinematic data is estimated via $\texttt{kinemetry}$ package (\citealt{Krajnovic2006}), yeilding a typical uncertainty for our 
galaxies of 1\%.  The disk scale length ($h_d$) and effective radius of the bulge ($r_b$) are taken from G09 including their typical 
uncertainties of 5\% and 15\%, respectively. We then calculate the radial extent ($R_{\mathrm{max}}/h_d$) of the stellar kinematic data 
in terms of disk scale lengths with an associated typical uncertainty of 5 \% (see Table~\ref{tab:prop}).

\subsection{\sauron\ integral-field spectroscopy}
\label{SS:sauronifs}
The sample galaxies were observed with the integral-field unit (IFU) spectrograph \sauron\ at the 4.2-m William Herschel Telescope of the Observatorio del Roque de los Muchachos on La Palma, Spain. The \sauron\ IFU  \citep{Bacon2001} has a $33\arcsec\times 41 \arcsec$ field-of-view (FoV), sampled by an array of $0.94\arcsec \times 0.94 \arcsec$ square lenses. The FoV corresponds to a typical radial extent of 1/5 to 1/3 of the galaxy's half-light radius ($R_e$). The spectral resolution is $\sim$ 4.2 $\angstrom$ (FWHM), corresponding to an instrumental velocity dispersion of $\sim 100$ \kms in the observed spectral range 4800-5380 $\angstrom$ (at 1.1 $\angstrom$ per pixel). This range includes a number of absorption features -- Fe, Mgb and H$\beta$, which we use to measure the stellar kinematics. Emission lines in this range, such as  [OIII], [NI] and H$\beta$, can be used to probe the ionised gas properties (G06).

The observations were reduced by G06 using the dedicated software $\texttt{XSAURON}$ \citep{Bacon2001}. To obtain a sufficient signal-to-noise ratio (S/N), we spatially binned the data cubes using the Voronoi 2D binning algorithm of \citet{Cappellari2003}. We created compact bins with a minimum S/N~$\sim60$ per spectral resolution element. In the central regions many individual spectra have S/N $>$ 60 and thus remained un-binned.
%
\subsection{Near-infrared imaging }
\label{SS:nirimaging}
We parametrize the light distribution of the galaxies with the surface brightness profiles obtained by G09.  They used archival ground-based $H$-band images from Two-Micron All Sky survey (hereafter, 2MASS)\footnote{http://www.ipac.caltech.edu/2mass/} complemented with near-infrared HST NICMOS/F160W images for 11 cases and optical HST WFPC2/F814W for the remaining 7 galaxies (NGC\,1042, NGC\,2805, NGC\,3346, NGC\,3423, NGC\,4487, NGC\,4775, NGC\,5668). DSS data were used for the outer parts of the same galaxies to obtain an accurate determination of the sky level and the galaxy disc geometry.

G09 derived the light distribution parameters using the \texttt{ellipse} task in Imaging Reduction and Analysis Facility (IRAF)\footnote{http://iraf.noao.edu/}. They first fitted elliptical isophotes to galaxy images with the centre, position angle and ellipticity left as free parameters in order
to obtain the centre coordinates. G09 then fit again ellipses to the images with the centre fixed, where the position angle and ellipticity were set as free parameters. In the end, there were three photometric profiles and their combination gave a single near-infrared $H$-band profile with the maximum extent and inner spatial resolution allowed by the data. The error introduced by this combination of optical and infrared images was negligible (G09, Sec.3.3) and does not affect our analysis.

\section{Mass modelling methods}
\label{S:method} 

For most of the 18 late-type spiral galaxies, the \sauron\ stellar mean velocity and velocity dispersion maps are, within the observational uncertainties,  consistent with axisymmetry.  The non-axisymmetric features appear to be mainly due to dust obscuration.  
In half of the galaxies, the effect of bars seems to be weak or more dominant in the outer parts.  These trends support the assumption of a stationary axisymmetric stellar system for the inner parts of the galaxies that we study here.

The methodology of \citet{Schwarzschild1979} has proven to be powerful in building detailed models for spherical, axisymmetric \citep[e.g.][]{Rix1997, Cappellari2006}, and triaxial nearby galaxies \citep{vandeVen2006, vandenBosch2006}, as well as globular clusters \citep[e.g.][]{vandeVen2006, vandenBosch2006}.  The method finds the set of weights of orbits computed in an arbitrary gravitational potential that best reproduces all available photometric and kinematic data at the same time.  Since higher-order stellar kinematic measurements are necessary to constrain the large freedom in this general modelling method, a less computationally intensive approach has been to construct simpler -- but still realistic -- dynamical models based on the solution of the axisymmetric Jeans equations (\citealt{Cappellari2008, vandeVen2010}).  We adopt the latter approach here.

\subsection{Jeans equations}
\label{SS:axijeans}

In the case of steady state axisymmetry both the potential $\Phi(R,z)$ and distribution function (DF) are independent of azimuth $\phi$ and time. By Jeans' theorem \citep{Jeans1915} the DF depends only  on the isolating integrals of motion: $f(E,L_z,I_3)$, with energy $E =(v_R^2+v_\phi^2+v_z^2)/2 + \Phi(R,z)$, angular momentum $L_z = Rv_\phi$ parallel to the symmetry $z$-axis, and a third integral $I_3$ for which in general no explicit expression is known. However,  $I_3$ usually is invariant under the change $(v_R,v_z) \to (-v_R,-v_z)$, though $I_3$ may loose this symmetry if resonances are present.  Such symmetry implies that the mean velocity is in the azimuthal direction ($\overline{v_R}=\overline{v_z}=0$) and that the velocity ellipsoid is aligned with the rotation direction ($\overline{v_R  v_\phi} = \overline{v_\phi v_z} = 0$)\footnote{This is not explicitly true in the solar neighbourhood, where the vertex deviations are $\sim$20 degrees (\citealt{Fuchs2009}), and these vertex deviations may also correlate with morphological features (\citealt{Vorobyov2008})}.

When we multiply the collisionless Boltzmann equation in cylindrical coordinates by $v_R$ and $v_z$ respectively and integrate over all velocities, 
we obtain two of the Jeans equations \citep{Jeans1915},
\begin{eqnarray}
  \label{eq:jeansaxiR}
  \frac{\partial(R\nu\overline{v_R^2})}{\partial R}
  + R\frac{\partial(\nu\overline{v_R v_z})}{\partial z}
  - \nu\overline{v_\phi^2} 
  + R\nu\frac{\partial \Phi}{\partial R} 
  & = & 0,
  \\
  \label{eq:jeansaxiz}
  \frac{\partial(R\nu\overline{v_R v_z})}{\partial R}
  + R\frac{\partial(\nu\overline{v_z^2})}{\partial z}
  + R \nu \frac{\partial \Phi}{\partial z} 
  & = & 0,
\end{eqnarray}
where $\nu(R,z)$ is the intrinsic luminosity density\footnote{Here $\nu$ has the same meaning as the luminosity density $j(R,z)$ in \cite{Binney2008}}. Due to the assumed axisymmetry, all terms in the third Jeans equation, that follows from multiplying by $v_\phi$, vanish.

\subsection{JAM Circular-Speed Calculation}
\label{SS:jam}

The \emph{Jeans Axisymmetric Model} (JAM) is based on solving the  above two Jeans equations~\eqref{eq:jeansaxiR} and~\eqref{eq:jeansaxiz}. However, since there are  four unknown second-order velocity moments $\overline{v_R^2}$, $\overline{v_z^2}$, $\overline{v_\phi^2}$ and $\overline{v_R v_z}$ this is an underconstrained system, and we have to make two assumptions about the velocity anisotropy, or in other words about the shape and alignment of the velocity ellipsoid.  
First, we assume the velocity ellipsoid is aligned with the cylindrical $(R,\phi,z)$ coordinate system, which gives $\overline{v_R v_z} = 0$.  We can then readily solve equation~\eqref{eq:jeansaxiz} for $\overline{v_z^2}$. 
Second, we also assume a constant flattening of the velocity ellipsoid in the meridional plane, so we can write $\overline{v_R^2} = \overline{v_z^2}/(1-\beta_z)$ and solve equation~\eqref{eq:jeansaxiR} for $\overline{v_\phi^2}$.  \citet{Cappellari2008} argue that this second assumption provides a good, general description for the kinematics of real disc galaxies. When $\beta_z = 0$, the velocity distribution is isotropic in the meridional plane, corresponding to the well-known case of a two-integral distribution function $f(E,L_z)$ \citep[e.g.][]{Lynden-Bell1962,Hunter1977}.

Given the intrinsic second-order velocity moments, we can then calculate the observed second-order velocity moment by integrating along the line-of-sight through the stellar system viewed at an inclination $i > 0$ away from the $z$-axis:
\begin{eqnarray}
  \label{eq:vlosintegral}
  \overline{v_\mathrm{los}^2} & = & \frac{1}{I(x',y')}
  \int_{-\infty}^{+\infty} \nu \biggl[
    \left(\overline{v_R^2}\sin^2\phi +
      \overline{v_\phi^2}\cos^2\phi\right) \sin^2i
  \biggr.
  \nonumber \\ 
  && 
  \biggl. 
  + \overline{v_z^2} \cos^2i 
  \biggr]\, \du z',
\end{eqnarray}
where $I(x',y')$ is the (observed) surface brightness with the $x'$-axis along the projected major axis.  For each position $(x',y')$ on the sky-plane, $\overline{v_\mathrm{los}^2}$ yields a prediction of the (luminosity weighted) combination $\Vrms^2 = V^2 + \sigma^2$ of the (observed) mean line-of-sight velocity $V$ and dispersion $\sigma$.

Under the above assumptions, the only unknown parameters are the anisotropy parameter $\beta_z$, the inclination $i$, and the gravitational potential $\Phi(R,z)$, which is related to the total mass density $\rho(R,z)$ through Poisson's equation.  To estimate $\rho(R,z)$, we derive the intrinsic luminosity density $\nu(R,z)$ by deprojecting the observed surface brightness $I(x',y')$.  We then assume a total mass-to-light ratio $\Upsilon$ to derive $\rho(R,z)=\Upsilon \nu(R,z)$.  Dynamical studies of the inner parts of galaxies typically consider $\Upsilon$ as an additional parameter and further assume its value to be constant, i.e., mass follows light \citep[e.g.][]{Cappellari2006}.  Since $\Upsilon$ may be larger than the stellar mass-to-light ratio $\MLstar$, this assumption still allows for possible dark matter contribution, albeit with a constant fraction.

A convenient way to arrive at $\Phi(R,z)$ is through the Multi-Gaussian Expansion method \citep[MGE;][]{Monnet1992,Emsellem1994}, which models the observed surface brightness as a sum of $N$ Gaussian components,
\begin{equation}
  \label{eq:mgeSB}
  I(x',y') = \sum_{j=0}^{N} I_{0,j} \exp\left\{ -\frac{1}{2{
{\bf \xi}'_j}^2} \left[ x'^2 + \frac{y'^2}{{q'_j}^2} \right] \right\},
\end{equation}
each with three parameters: the central surface brightness $I_{0,j}$, the dispersion $ \xi'_j$ along the major $x'$-axis and the flattening $q'_j$.  

The MGE approach has several advantages.  
Even though Gaussians do not form an orthogonal set of functions, surface density distributions are accurately reproduced. 
When the point spread function (PSF) of the instrument is also represented as a sum of Gaussians, the convolution with the PSF becomes straightforward.
Given the viewing direction, the MGE parametrization can be deprojected analytically into an intrinsic luminosity density $\nu(R,z)$.
Furthermore, the calculation of $\overline{v_\mathrm{los}^2}$ in Equation~\eqref{eq:vlosintegral} reduces from the (numerical) evaluation of a triple integral to a straightforward single integral \citep[][eq.~27]{Cappellari2008}.
Similarly, the gravitational potential $\Phi(R,z)$ can be calculated by means of one-dimensional integral \cite[][eq.~39]{Emsellem1994}.

Given the latter, the circular velocity from the JAM model in the equatorial plane then follows upon (numerical) evaluation of 
\begin{multline}
  \label{eq:mgevcirc}
  v_{c,\mathrm{JAM}}^2(R) = \sum_{j=0}^{N} 
  \frac{2 G L_j \Upsilon_j}{\sqrt{2\pi}\xi_j} \frac{R^2}{\xi_j^2} 
   \times \\
 \int_0^1 \exp\left\{ -\frac{u^2 R^2}{2\xi_j^2} \right\}
\frac{u^2\,\du u}{\sqrt{1-(1-q_j^2)u^2}},
\end{multline}
where $L_j \equiv 2 \, \pi \xi_j^2 q'_j I_{0,j}$ and $\Upsilon_j$ are the total luminosity and the mass-to-light ratio of the $j$th Gaussian. The intrinsic dispersion and flattening, $\xi_j$ and $q_j$, are related to their observed quantities, as
\begin{equation}
  \label{eq:qintr}
  	\xi_j = \xi'_j
  	\quad \mathrm{and} \quad
  	{q'_j}^2 = \cos^2i + q_j^2 \sin^2i, 
\end{equation}
with inclination $i$, ranging from $i=0^\circ$ for face-on viewing to $i=90^\circ$ for edge-on viewing.
Note here that the oblate MGE deprojection is valid if
\begin{equation}
\label{eq:ilim}
\cos(i)^2<{q'_j}^2 
\end{equation}
for all Gaussians (\citealt{Cappellari2002}).
Therefore, the flattest Gaussian in the MGE fit defines the minimum possible inclination ($\emph{i}^{MGE}$) for which the 
MGE model can be applied within the JAM approach.

In what follows, we assume mass follows light, so that the mass-to-light ratio is the same for each Gaussian: $\Upsilon_j=\Upsilon$. 
We fix the galaxies' inclinations to the values shown in Table \ref{tab:prop}, and
we are left with two free parameters in JAM: the total or dynamical mass-to-light ratio $\Upsilon$ 
and the velocity anisotropy $\beta_z^\mathrm{JAM}$. For each galaxy, we obtain the values of these two parameters by fitting 
the observed second-order velocity moment $\Vrms^2 = V^2 + \sigma^2$, after which the JAM circular speed follows from equation~\eqref{eq:mgevcirc}.

\subsection{ADC Circular-Speed Calculation}
\label{SS:adc}

In this work, we compare the JAM-derived circular speed curves to those from the commonly adopted ADC method.  Here we relate the ADC approach back to the Jeans equations.  Instead of solving both Jeans equations~\eqref{eq:jeansaxiR} and~\eqref{eq:jeansaxiz}, we can instead evaluate the first equation 
in the equatorial plane ($z=0$) and use $v_c^2 = R\,(\partial\Phi/\partial R)$ and $\partial\nu/\partial z = 0$ (by symmetry) to rewrite Eq.~\eqref{eq:jeansaxiR} as
\begin{multline}
  \label{eq:jeansRtovc}
  v^2_{c,\mathrm{ADC}}(R) = 
  	\overline{v_\phi}^2 
  	+ \sigma_R^2 \left[
    	\frac{\partial\ln\left(\nu\sigma_R^2\right)^{-1}}{\partial\ln R}
	\right. \\ \left.
    	+ \left( \frac{\sigma_\phi^2}{\sigma_R^2} - 1 \right)
    	- \frac{R}{\sigma_R^2} \frac{\partial\overline{v_R v_z}}{\partial z}
  	\right] .
\end{multline}
Here, $\overline{v_\phi}$ is the intrinsic mean velocity, and 
$\sigma_\phi^2 = \overline{v_\phi^2} - \overline{v_\phi}^2$, $\sigma_R^2 = \overline{v_R^2}$ are (the square of) 
the intrinsic mean velocity dispersions. 

In case of a dynamically cold tracer such as neutral gas, observed through \HI\, and CO emission at radio wavelengths, the mean velocity is typically much larger than the velocity dispersion ($V \gg \sigma$), so that the circular velocity can be inferred from the deprojected observed rotation, $v_c \simeq \overline{v_\phi}$ (e.g., \citealt{Ianjamasimanana2015,Colombo2014b,Caldu-Primo2013,deBlok2008}). However, the velocity dispersion can be non-negligible or even dominant for dynamically hot tracers like stars where it is possible that $V \ll \sigma$.  In this case, the asymmetric drift must be taken into account since the observed rotation only captures part of the circular velocity (e.g., \citealt{Blanc2013,weijmans2008}).

From equation~\eqref{eq:jeansRtovc}, this asymmetric drift correction depends on the two intrinsic velocity dispersions $\sigma_R$ and $\sigma_\phi$, plus the cross term $\overline{v_R v_z}$, which together define the velocity ellipsoid of the galaxy.
To allow for a direct comparison with the JAM model, we adopt the same assumption of alignment of the velocity ellipsoid with the cylindrical coordinate systems, so that $\overline{v_R v_z}=0$.
Assuming that the velocity ellipsoid is furthermore symmetric around $v_\phi = \overline{v_\phi}$, it follows from 
Appendix~A\footnote{Note here that the relation $\sigma_\phi^2/\sigma_R^2$ from \cite{weijmans2008} is efectively the epicycle approximation, which is applied only in ADC and not in JAM.} of \citet{weijmans2008} that $\sigma_\phi^2/\sigma_R^2 = (1+\alpha_R)/2$ with $\alpha_R$ the radial logarithmic gradient of the intrinsic mean velocity: $\alpha_R \equiv \partial \ln \overline{v_\phi} / \partial \ln R$.

Under the assumption of a 'thin disc' with $\partial\ln\nu/\partial\ln R \approx d\ln I/d\ln R$, the circular velocity from the ADC approach in the equatorial plane then reduces to
\begin{multline}
  \label{eq:adcvcirc}
  v^2_{c,\mathrm{ADC}}(R) = 
  	\overline{v_\phi}^2 + \sigma_R^2 
	\biggl[
    	\frac{\partial\ln\left(I\,\sigma_R^2\right)^{-1}}{\partial\ln R}
 	+ \frac{(\alpha_R -1)}{2}
	\biggr].
\end{multline}
This final equation provides an estimate of $v_{c,\mathrm{ADC}}$ that uses: (1) the surface brightness profile $I$ from the MGE parametrization, (2) $\overline{v_\phi}$ and $\alpha_R$ from the observed mean line-of-sight velocity $V$, and (3) $\sigma_R$ from the observed line-of-sight velocity dispersion $\sigma$. The latter two inferences also involve the velocity anisotropy $\beta_z^\mathrm{ADC}$ and 
inclination $i$ as described in Sec.~\ref{SS:radialprofiles} below.

\section{Analysis}
\label{S:analysis}
Here, we describe the application of our model analyses to the observational data.  In Sec. \ref{SS:surfacebrightness}, we discuss the Multi-Gaussian Expansion model adopted to the photometry of the \sauron\ galaxies in order to derive a smooth, analytic representation of their surface brightness.  We explain the extraction of the stellar kinematic maps in Sec. \ref{SS:stellarkinematics}.  The performance of the Markov Chain Monte Carlo analysis to the JAM and ADC models is shown in Sec. \ref{SS:MCMC}. ADC holds a thin disc assumption, whereas JAM does a full line-of-sight integration through the luminosity distribution to model the observed velocity moments. 
\subsection{Surface brightness parametrization}
\label{SS:surfacebrightness}

\begin{figure*}
{\includegraphics[width=0.255\textwidth]{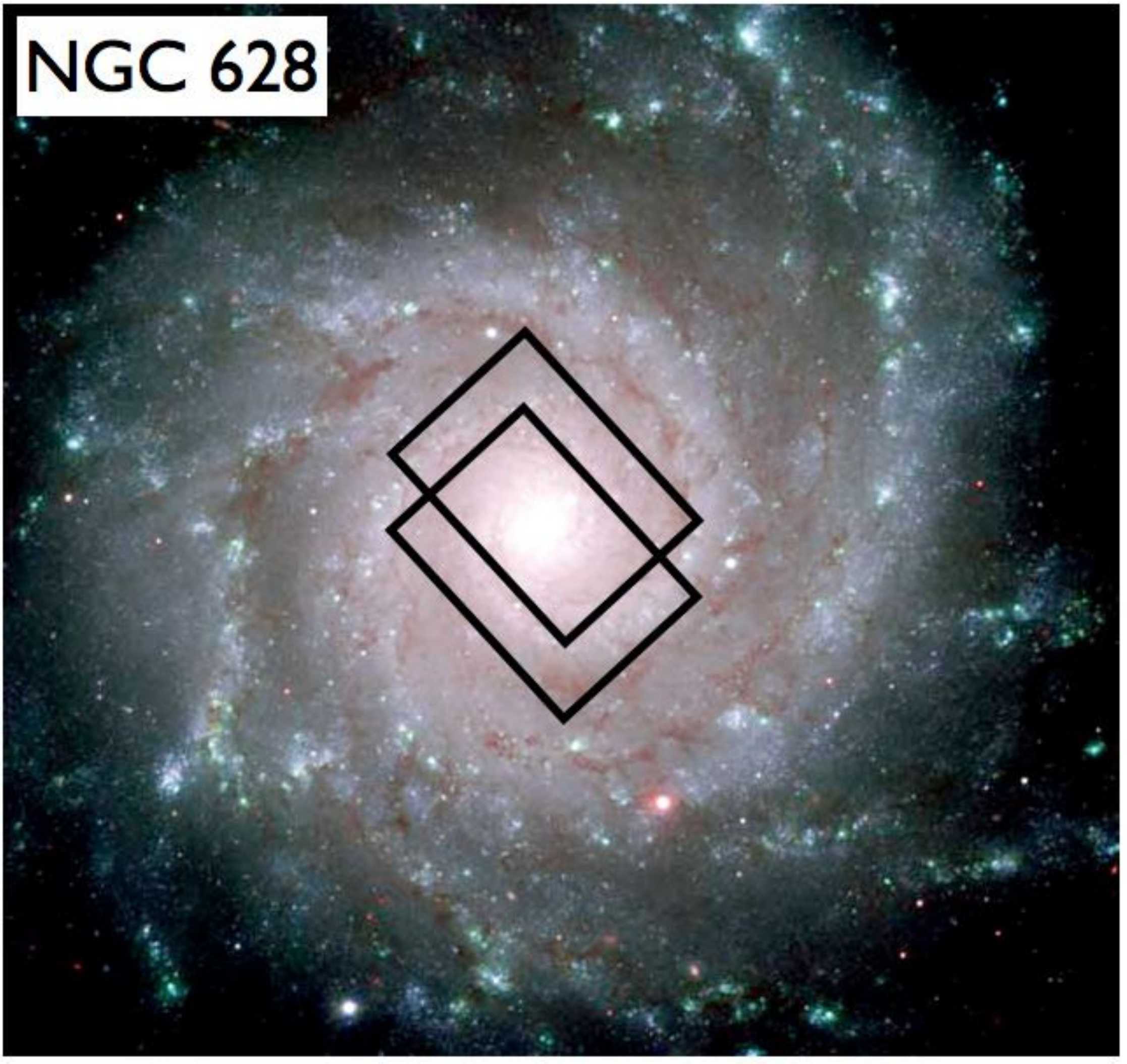}
\includegraphics[width=0.235\textwidth]{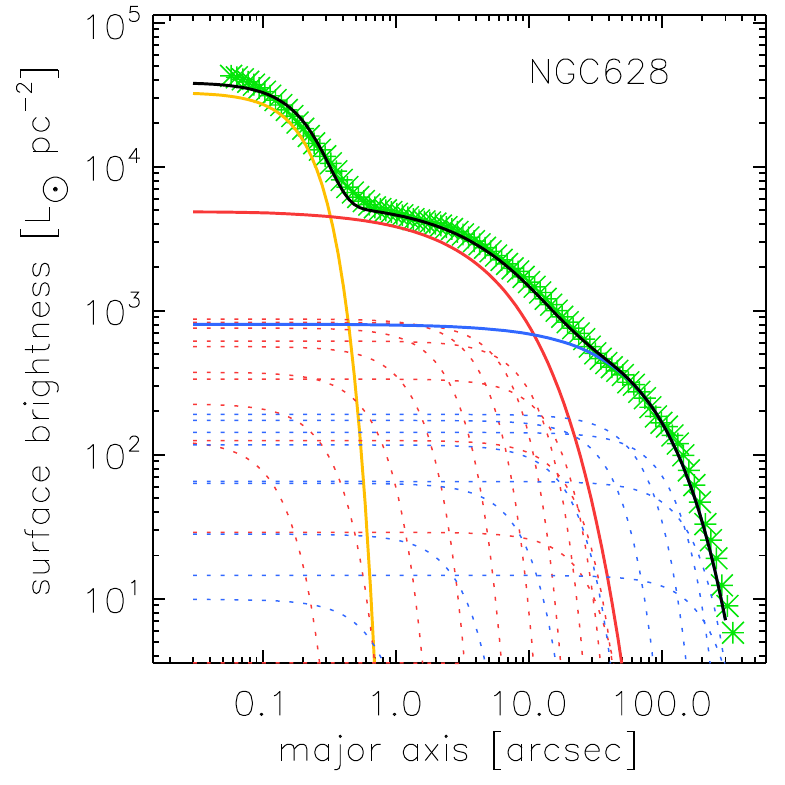}
\includegraphics[width=0.255\textwidth]{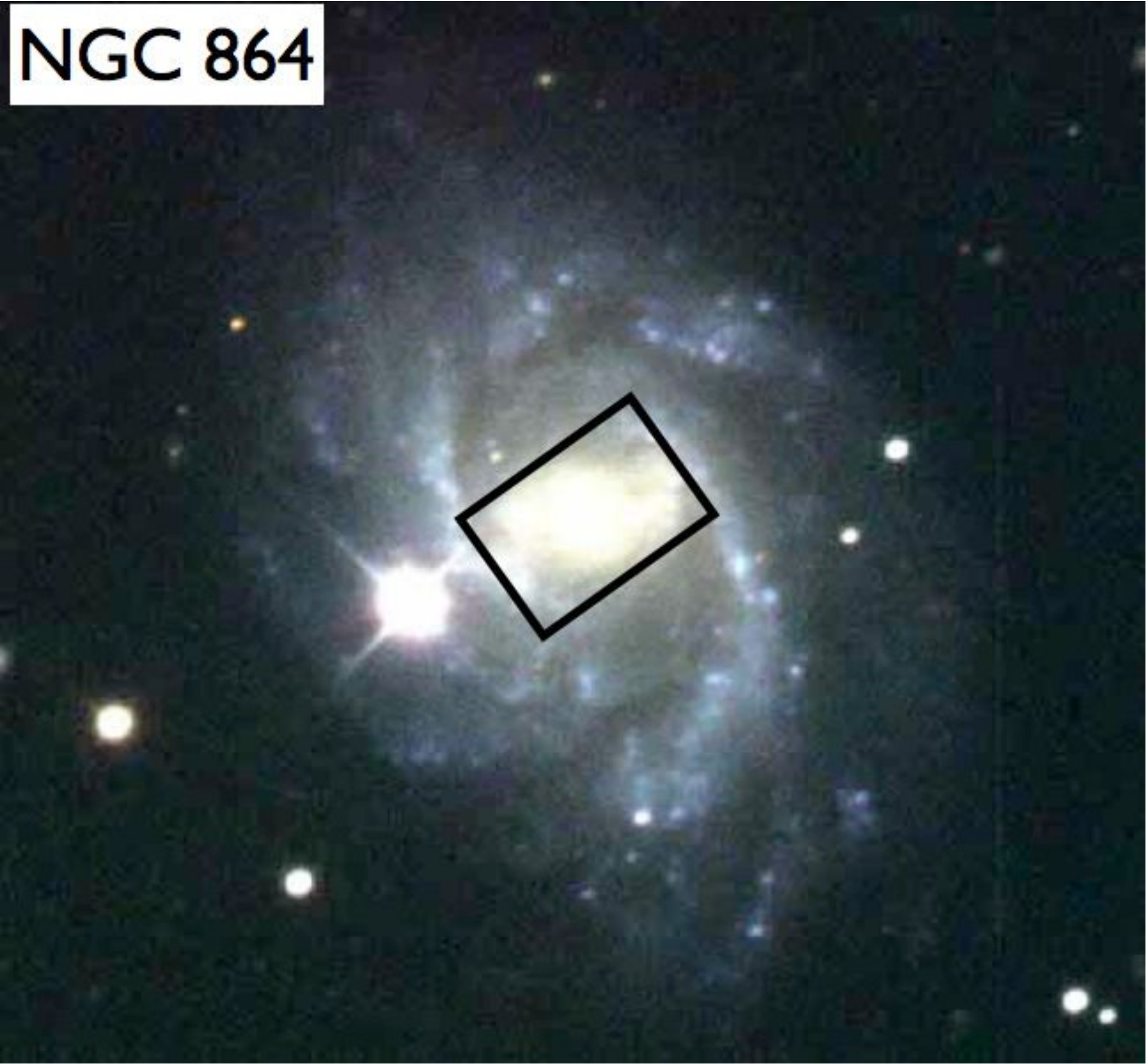}
\includegraphics[width=0.235\textwidth]{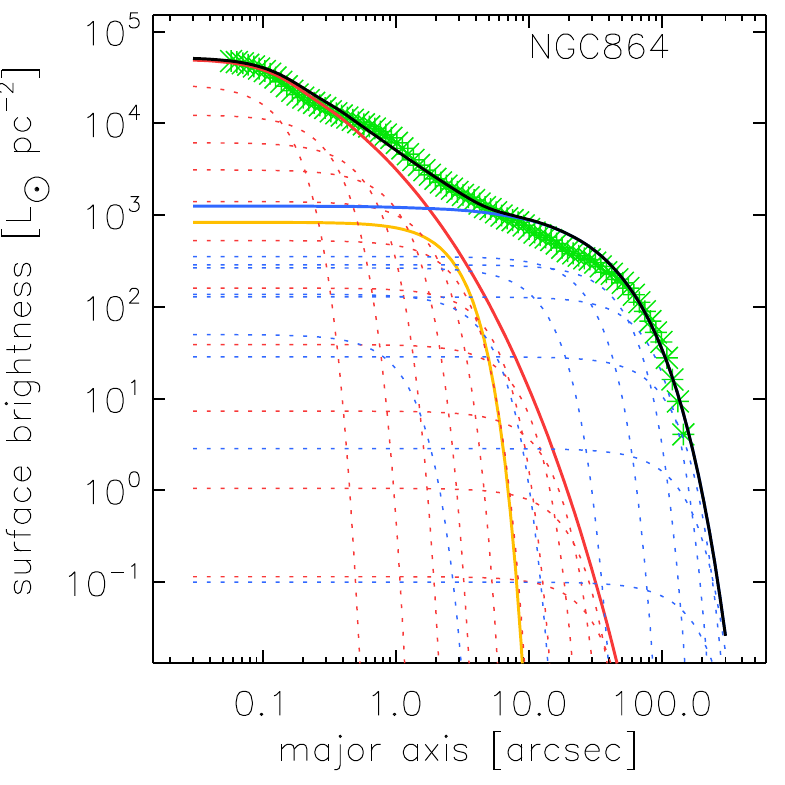}
}
\caption{ Multi-colour SDSS images, with the \sauron\ fields-of-view overlaid (black rectangles), and multi-Gaussian expansion models of the galaxies NGC\,628 (left two panels) and NGC\,864 (right two panels). Green asterisks represent the observed luminosity profiles. The black thick curves show the sum of the individual Gaussians (dotted curves) of the three components: nucleus (yellow), S{\'e}rsic bulge (red) and exponential disc (blue). }
\label{fig:mge}
\end{figure*}

To infer the mass distributions of the 18 \sauron\ galaxies, we use the photometric bulge-disc decompositions of the total H-band surface brightness profile, as presented in \citet{Ganda-PhDT2007} and G09 (see also Sec. \ref{SS:nirimaging}).  These models represent each galaxy's profile as the superposition of an exponential disc and a S{\'e}rsic bulge. 

The galaxies in our sample can contain a significant amount of interstellar dust, which is mostly transparent at near-infrared wavelengths.  The extinction $A_H$ is roughly a factor of 8 lower than the extinction $A_V$ \citep{Rieke1985}, so we do not apply an internal extinction correction. Additionally, the near-infrared is considered as a good tracer of the stellar mass in galaxies as the light is dominated by old stars \citep{Frogel1988,Rix1995,Lilly1989}, although there is some concern that this might be not true due to uncertain influence of intermediate age populations (\citealt{Maraston2005, Conroy2009, Melbourne2012, Zibetti2009,Zibetti2013}).

\cite{Ganda-PhDT2007} and G09 use a one-dimensional MGE decomposition \citep{Emsellem1994} of the bulge and disk profiles, using the implementation of \cite{Cappellari2002}. They adopt 10 Gaussians to represent the discs and between 13 and 19 Gaussians to represent the bulges.  The fits to the one-dimensional profiles are excellent for all galaxies, but the MGE is not applied to the two-dimensional images.

Most of the galaxies (16 out of 18) display a clear light excess above the S{\'e}rsic fit to the bulge, which can be attributed to a bright nuclear star cluster \citep{Boeker2002}, where a luminous mass contributes to the kinematics in the inner regions. For these galaxies, we amend the MGE decomposition of \citep[][Chapter 5; Sec.5.4.]{Ganda-PhDT2007} by including one additional, circular Gaussian to accurately represent the light excess in the centre.  

Figure~\ref{fig:mge} shows the MGE light models of two representative galaxies in our sample. The green asterisks indicate their observed light profiles.  The black thick curves represent the sum of the individual gaussians (dotted lines) of the three components: nucleus (yellow), S{\'e}rsic bulge (red), and exponential disc (blue), which are marked in Appendix \ref{A:mgeTab} with index of 0, 1 and 2, respectively.
 NGC\,628 (left panels) is an example of a very good fit to the data, which is typical for the majority of our galaxies. However, there are a few exceptions with mismatches due to bars or prominent spiral arms. These are NGC\,864 (Figure~\ref{fig:mge}, right panel), NGC\,772, and NGC\,1042. Nevertheless, we considered the MGE fits to these three galaxies to be satisfactory for our needs, because fitting non-symmetric features could lead to uneven representation of the galaxies' surface brightness profiles, and hence, gravitational potentials.
 We converted the resulting peak surface brightnesses of the Gaussians into physical units of $L_{\odot}~\mbox{pc}^{-2}$, taking the absolute magnitude of the Sun $M_{H,\sun}=3.32$ \citep{Binney1998}.  
\subsection{Stellar kinematic maps}
\label{SS:stellarkinematics}
We measured the stellar kinematics using the penalised pixel-fitting (pPXF) method of \citet{Cappellari2004}.  For spectral templates, we used a sub-sample of the MILES stellar library \citep{Sanchez-Blazquez2006, Falcon-Barroso2011}, containing $\sim$\,115 stars that span a large range in atmospheric parameters such as surface gravity, effective temperature and metallicity. We convolve the MILES models of stars from their original spectral resolution to that of the data. This is done by convolving with a Gaussian with dispersion equal to the difference in quadrature between the final and starting resolution. The pPXF method fits a non-negative linear combination of these stellar template spectra, convolved with the Gaussian velocity distribution, to each galaxy spectrum by chi-square minimisation.  The higher-order Gauss-Hermite moments h3 and h4 are not included in this fit as free parameters. The spectral regions affected by emission lines were  masked out during this process. A low-order polynomial (typically sixth degree) is included in the fit to account for small differences in the flux calibration between the galaxy and the template spectra.  This analysis yields the stellar mean velocity ($V$) and stellar velocity dispersion ($\sigma$) for each bin. Their errors are estimated through Monte-Carlo simulations with noise added to the galaxy's best fitting model spectrum.

The \sauron\ instrumental resolution is 105 \kms while the measured velocity dispersions of our galaxies can be significantly below that level. \citet{Emsellem2004} used Monte-Carlo simulations to show that the intrinsic velocity dispersion is still well recovered.  For a spectrum with signal-to-noise of $\approx{60}$ and $\sigma\approx{50}$ \kms, the pPXF method will give velocity dispersions differing from the intrinsic ones by $\sim 10$ \kms, consistent with the measured errors. We might even expect larger uncertainties for some of our galaxies (e.g., NGC\,3346, NGC\,4487 and NGC\,5585), which own central velocity dispersions below this limit. 
On the other hand, recent papers put forward evidence that our adopted approach could accurately recover the intrinsic velocity dispersion to better than 10\% precision given the simulations of \citealt{Toloba2011} and \citealt{Rys2013} and our high S/N ($\sim 60$) in each Voronoi bin. 

In Fig.~\ref{fig:maps1} we show the stellar kinematic maps of our sample of 18 late-type spiral galaxies resulting from the pPXF fits. The first column shows the stellar flux in arbitrary units on a logarithmic scale. The next two columns display $V$ and $\sigma$ maps in \kms\ respectively. We over-plotted the surface brightness contours\footnote{ Note here that the surface brightness contours are not at the same levels as the MGE contours.} of the galaxies as derived from their intensity maps.

For some of our galaxies, the centre position was not accurately determined during the data reduction due to foreground stars and dust lanes, which in some cases caused a significant offset in the measurement of the systemic velocity V$_{\mathrm{sys}}$. Therefore, we use the velocity field symmetrisation method described in Appendix A of \citet{Bosch2010} to estimate a robust V$_{\mathrm{sys}}$. This method assumes that the velocity field (1) is symmetric with respect to the galaxy centre, (2) is uncorrelated and (3) varies linearly along the spatial coordinates.   Then, for each position that has a counterpart, it computes their weighted mean velocities and combined errors (rejecting data with errors $>2.0$~km~s$^{-1}$). In this way we obtain a robust estimate of V$_{\mathrm{sys}}$ for each galaxy (see Table~\ref{tab:prop}).

Galaxies with more concentrated light distributions generally have larger central peaks in their velocity dispersion fields. Elliptical galaxies usually have radially decreasing $\sigma$ \citep[][]{D'Onofrio1995}, and this is also the case for many early-type spirals, as a result of a centrally concentrated bulge. But for very late-type spirals, we expect the $\sigma$ field to be either flat (due to their lower bulge-to-disc ratios) or with a central decrease, because of cold components or counter-rotating discs (\citealt{Falcon-Barroso2006}). In comparison with early-type galaxies, the bulges of the late-type spirals are smaller and have lower surface brightness \citep[][G09]{Yoachim2006}.
\cite{Martinsson2013-130} also find that $\sigma$ generally decreases with radius with an $e$-folding length that is twice the photometric disc scale length. Thus, the variety seen in $\sigma$ profiles of \sauron\ galaxies might be also due to their limited radial coverage, ranging between 0.4 and 1.2
scale lengths.

%
\begin{figure*}
{\includegraphics[width=1.0\textwidth]{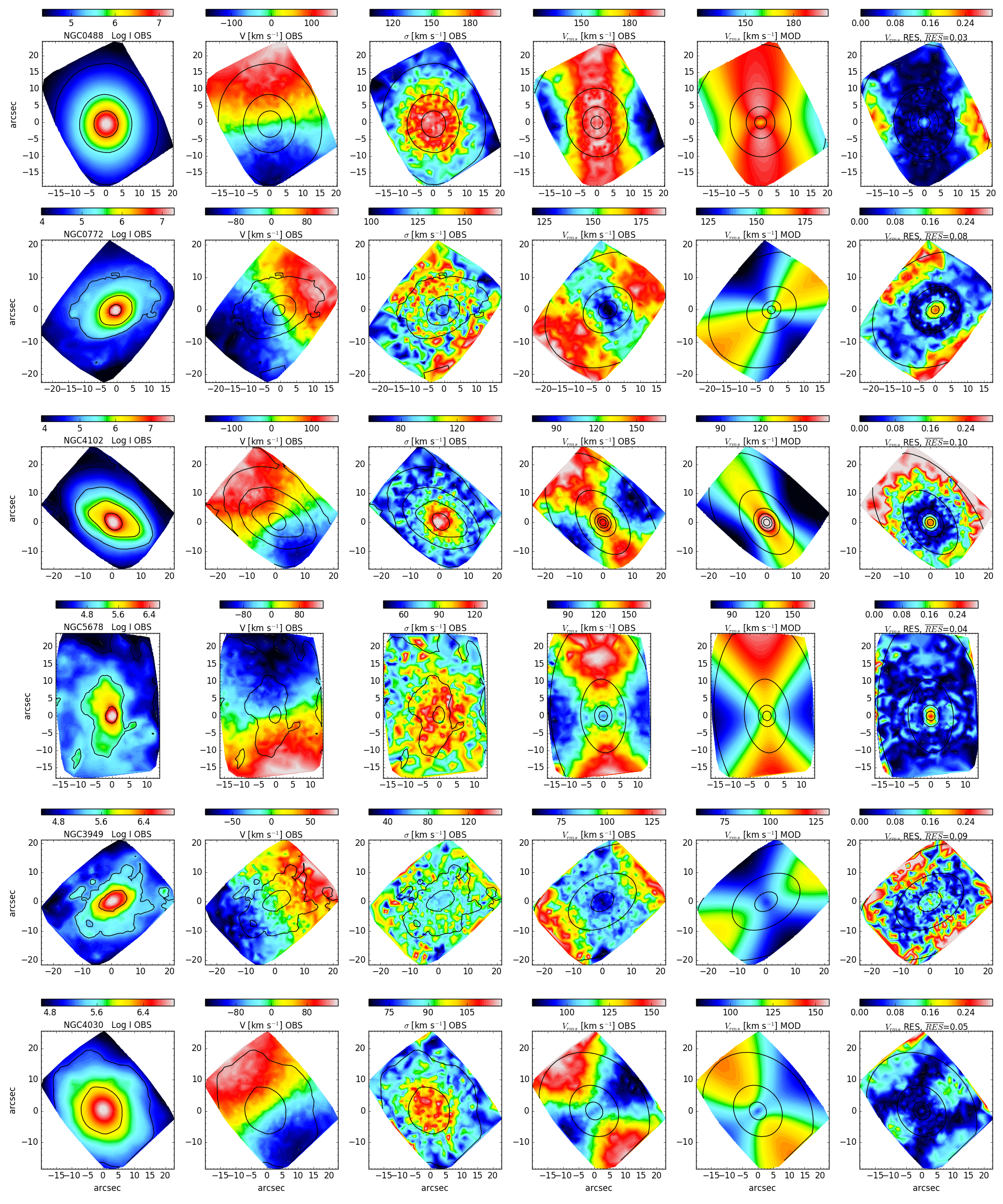}}
\caption{Stellar kinematics and second moment maps of our sample of 18
 galaxies. First column: stellar flux derived from the collapsed data cubes; 
 second and third column: observed mean line-of-sight velocity and velocity
 dispersion; fourth-sixth columns with overplotted MGEs contours: 
 second moment maps $\Vrms = \sqrt{V^2+\sigma^2}$ of the data and
 the model, and the residual maps $\emph{RES}$=$|(V_{\mathrm{rms,obs}}/V_{\mathrm{rms,mod}})-1|$ with median values
 $\overline{\emph{RES}}$.
 }
\label{fig:maps1}
\end{figure*}

\addtocounter{figure}{-1}
\addtocounter{subfigure}{1}

\begin{figure*}
{\includegraphics[width=1.0\textwidth]{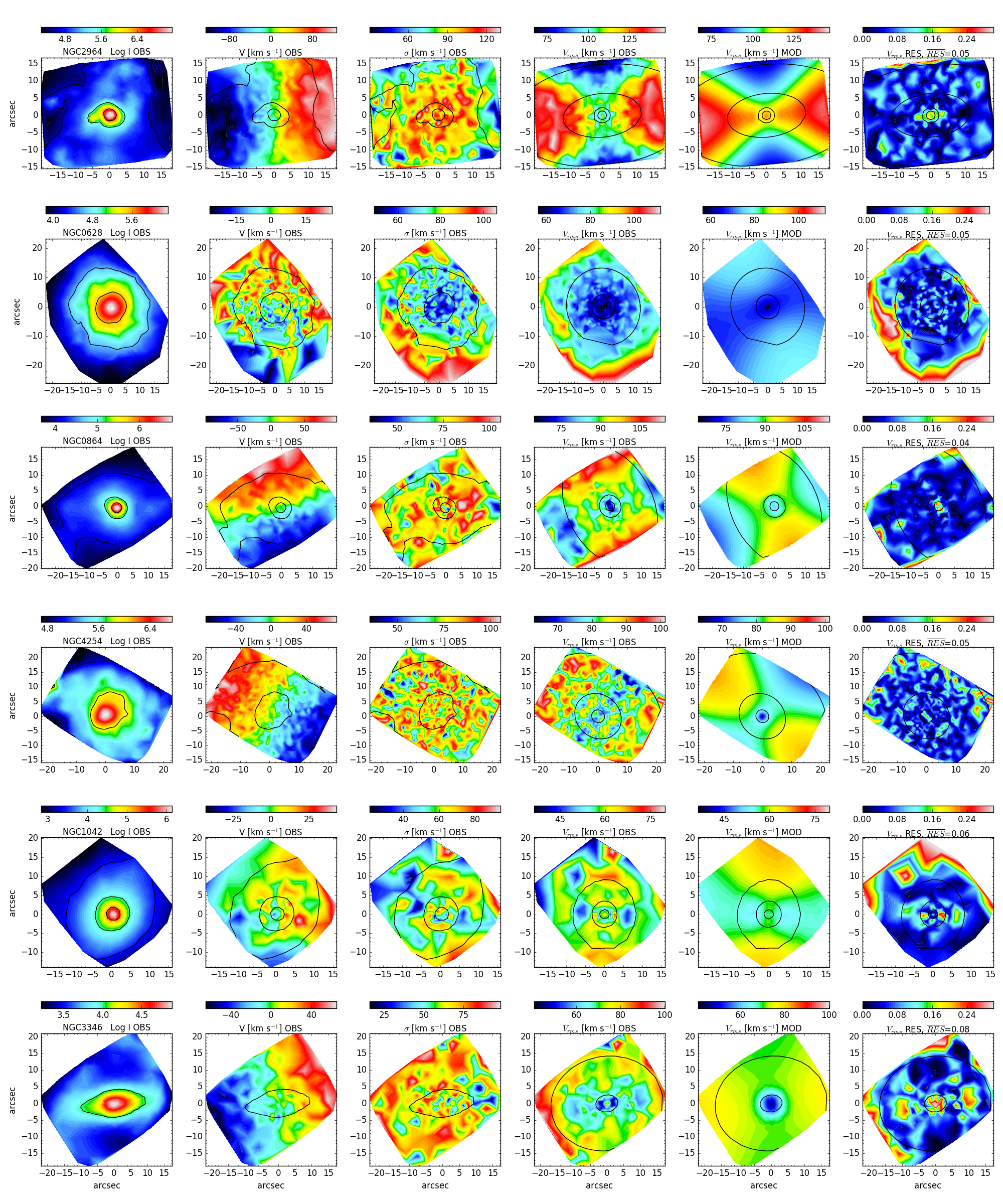}}
\caption{{\it -- continuation}}
\label{fig:maps2}
\end{figure*}

\addtocounter{figure}{-1}
\addtocounter{subfigure}{1}

\begin{figure*}
{\includegraphics[width=1.0\textwidth]{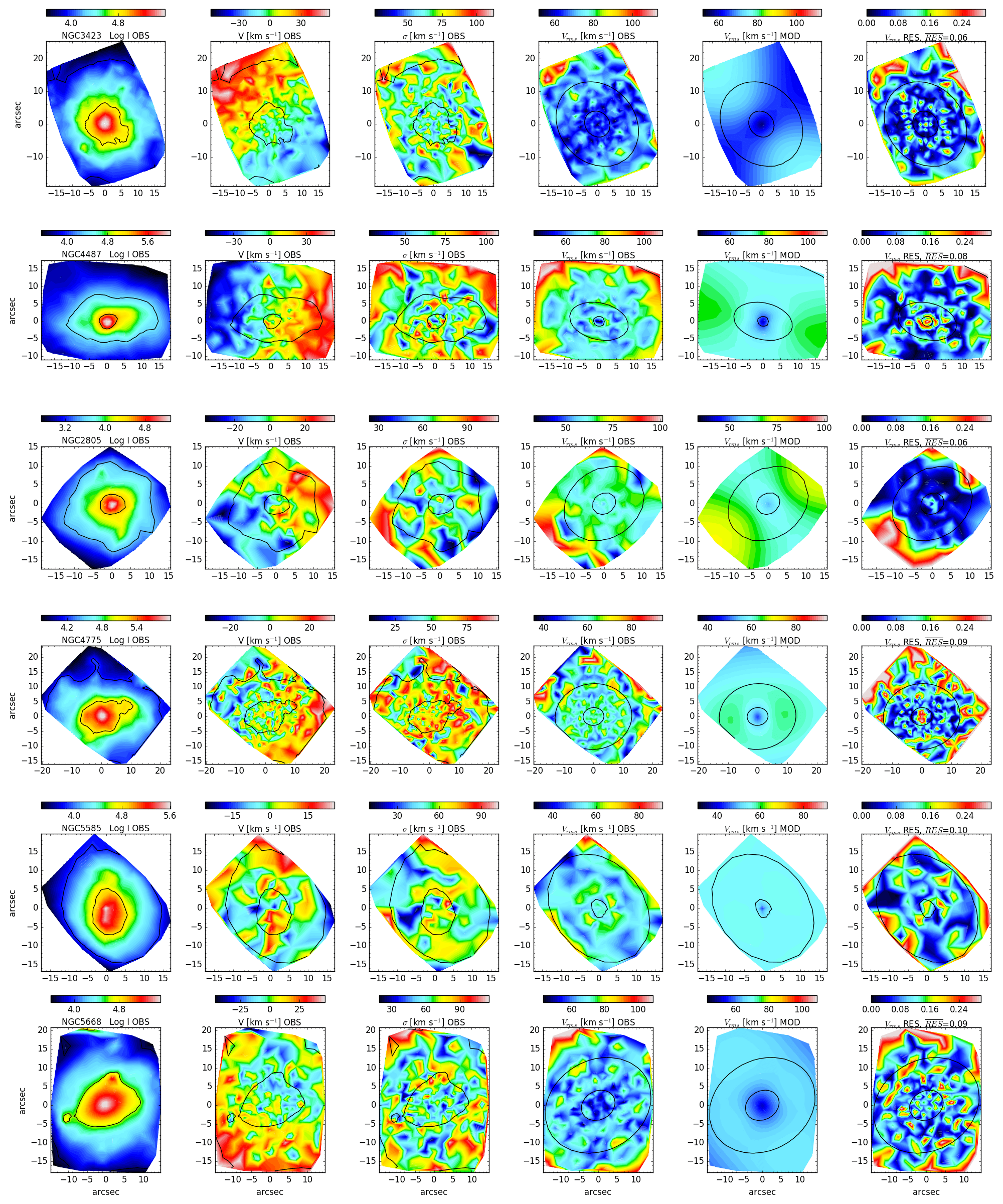}}
\caption{ {\it -- continuation}}
\label{fig:maps3}
\end{figure*}
\begin{figure*}
{\includegraphics[width=1.0\textwidth]{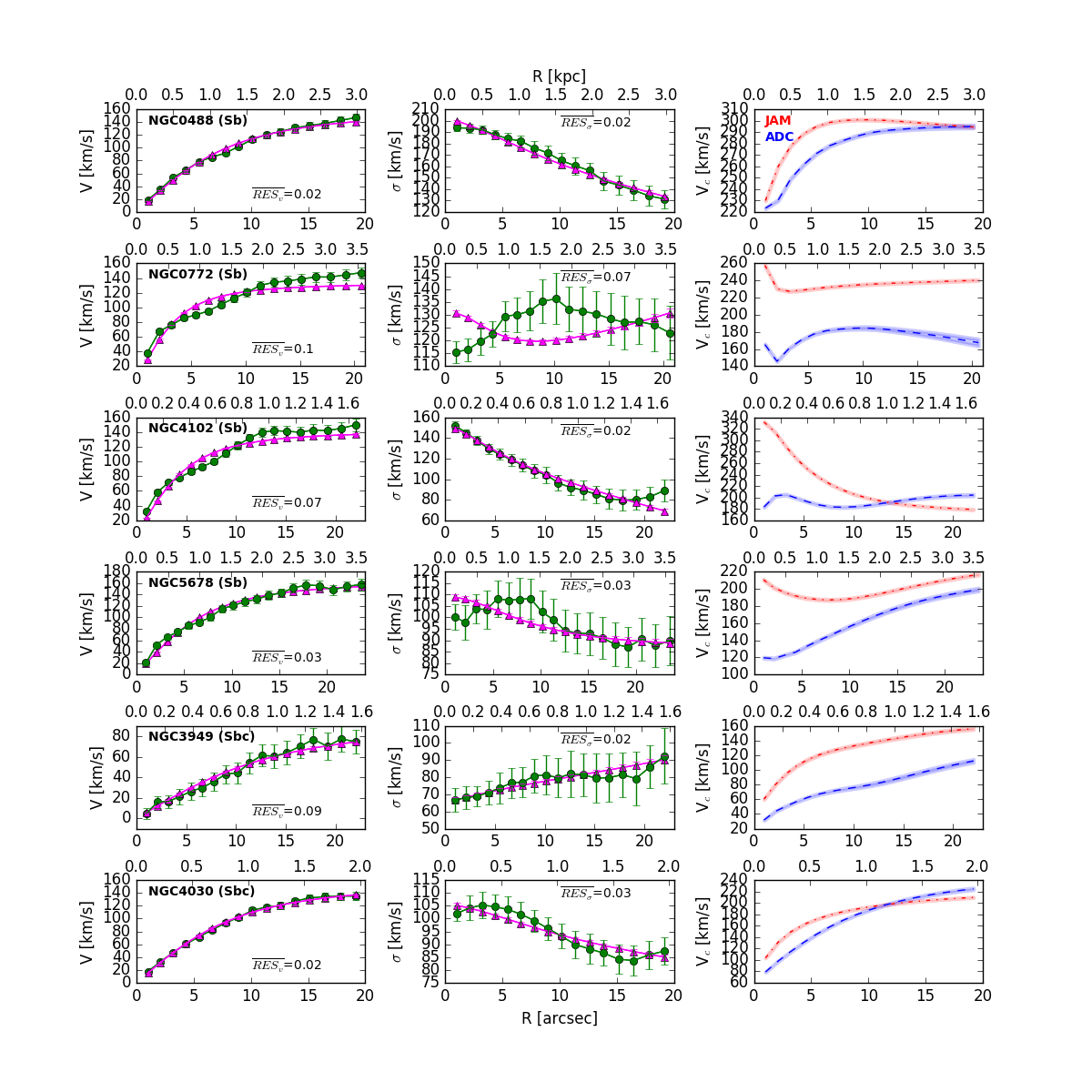}}
\caption{Rotation curves of our sample of 18
 galaxies using ADC-MCMC and JAM-MCMC methods. First column: stellar mean line-of-sight velocity $V$ (green filled circles) and its MCMC fit of simple analytic functions (magenta filled triangles); second column: stellar line-of-sight velocity dispersion $\sigma$ (green filled circles) and its MCMC fit of simple analytic functions (magenta filled triangles); third column: circular velocity curves derived from ADC-MCMC (blue dashed line) and JAM-MCMC (red dashed-dot line) methods. The bands indicate the uncertainties of the presented circular curves. There is a significant and systematic offset between the two models, where ADC gives lower values of the circular velocity curves than JAM.
 }
\label{fig:Vc_a}
\end{figure*}

\addtocounter{figure}{-1}
\addtocounter{subfigure}{1}

\begin{figure*}
{\includegraphics[width=1.0\textwidth]{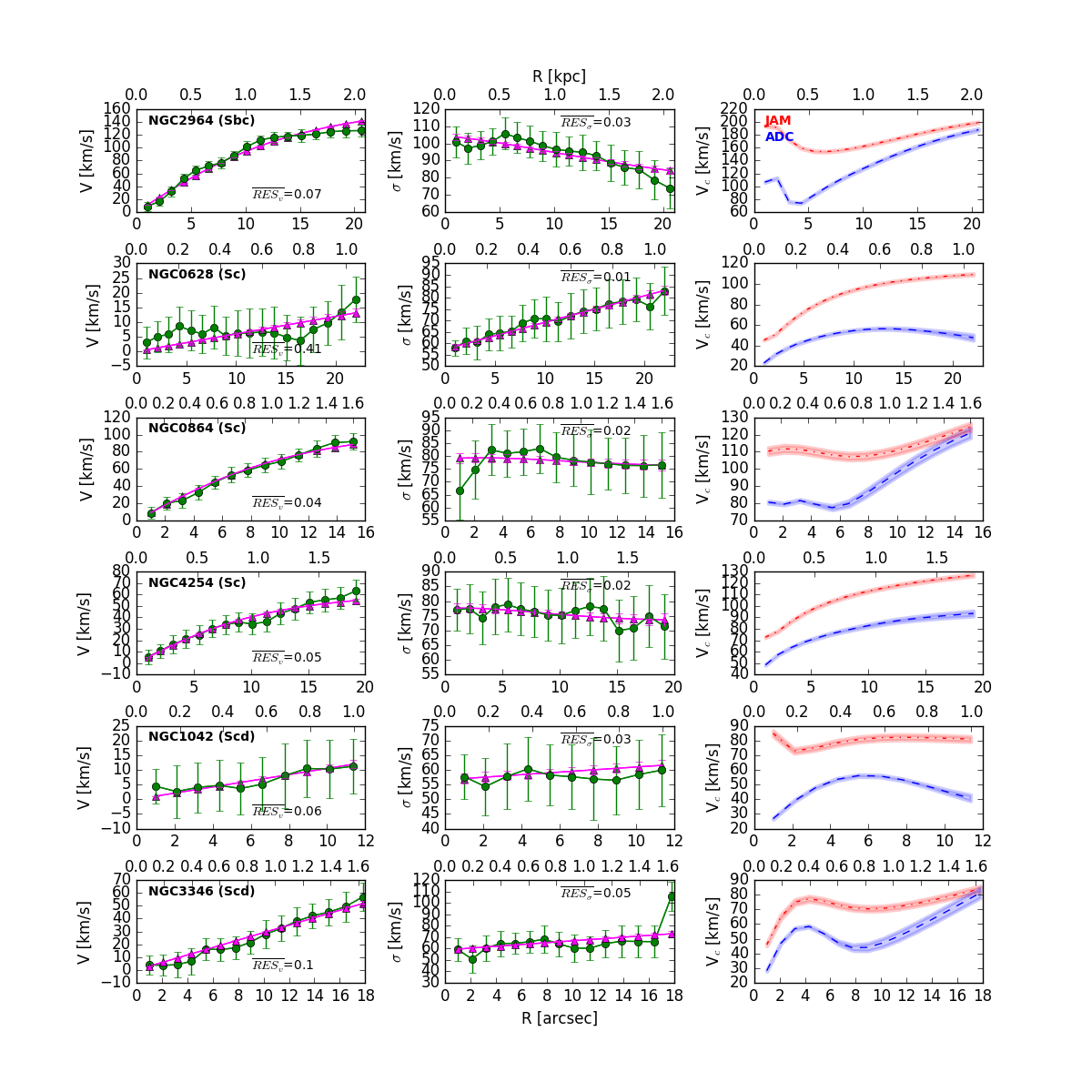}}
\caption{{\it -- continuation}}
\label{fig:Vc_b}
\end{figure*}

\addtocounter{figure}{-1}
\addtocounter{subfigure}{1}

\begin{figure*}
{\includegraphics[width=1.0\textwidth]{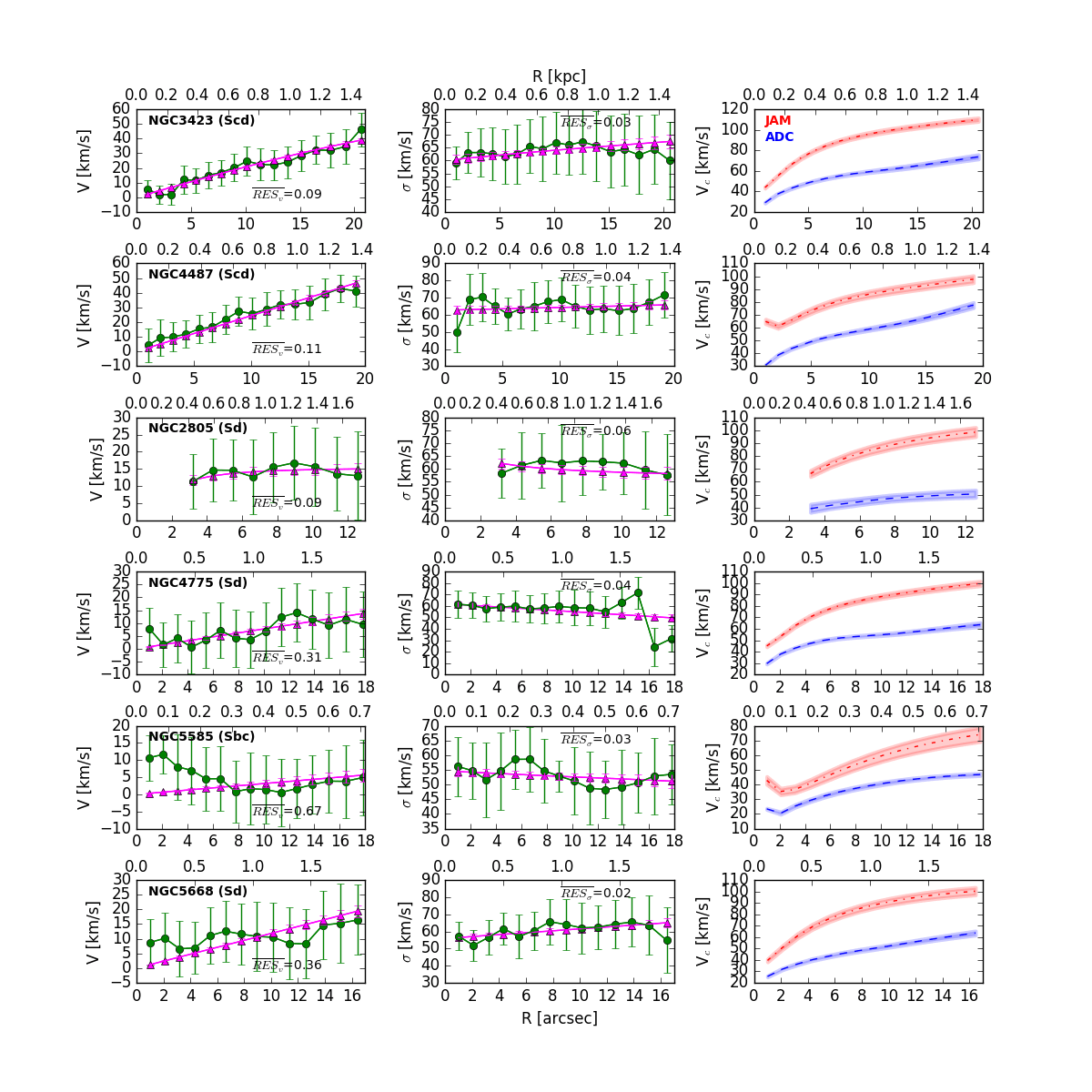}}
\caption{ {\it -- continuation}}
\label{fig:Vc_c}
\end{figure*}
\subsection{Markov Chain Monte Carlo}
\label{SS:MCMC}
To obtain reliable circular velocity curves and associated uncertainties, we assume these properties are random variables viewed in a Bayesian framework. We then applied Markov chain Monte Carlo (MCMC) to estimate these variables under both the ADC and JAM model approaches. We used the
$\texttt{EMCEE}$ code of \cite{Foreman-Mackey2013}, an implementation of an affine invariant ensemble sampler for the MCMC method of parameter estimation, together with JAM code (\citealt{Cappellari2008}) and our own ADC routines all of which are implemented in $\texttt{PYTHON}$.  This approach allows a robust understanding of the uncertainties in the modelling combined with their dependance on assumptions, which is particularly important since we ultimately compare ADC and JAM results.

\subsubsection{ADC-MCMC modeling}
\label{SS:radialprofiles}

\begin{table*}
 \begin{minipage}{170mm}
 \caption{Summaries of the posterior distribution for the ADC model.  The value is the median of the distribution and the uncertainties are the 25th and 75th percentiles of the data corresponding to the median absolute deviation. Columns list: (1) Galaxy
  identifier; (2) Hubble type (NED); (3) and (4) $\overline{v_\phi}$ power-law fit with  $v_\infty$ in \kms\ and $R_c$ in arcsec 
 (see Eq.~\ref{eq:modvphisimple}); 
(5) $\overline{v_\phi}$ linear fit with slope $k_v$ (see Eq.~\ref{eq:vphilin});
(6) and (7) $\sigma_R$ linear fit to the deprojected line-of-sight velocity dispersion
 with y-intercept $\sigma_0$ in \kms\ and slope $k_\sigma$ (see Eq.~\ref{eq:fit1});
 (8) azimuthal velocity anisotropy; (9)  Goodness of ADC-MCMC fits (Sec. \ref{SS:radialprofiles}})
  \begin{center}
   \tabcolsep=1.5mm 
      \begin{tabular}{|*{2}{l}|*{6}{r}|*{1}{r}|}
      \hline
  \multicolumn{9}{|c|}{Best fit parameters}      \\
  \hline  
 \multicolumn{2}{|c|}{}  & \multicolumn{7}{c|}{ADC-MCMC method}  \\ 
\hline
NGC  & Type &  $v_\infty$ &  $R_c$  &  $k_v$  & $\sigma_0$  & $k_\sigma$ & $\beta_z^{\mathrm{ADC}}$  & Good Fit \\
 (1)      &  (2)  &  (3)  &      (4) &    (5) &    (6) &  (7)   & (8)    &   (9)   \\
\hline
%
NGC0488 & SA(r)b & $234.648^{+2.532}_{-2.546}$ & $9.572^{+0.189}_{-0.194}$ & $--$ & $240.755^{+0.652}_{-0.657}$ & $-2.972^{+0.149}_{-0.150}$ & $0.529^{+0.015}_{-0.014}$ & Yes \\
NGC0772 & SA(s)b & $170.975^{+1.610}_{-1.670}$ & $4.484^{+0.119}_{-0.117}$ & $--$ & $153.893^{+0.672}_{-0.646}$ & $2.256^{+0.208}_{-0.200}$ & $0.724^{+0.030}_{-0.031}$ & Yes \\
NGC4102 & SAB(s)b & $165.69^{+1.750}_{-1.626}$ & $6.008^{+0.145}_{-0.139}$ & $--$ & $145.888^{+0.674}_{-0.667}$ & $-3.031^{+0.116}_{-0.118}$ & $-0.398^{+0.067}_{-0.069}$ & Yes \\
NGC5678 & SAB(rs)b & $187.225^{+2.392}_{-2.344}$ & $8.57^{+0.260}_{-0.247}$ & $--$ & $111.533^{+0.656}_{-0.649}$ & $0.059^{+0.166}_{-0.161}$ & $0.185^{+0.121}_{-0.125}$ & Yes \\
NGC3949 & SA(s)bc & $115.368^{+3.287}_{-3.250}$ & $15.897^{+0.572}_{-0.585}$ & $--$ & $61.838^{+0.667}_{-0.667}$ & $1.51^{+0.208}_{-0.197}$ & $-0.296^{+0.158}_{-0.166}$ & Yes \\
NGC4030 & SA(s)bc & $226.882^{+2.808}_{-2.731}$ & $10.02^{+0.230}_{-0.230}$ & $--$ & $108.476^{+0.651}_{-0.653}$ & $-0.609^{+0.112}_{-0.104}$ & $0.089^{+0.039}_{-0.041}$ & Yes \\
NGC2964 & SAB(r)bc & $210.274^{+4.006}_{-3.998}$ & $16.415^{+0.496}_{-0.506}$ & $--$ & $93.049^{+0.666}_{-0.677}$ & $-0.497^{+0.161}_{-0.148}$ & $-0.996^{+0.195}_{-0.185}$ & Yes \\
NGC0628 & SA(s)c & $--$ & $--$ & $0.961^{+0.107}_{-0.105}$ & $59.047^{+0.680}_{-0.708}$ & $1.197^{+0.142}_{-0.138}$ & $0.086^{+0.072}_{-0.078}$ & No \\
NGC0864 & SAB(rs)c & $152.711^{+4.104}_{-4.050}$ & $12.886^{+0.511}_{-0.506}$ & $--$ & $78.378^{+0.682}_{-0.682}$ & $0.322^{+0.273}_{-0.262}$ & $-0.032^{+0.136}_{-0.146}$ & Yes \\
NGC4254 & SA(s)c & $92.836^{+3.071}_{-3.164}$ & $12.984^{+0.603}_{-0.600}$ & $--$ & $76.031^{+0.691}_{-0.664}$ & $0.17^{+0.194}_{-0.194}$ & $-0.078^{+0.105}_{-0.109}$ & Yes \\
NGC1042 & SAB(rs)cd & $--$ & $--$ & $1.432^{+0.169}_{-0.175}$ & $53.823^{+0.714}_{-0.681}$ & $0.422^{+0.223}_{-0.224}$ & $-0.225^{+0.170}_{-0.186}$ & Yes \\
NGC3346 & SB(rs)cd & $--$ & $--$ & $4.387^{+0.160}_{-0.160}$ & $57.403^{+0.701}_{-0.687}$ & $0.782^{+0.217}_{-0.220}$ & $-0.082^{+0.134}_{-0.150}$ & Yes \\
NGC3423 & SA(s)cd & $125.645^{+4.941}_{-4.882}$ & $39.504^{+0.641}_{-0.615}$ & $--$ & $56.791^{+0.710}_{-0.700}$ & $0.409^{+0.191}_{-0.184}$ & $-0.215^{+0.135}_{-0.141}$ & Yes \\
NGC4487 & SAB(rs)cd & $--$ & $--$ & $3.025^{+0.125}_{-0.126}$ & $60.728^{+0.649}_{-0.638}$ & $0.156^{+0.212}_{-0.200}$ & $-0.183^{+0.229}_{-0.245}$ & Yes \\
NGC2805 & SAB(rs)d & $22.525^{+2.492}_{-2.490}$ & $2.635^{+0.629}_{-0.627}$ & $--$ & $67.312^{+0.657}_{-0.662}$ & $-0.097^{+0.288}_{-0.295}$ & $-0.001^{+0.122}_{-0.136}$ & Yes \\
NGC4775 & SA(s)d & $--$ & $--$ & $1.380^{+0.198}_{-0.201}$ & $58.592^{+0.696}_{-0.704}$ & $-0.654^{+0.208}_{-0.205}$ & $-0.176^{+0.141}_{-0.150}$ & No \\
NGC5585 & SAB(s)d & $--$ & $--$ & $0.396^{+0.131}_{-0.126}$ & $49.839^{+0.639}_{-0.638}$ & $-0.173^{+0.174}_{-0.152}$ & $-0.558^{+0.231}_{-0.204}$ & No \\
NGC5668 & SA(s)d & $--$ & $--$ & $1.966^{+0.201}_{-0.203}$ & $53.332^{+0.696}_{-0.666}$ & $0.538^{+0.236}_{-0.235}$ & $-0.153^{+0.129}_{-0.135}$ & No \\
\hline
 \end{tabular}
  \end{center}
  \label{tab:adcMC}
\end{minipage}
\end{table*}
To infer circular velocity curves for our sample of galaxies using the ADC method (Sect.~\ref{SS:adc}),  
we first need their kinematic profiles along the projected major axis. To this end, we use the \texttt{kinemetry} package of \citet{Krajnovic2006} 
that is based on harmonic expansion of two-dimensional maps along ellipses. We extract the observed mean velocity $V_\mathrm{maj}$ and velocity 
dispersion $\sigma_\mathrm{maj}$ profiles along ellipses with  fixed axis ratio of the galaxies ($q=1-\epsilon$) and photometric position angle, where $\epsilon$ and $PA$ are shown in 
Table \ref{tab:prop}.
However, for galaxies NGC3346, NGC4775 and NGC5668, we calculate the axis ratio $q$ from the adopted inclination $\emph{i}=\emph{i}^{MGE}$ using eq. \ref{eq:incl} (see Sec. \ref{SS:sample}) .

Then under the 'thin-disc' assumption we obtain the intrinsic mean velocity $\overline{v_\phi}$ and radial velocity dispersion $\sigma_R$ profiles from the observed profiles as:
\begin{eqnarray}
  \label{eq:obsV}
 V_\mathrm{maj}  & = & V_\mathrm{sys} + \overline{v_\phi}\sin i,
  \\ 
  \label{eq:obssig}
  \sigma_\mathrm{maj}^2 & = & \sigma_\phi^2\sin^2i + \sigma_z^2\cos^2i
  \nonumber
  \\
  & = & \sigma_R^2 \left[ \frac{(1+\alpha_R)}{2}\sin^2i +(1- \beta_z) \cos^2i \right],
 \end{eqnarray}
with systematic velocity $V_\mathrm{sys}$ and adopted inclination $i$ from Table \ref{tab:prop}. To get to equation~\eqref{eq:obssig}, we have used $\sigma_\phi^2/\sigma_R^2 = (1+\alpha_R)/2$ under the assumptions that the velocity ellipsoid is aligned with the cylindrical coordinate system and symmetric around $v_\phi = \overline{v_\phi}$ (see Section~\ref{SS:adc}). As before, $\beta_z = 1 - \sigma_z^2/\sigma_R^2$ is the velocity anisotropy in the meridional plane, and $\alpha_R = \partial \ln \overline{v_\phi} / \partial \ln R$ is the radial logarithmic gradient of the intrinsic mean velocity.\\

Given the assumptions of the ADC approach, we relate the model parameters to the observed data as follows:
\begin{eqnarray}
   \label{eq:obsVMC}
  V = \overline{v_\phi}\sin i,
   \\
   \label{eq:obssigMC}
   \sigma = \sigma_R \sqrt{0.5(1+\alpha_R)\sin^2i+(1 - \beta_z^{\mathrm{ADC}}) \cos^2i},
 \end{eqnarray}
where $V = V_\mathrm{maj} - V_\mathrm{sys}$ and $\sigma  =\sigma_\mathrm{maj}$.
To avoid numerical noise in the derivatives, we express $\overline{v_{\phi}}$ and $\sigma_R$ using a smooth, analytic functional forms and then compute the derivatives explicitly. 

As an analytical representation for $v_{\phi}$, we use the 'power-law' prescription of \citet[][eq.2.11]{Evans1994}, which in the equatorial plane ($z = 0$) and assuming flat rotation curve, becomes
\begin{equation}
  \label{eq:modvphisimple}
  \overline{v_\phi}(R) = \frac{v_{\infty} R}{\sqrt{R_c^2+R^2}},
\end{equation}
where $v_{\infty}$ is the asymptotic velocity and $R_c$ is the core
radius. This model describes a rotation curve that increases linearly
with radius $\propto (v_{\infty}/R_c)\, R$ when $R \ll R_c$, and
flattens to the value $v_{\infty}$ when $R \gg R_c$. 
However, in several cases the velocity profile shows little indication of flattening the observed region.  In such cases, the core radius $R_c$ is not well constrained and fitting Eq.~\eqref{eq:modvphisimple} would lead to unphysical values for both $R_c$ and $v_{\infty}$. Such galaxies are in the regime of $R \ll R_c$  and we adopt a linear solid-body rotation instead:
\begin{equation}
  \label{eq:vphilin}
  \overline{v_\phi}(R) = k_v \, R,
\end{equation}
where $k_v$ is the linear coefficient. \\
For $\sigma_R$, we use a linear expression of the form
\begin{equation}
\label{eq:fit1}
\sigma_R(R) = \sigma_0 + k_\sigma \, R
\end{equation}
where $\sigma_0$ and $k_\sigma$ are free parameters. We find the linear fits to the $\sigma_R$ profiles are broadly representative, although the model does not reproduce the central part of the $\sigma_R$ profiles for some of the galaxies (e.g., NGC\,772, NGC\,5678 and NGC\,864) due to the marked dips in $\sigma$ in the central parts of the galaxies (\citealt{Falcon-Barroso2006}). 
Such a descrease in the observed stellar velocity dispersion might be the result of a small dynamically cold component or counter-rotating disc of high surface brightness that obscures the bulge, dominating the luminosity-derived kinematic measurements (\citealt{Falcon-Barroso2006}) or due to $e$-folding length of our galaxies (see \citealt{Martinsson2013-130}).

To apply a Bayesian framework to the problem, we consider the free parameters of the model to be random variables and then calculate their posterior distributions using MCMC. We model 5 free parameters for the galaxies in the power-law model for $v_{\phi}$ ($v_\infty$, $R_c$, $\sigma_0$, $k_\sigma$,  $\beta_z^{\mathrm{ADC}}$) and 4 free parameters ($k_v$, $\sigma_0$, $k_\sigma$, 
$\beta_z^{\mathrm{ADC}}$) for the galaxies with linear models for $v_{\phi}$. For these vectors of parameters 
$\boldsymbol\theta$, we compute model profiles for $V_{\mathrm{MOD}}$ and  $\sigma_{\mathrm{MOD}}$ using Eq.(\ref{eq:obsVMC}) and (\ref{eq:obssigMC}). These profiles are compared to observed data using a chi-squared statistic, summing over radius positions with normalization set by the uncertainties in the observed data ($\delta V_{\mathrm{OBS}}$ and $\delta \sigma_{\mathrm{OBS}}$):
\begin{eqnarray}
\chi^2 &=& \sum_k [V_{\mathrm{MOD},k}(\boldsymbol\theta) - V_{\mathrm{OBS},k}]^2/(\delta V_{\mathrm{OBS},k}^2)\nonumber\\
&+& \sum_k [\sigma_{\mathrm{MOD},k}(\boldsymbol\theta) - \sigma_{\mathrm{OBS},k}]^2/(\delta \sigma_{\mathrm{OBS},k}^2).
\label{eq:chisq}
\end{eqnarray}
We adopt a log-probability function for the parameters given the data as $\mathcal{L}(\boldsymbol\theta|V_{\mathrm{OBS}},\sigma_{\mathrm{OBS}}) 
= -\chi^2 + \mathbf{p}$, where $\mathbf{p}$ is a set of priors.  We adopt uniform prior distributions for our parameters on fixed ranges:  
$v_\infty \in [0,400]$~km~s$^{-1}$, $R_c\in [0',50'']$ , $k_v \in [0,10]$ km~s$^{-1}$~arcsec$^{-1}$, $\sigma_0 \in [0,300]$~km~s$^{-1}$, 
$k_\sigma \in [-5,5]$~km~s$^{-1}$~arcsec$^{-1}$ and $\beta_z^{\mathrm{ADC}}\in [-1,+1]$. Only for one galaxy (NGC2964) we had to expand 
the range of the walkers to $\beta_z^{\mathrm{ADC}}\in [-1.5,+1]$, since the distribution peaked at $\beta_z^{\mathrm{ADC}}=-0.996$.

We run the ADC-MCMC code with 60 walkers (i.e., members of the ensemble sampler), which have initial position set by the prior information and physical information. 
We use a small random distribution around the parameters: $v_\infty$, $R_c$, $k_v$, $\sigma_0$, $k_\sigma$, 
which are adopted from the least-squares minimisation routine \texttt{mpfit}\footnote{http://www.physics.wisc.edu/~craigm/idl/down/mpfit.pro} in IDL to find the values that minimize Eq. \ref{eq:chisq}. 
We then start the walkers around the optimized values in a normal distribution with scale determined by the uncertainties in the fit. 
The walkers of the parameter $\beta_z^{\mathrm{ADC}}$ were allowed to move freely in the parameter space. On the other hand, the inclination of the galaxies is fixed to the value presented in column 6 of Table \ref{tab:prop}.
The MCMC code samples the posterior distribution, using 400 steps for burn-in and 1000 steps for sampling.   
We find that all walkers converge after $\sim 50$ steps of burn-in to sample a similar distribution indicating a well-sampled unimodal posterior distribution.
We summarize the posterior distributions for the parameters in Table \ref{tab:adcMC} of this paper.
Additionally, we assign a quality flag to each ADC-MCMC model in Table \ref{tab:adcMC}
 depending on the residual of the V profile 
($RES_{v}=|(V_{\mathrm{obs}}/V_{\mathrm{mod}})-1|$) and 
$\sigma$ profile ($RES_{\sigma}=|(\sigma_{\mathrm{obs}}/\sigma_{\mathrm{mod}})-1|$), where the label "Yes/No" correspond to "Good/Bad" fit, respectively. Bad fit label is associated to galaxies (i.e., NGC0628, NGC4775, NGC5585 and NGC5668) if the median value of one of their residuals ($\overline{RES_{v}}$ or $\overline{RES_{\sigma}}$) is larger than 0.3.
The burn-in chains, posterior chains and corner plots of ADC-MCMC method are shown in Appendix \ref{C:chains}.

We present the results of the analysis in Fig.~\ref{fig:Vc_c}.  
In the first column, we show the observed mean line-of-sight velocity 
$V$ (filled green circles) and its fiting function (Eq. \ref{eq:obsVMC}, 
filled magenta triangles). The galaxies are ordered by their morphological 
type from Sb to Sd. We observe that Sb--Sc galaxies (at $R_e/5$) have higher 
observed velocity in contrast to Scd--Sd. We adopt a power-law model function 
to most of the $v_{\phi}$ radial profiles, except for the galaxies --  NGC\,0628, NGC\,1042, NGC\,3346, NGC\,4487, NGC\,4775, NGC\,5585 and NGC\,5668, which require a linear function. 

In the second column of Fig.~\ref{fig:Vc_c}, we present the observed line-of-sight velocity dispersion $\sigma$ profiles 
(filled green circles) and its fiting function (Eq.~\ref{eq:obssigMC}, filled magenta triangles). 
The $\sigma$ profiles are more varied than the $V$ profiles. Some velocity dispersion profiles show an almost-linear 
decrease or increase with radius on the whole radial range available, some are flat, and others have a more complex 
behaviour that seems difficult to reproduce with a linear function.  Sb--Sbc galaxies (at $R_e/5$) are characterised 
by high observed velocity dispersion $\sigma$ with respect to Sc--Sd.  Most of our Sb-Sc galaxies have regular 
velocity fields with well-defined axisymmetric rotation and high amplitude, while Sd velocity fields show more 
complex structure and lower amplitude of rotation. In our sample we have four types of $\sigma$ behaviour: 
decreasing outward (NGC\,4102), increasing outward (NGC\,3949), a central $\sigma$-dip (NGC\,0772) or flat (NGC\,4775).

We estimate the uncertainty of the velocity and velocity dispersion profiles at each radius (see Fig.~\ref{fig:Vc_c}) from the corresponding error maps since the provided uncertainties from \texttt{kinemetry} routine are unrealistically small ($\sim$ 1 km$^{-1}$). Thus, we took the absolute value of the median of the velocity bins and the median of the velocity dispersion bins within each annulus defined by \texttt{kinemetry} routine.
\subsubsection{JAM-MCMC modeling}
\label{SS:axijeansmodels}
As described in Sect.~\ref{S:method}, we use a solution of the axisymmetric Jeans equations based on a Multi-Gaussian Expansion (MGE) of the intrinsic luminosity density to predict the observed second velocity moment $\Vrms = \sqrt{V^2+\sigma^2}$ (\citealt{Cappellari2008}). The solution has a constant meridional plane velocity anisotropy $\beta_z^{\mathrm{JAM}}$ and constant mass-to-light ratio \MLtot\ within the galaxy. Our analysis does not account for the variation of \Vrms\ within each Voronoi bin.  Such variations will cause the averaged $\Vrms$ in a bin to be larger than the intrinsic \Vrms\ along a ray, as evaluated by the JAM code.  We do not anticipate this effect to bias our results significantly: since the surface brightness decreases with $R$, the larger Voronoi bins are found in the outer regions of the galaxies where the variation in the \Vrms\ surface is low.  At small $R$, where the variation with position can be significant, the higher surface brightness means the bins are smaller. 

We symmetrize our observed \Vrms\ fields in order to avoid outliers in the data (see \citealt{Cappellari2015}). In this way, the results will be not influenced by a few deviant values. This is especially important in the case of the MCMC method application, where the best fit \Vrms\ model rely on $\chi^2$-statistics. Moreover, the symmetrization renders the \Vrms\ field axisymmetric which is the approximation consider by JAM.
At the same time we symmetrize the variance of the \Vrms\ field:

\begin{equation}\label{eq:dVrms}
\Delta \Vrms = 2 \sqrt{\mathrm{sym}\left\{\frac{(\Delta V\cdot V)^2 + (\Delta \sigma \cdot \sigma)^2)}{\Vrms^2}\right\}},\\
\end{equation}
where the term $\mathrm{sym}\left\{...\right\}$ indicates the symmetrized variance of \Vrms\ field.\\  

\noindent For this purpose we use the code presented in \cite{Cappellari2015}.
 The procedure interpolates the bin values over a new grid generated by permuting the $j$th-bin coordinates as follow:

\begin{align}\label{eq:xjyj}
x_{j} \rightarrow (x_{j},-x_{j},x_{j},-x_{j}),
\nonumber
\\
y_{j} \rightarrow (y_{j},y_{j},-y_{j},-y_{j}).
\end{align}

\noindent This operation generates three additional velocity fields (the first permutation obtained by $x_{j},y_{j}$ corresponds to the original velocity field). Finally, the symmetrized velocity field is the mean of the four coordinate-permuted velocity fields. Given this, 
equation (\ref{eq:dVrms}) reduces to the common addition of the error in quadrature:


\begin{eqnarray}\label{eq:sqrtVrms}
\Delta \Vrms = 2 \sqrt{ \frac{Q}{4} },
\end{eqnarray}
where $Q=\Delta \Vrms^{2}(x_{j},y_{j})+\Delta \Vrms^{2}(-x_{j},y_{j})+\Delta \Vrms^{2}(x_{j},-y_{j})+\Delta \Vrms^{2}(-x_{j},-y_{j})$, and\\
$\Delta \Vrms^{2} = (\Delta V\cdot V)^2 + (\Delta \sigma \cdot \sigma)^2$ of a certain permutation.\\

\noindent Although the interpolation is precise, the code does not consider the systematic effect that a given bin at $(x,y)$ position not always have a corresponding bin atat the other positions, e.g., (-x,y). To account for this, the final \Vrms\ field error is set by the maximum of either the values from equation (\ref{eq:dVrms}) or half the deviation between neighbors \Vrms\ field bin values (M. Cappellari private communication, see also \citealt{Morganti2013}).\\

We formulate a parallel MCMC approach for determining the posterior parameter distributions of the JAM model.  
In this case, there are only 2 free parameters for all galaxies ($\beta_z^{\mathrm{JAM}}$ and $\Upsilon$). 
Because of the large computational expense of evaluating a JAM model, we run the JAM-MCMC code with only 30 walkers, though the model still converges using 100 steps for a burn-in and 200 steps for sampling. The prior distributions are taken to be uniform with, e.g., 
 $\beta_z^{\mathrm{JAM}}\in [-1,+1]$ and $\Upsilon\in [0,10]~M_{\odot}/L_{\odot}$.

In JAM-MCMC method, we fix the inclination to the values in the column 6 of Table \ref{tab:prop}. 
For galaxies NGC3346, NGC4775, and NGC5668, we adopt $\emph{i}=\emph{i}^{\mathrm{MGE}}$ regarding to 
the minimum possible inclination condition of JAM model (see Sec.~\ref{SS:jam}; eq. {\ref{eq:ilim}}).

We use a similar log-likelihood statistic as we did for the ADC, except in this case, only the model and observed \Vrms\ are compared.  
Summaries of the posterior distribution of parameters for JAM-MCMC model are presented in Table \ref{tab:jamMC}.
 The burn-in chains, posterior chains and corner plots are shown in Appendix \ref{B:chains}.

In the fourth to fifth columns of Fig. \ref{fig:maps1}, we show the observed and the best fitting \Vrms\/ maps with over-plotted MGE contours. 
For all galaxies, the MGE contours are aligned with the surface brightness contours in position angle and ellipticity.  
In the sixth colum, we calculate the residual maps $\emph{RES}$ between the observations ($V_{\mathrm{rms,obs}}$) and 
the models ($V_{\mathrm{rms,mod}}$), where
$RES$=$|(V_{\mathrm{rms,obs}}/V_{\mathrm{rms,mod}})-1|$. The median values of these residual maps 
$\overline{\emph{RES}}$ indicate that the uncertainties of JAM fit ranges between 3 \% and 10 \%. 
Note here that we do not count for systematics errors in these uncertainties. From the posterior distributions of 
JAM-MCMC method, we define the goodness of the fit with the label "Yes/No" for "Good/Bad" fit, respectively. Bad fit labels are associated to galaxies (i.e., NGC\,4030, NGC\,1042, NGC\,3346), which do not have
well defined posterior distributions of the velocity anisotropy $\beta^{\mathrm{JAM}}_z$ (see Appendix \ref{B:chains} and Table \ref{tab:jamMC}).

Further, we find that the JAM produces the best fits to the second velocity moment $\Vrms$ for Sb and Sc spirals 
(e.g.,  NGC\,488, NGC\,5678, and NGC\,2964, characterised by fast-rotating discs) and the model gets worse for Scd and Sd galaxies.   
Given the observed values of $V$ and $\sigma$ show evidence for non-monotonic behaviour (Figure \ref{fig:maps1}), 
the model provides an average representation of the data in the late-type galaxies.  
We stress that the values derived for the parameters are the posterior distributions given the data, subject to the assumptions of the model. 

\begin{table}
 \caption{Summaries of the posterior distribution for the JAM model.  The value is the median of the distribution and the uncertainties are the 
 25th and 75th percentiles of the data corresponding to the median absolute deviation.  Columns are:  
 (1) Galaxy identifier; (2) Hubble type (NED); 
  (3) Azimuthal velocity anisotropy; (4)  Dynamical mass-to-light ratio in \emph{H}-band;
 (5) Goodness of JAM-MCMC fit (Sec. \ref{SS:axijeansmodels})}
  \begin{center}
   \tabcolsep=1.5mm 
      \begin{tabular}{|*{2}{l}|*{2}{r}|*{1}{r}|}
      \hline
  \multicolumn{5}{|c|}{Best fit parameters}      \\
  \hline  
 \multicolumn{2}{|c|}{}  &   \multicolumn{3}{c|}{JAM-MCMC method}  \\ 
\hline
NGC  & Type  & $\Upsilon$  & $\beta^{\mathrm{JAM}}_z$ & Good Fit \\
 (1)      &  (2)  &  (3)  &      (4) &    (5)    \\
\hline
0488 & SA(r)b &     $1.148^{+0.003}_{-0.003}$ & $-0.223^{+0.018}_{-0.017}$  & Yes \\
0772 & SA(s)b &     $1.024^{+0.005}_{-0.005}$ & $-0.628^{+0.035}_{-0.036}$ & Yes \\
4102 & SAB(s)b &   $0.618^{+0.002}_{-0.002}$ & $0.157^{+0.009}_{-0.009}$ & Yes \\
5678 & SAB(rs)b &  $1.237^{+0.008}_{-0.008}$ & $-0.209^{+0.031}_{-0.031}$ & Yes \\
3949 & SA(s)bc &   $1.380^{+0.020}_{-0.018}$ & $0.137^{+0.030}_{-0.032}$ & Yes \\
4030 & SA(s)bc &   $0.743^{+0.002}_{-0.002}$ & $-0.997^{+0.003}_{-0.002}$ & No \\
2964 & SAB(r)bc &  $1.371^{+0.017}_{-0.016}$ & $0.164^{+0.025}_{-0.027}$ & Yes\\
0628 & SA(s)c &     $1.153^{+0.016}_{-0.018}$ & $-0.692^{+0.133}_{-0.134}$ & Yes\\
0864 & SAB(rs)c &   $1.466^{+0.056}_{-0.058}$ & $0.215^{+0.065}_{-0.073}$ & Yes \\
4254 & SA(s)c &    $0.451^{+0.007}_{-0.008}$ & $-0.105^{+0.057}_{-0.060}$ & Yes \\
1042 & SAB(rs)cd & $1.800^{+0.083}_{-0.103}$ & $-0.480^{+0.274}_{-0.265}$ & No \\
3346 & SB(rs)cd &  $2.201^{+0.106}_{-0.133}$ & $-0.480^{+0.285}_{-0.288}$ & No \\
3423 & SA(s)cd &   $2.008^{+0.064}_{-0.069}$ & $-0.310^{+0.152}_{-0.187}$ & Yes \\
4487 & SAB(rs)cd & $2.068^{+0.118}_{-0.119}$ & $0.466^{+0.056}_{-0.059}$ & Yes \\
2805 & SAB(rs)d &  $1.945^{+0.145}_{-0.162}$ & $-0.141^{+0.277}_{-0.332}$ & Yes \\
4775 & SA(s)d &    $1.043^{+0.042}_{-0.040}$ & $0.331^{+0.087}_{-0.119}$ & Yes \\
5585 & SAB(s)d &   $3.538^{+0.482}_{-0.491}$ & $0.317^{+0.163}_{-0.193}$ & Yes \\
5668 & SA(s)d &    $1.222^{+0.081}_{-0.069}$ & $0.684^{+0.036}_{-0.045}$ & Yes \\
\hline
 \end{tabular}
  \end{center}
  \label{tab:jamMC}
\end{table}
\section{Comparing circular velocity curves}
\label{S:vcirc}
\begin{figure*}
  \begin{center}
    \includegraphics[width=1.0\textwidth]{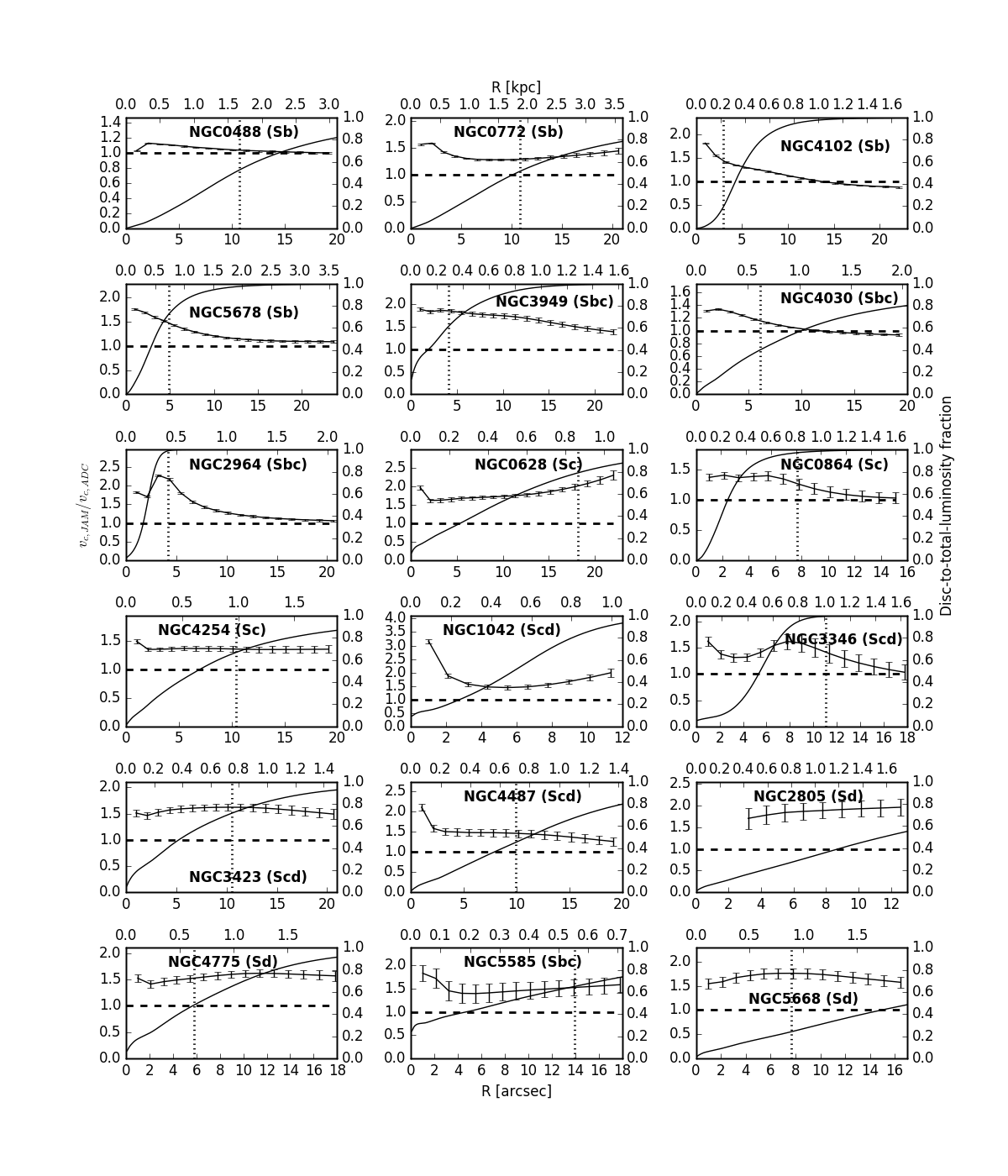}
  \end{center}
  \caption{Velocity ratio between $v_{\mathrm{c,JAM}}$ best fit and  $v_{\mathrm{c,ADC}}$, where the bar indicates the typical uncertainty. Over-plotted are the disc-to-total luminosity fraction (thin solid line), the $R_e/5$ (dotted line) where $R_e$ is the MGE effective radius, and the one-to-one relation of the velocity ratio (dashed line). The significant differences between the two models come from the bulge in the central parts of Sb-Sbc ($<R_e/5$) and possibly a thick disc in the outer parts ($> R_e/5$) for some of the galaxies. 
   }
  \label{fig:resRatio}
\end{figure*}

In order to compare the two modelling approaches ADC and JAM, we first combine the measured deprojected stellar velocity profile $\overline{v_{\phi}}$ and radial dispersion profile $\sigma_R$. Next, we derive the circular velocity curve $\vcadc$ using the asymmetric drift correction (ADC) formula in equation~(\ref{eq:adcvcirc}).  These data are indicated by the blue-filled-pentagon curves with uncertainties derived from the transformed posterior distributions of ADC-MCMC results. Second, we construct axisymmetric Jeans models (JAM) that fit the combined observed stellar mean velocity and velocity dispersion fields, we use Equation~(\ref{eq:mgevcirc}) to obtain the circular velocity curve for each galaxy.  The resulting $\vcjam$ curves from the best-fit JAM model are plotted as red-filled-squares curves in Fig.~\ref{fig:Vc_a} with their uncertainty from JAM-MCMC code.
For all galaxies, ADC approach gives lower measurements of the circular velocity curve in comparison with the values obtained from JAM models for at least part of the radial range. To quantify these differences, we plot in Fig.~\ref{fig:resRatio} the velocity ratio between 
$v_{\mathrm{c,JAM}}$ and $v_{\mathrm{c,ADC}}$. 

Overall, the velocity discrepancy for Sb-Sbc galaxies is larger in the inner parts ($R<R_e/5$)  and decreases in the outer parts ($>R_e/5$), while for Scd-Sd galaxies the velocity discrepancy is constant or even increases in the outer parts.  

\begin{figure*}
\begin{center}
\includegraphics[width=1.0\textwidth]{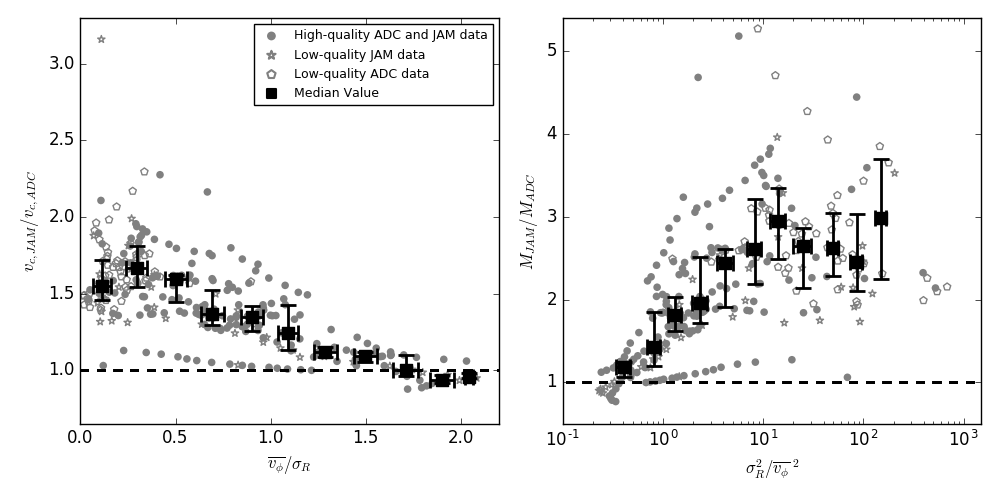}
\end{center}
\caption{Velocity and mass ratios of the two models. The grey filled circles correspond to the velocity and mass ratios of high-quality ADC and JAM model fits data, 
while the open pentagon and the open star correspond to low-quality ADC and JAM data, respectively. The black filled squares show the median of the distribution of the data per 
bin with error bar corresponding to the uncertainties at the 25th and 75th percentiles.
\emph{Left:} The ordered-over-random motion $\overline{v_{\phi}}/\sigma_R$ (from the deprojected rotation and radial velocity dispersion profiles) as derived at 
different radii versus the velocity ratio $v_{c,\mathrm{JAM}}/v_{c,\mathrm{ADC}}$ at the same radii. The discrepancy between the ADC and JAM velocities becomes 
gradually larger for decreasing $\overline{v_{\phi}}/\sigma_R$  $\lesssim 1.5$, consistent with the presence of a dynamically hot bulge in the inner 
parts and/or a dynamically hot(ter) and thick(er) disc in the latest-type spiral galaxies.
\emph{Right:} The ratio $\ensuremath{\sigma^{2}_R/\overline{v_\phi}^2}$  (from the radial velocity dispersion profiles and deprojected rotation) as 
measured at different radii versus the mass ratio $M_{\mathrm{JAM}}/M_{\mathrm{ADC}}$ at the same radii. The discrepancy between the ADC and JAM inferred 
enclosed masses becomes gradually larger for increasing  $\ensuremath{\sigma^{2}_R/\overline{v_\phi}^2}$, which has the same order of magnitude as the 
expected error in the epicycle approximation by \citet{Vandervoort1975}.
}
\label{fig:VSrel}
\end{figure*}

The differences we found between the circular velocity curves from the two modelling approaches can have a strong impact on the inferred total mass distribution and thus also on any follow-up inference like the amount of dark matter in the inner parts of these late-type spiral galaxies.
We list the different assumptions adopted in both the axisymmetric Jeans models (JAM) and the asymmetric drift correction (ADC), and discuss the possible reasons for their discrepancy.

\section{Discussion}
\label{S:discussion}

\subsection{Assumptions in the mass models}
\label{SS:assumptions}

\emph{Mass-follows-light (JAM):} 
Within the small radial range covered by the stellar kinematics, typically at $< R_e/3$, we expect the variations in the mass-to-light ratio to be small. \cite{Kent1986} provides evidence that in a sample of 37 Sb-Sc galaxies, most of their rotation curves computes from the luminosity profiles assuming constant mass-to-light ratio provide a good match to the observed curve out to the radius where the predicted curve turns over.  
Additionally, \citet{Kregel2005a} investigate the effects of realistic radially varying mass-to-light ratios and find the overall effect to only be $\sim$ 10\% variations on the derived kinematic properties within $R_e/2$.

\emph{Constant anisotropy in the meridional plane (JAM):}
Whereas the velocity anisotropy $\sigma_\phi/\sigma_R$ in the equatorial plane is inherent in the solution of the axisymmetric Jeans equations (Section~\ref{SS:axijeans}), the velocity anisotropy $\sigma_z/\sigma_R$ in the meridional plane is a free parameter, given as $\beta_z = 1 - \sigma^2_z/\sigma^2_R$.  
There is little evidence for strong variation of $\beta_z$.  For example, \citet{Bottema1993} already argued that for spirals the $\sigma_z/\sigma_R$ is constant at approximately 0.6, close to the measured value of $0.53\pm0.07$ in the solar neighbourhood \citep{Dehnen1998, Mignard2000}. This ratio of the anisotropy is measured in a few spirals of type Sa to Sbc \citep{Gerssen1997,Gerssen2000, Shapiro2003}, yielding slightly larger constant values between 0.6 and 0.8. Based on long-slit spectra for a sample of 17 edge-on Sb--Scd spirals, \citet{Kregel2005a} also adopt constant values, although slightly lower: 0.5 to 0.7.  While radial  variation of the anisotropy is not excluded, a constant value of $\beta_z$ should be sufficient for our analysis.

\emph{Shape of the velocity ellipsoid (JAM, ADC):}
 The ADC and JAM models in our study assume that the velocity ellipsoid in the meridional plane is aligned with the cylindrical coordinate system so that $\overline{v_R v_z}=0$. Additionally, the ADC approach may allow for a tilt of the velocity ellipsoid through the parameter $\kappa$ having intermediate values between $\kappa=0$  and $\kappa=1$, corresponding to the cylindrical and spherical coordinate system, respectively. However, we expect that the resulting circular velocity curve is only weakly dependent on this tilt, because most of the stellar mass is concentrated toward the equatorial plane,  particularly for late-type spiral galaxies.  In that case, the assumption of axisymmetry gives $\overline{v_R v_z}=0$. 

\emph{Dust (JAM, ADC):} 
The surface brightness distribution of the spiral galaxies that is used in both JAM and ADC modelling approaches can be strongly affected by extinction due to dust. We have tried to minimise the effects of dust in various ways (G09): (i) selecting galaxies with intermediate inclinations so that they are not edge-on (where dust extinction is strongest) or face-on (where the stellar velocity dispersion would be significantly below the spectral resolution), (ii) inferring the surface brightness distribution from images in the near-infrared where the extinction is significantly smaller than in the optical, and (iii) fitting smooth, analytical MGE profiles to the radial surface brightness profile after azimuthally averaging over annuli to suppress deviations caused by bars, spiral arms, and regions obscured by dust. The stellar kinematics are obtained from integral-field spectroscopy in the optical and thus could also be affected if the (giant) stars that contribute along the line-of-sight with different motions are affected by the dust in different ways. For example, if dynamically colder stars closer to the disc plane are relatively more obscured than dynamically hotter stars above the disc plane, the resulting combined ordered-over-random motion could be biased to lower values. The effects of dust are strongly reduced for lines of sight only a few degrees away from edge-on (\citealt{Baes2003b}) and the effects of extinction on the line-of-sight velocity are negligible: the change in $\sigma_z$ is only 1.3\% higher from the dustless case (\citealt{Bershady2010}). Given the selection in our sample, the effects of dust on the inferred circular velocity curves from both modelling approaches are expected to be minimal.

\emph{Thin-disc (ADC):} 
Circular velocity curves of spirals nearly always come from (cold) gas, which is naturally in a thin disc.  The stellar discs of these late-type spiral galaxies are also believed to be thin, with inferred intrinsic flattening $q \sim 0.14$ \citep[e.g.,][]{Kregel2002}. Even the bulges in late-type spiral galaxies are  different from the 'classical' bulges in lenticular galaxies. S\'ersic profile fits to their surface brightness ($I(R) \propto \exp(-R^{1/n})$) show that towards later-types, the bulges are smaller in size, have profiles closer to exponential ($n=1$) than de Vaucouleurs ($n=4$), and are flatter (e.g., G09).  Hence, the thin-disc assumption adopted in the ADC approach also seems to be reasonable.
\subsection{Bulges and thick discs}
\label{SS:bulgesandthickdiscs}
 Independent of the nature of the dynamically hot stellar (sub)system, it seems that the local value of the ordered-over-random motion is the key for understanding of the ADC and JAM discrepancy. 

To explore that, in Fig.~\ref{fig:resRatio}, we plot the radial velocity ratio $v_{\mathrm{c,JAM}}/v_{\mathrm{c,ADC}}$ of the galaxies, 
as well as the local luminosity fraction of the fitted exponential disc compared to the total luminosity as a thin solid curve.  
For the Sb--Sbc galaxies, the discrepancy in the estimated velocity ratio between the JAM and ADC models seems to be larger in the 
inner parts ($R<R_e/5$), where the luminosity is dominated by the presence of the bulge, and decreases outwards. For Scd--Sd galaxies 
the velocity ratio stays roughly constant and in a few cases increases towards larger radii, which could be due to the presence of a thick(er) disc component.

In the left panel of Fig.~\ref{fig:VSrel}, we plot the velocity ratios $v_{\mathrm{c,JAM}}/v_{\mathrm{c,ADC}}$ obtained from Fig.\ref{fig:resRatio} 
for all galaxies as measured at different radii versus the ordered-over-random motion $\overline{v_{\phi}}/\sigma_R$ (the deprojected rotation 
vs. radial velocity dispersion) at the same radii. In the right panel of Fig.~\ref{fig:VSrel}, we also show the ratio $\ensuremath{\sigma^{2}_R/\overline{v_\phi}^2}$  
(from the radial velocity dispersion profiles and deprojected rotation) as measured at different radii versus the mass ratio $M_{\mathrm{JAM}}/M_{\mathrm{ADC}}=\vcjam^2/\vcadc^2$ 
at the same radii. The discrepancy between the ADC and JAM inferred enclosed masses becomes gradually larger for increasing  $\ensuremath{\sigma^{2}_R/\overline{v_\phi}^2}$. 
The impact of the models' discrepancy is larger when is converted into mass ratio. It also has the same order of magnitude as the expected uncertainty of the epicycle 
approximation ($\Delta\sim \ensuremath{\sigma^{2}_R/\overline{v_\phi}^2}$, \citealt{Vandervoort1975}), which is actually applied in ADC (but not in JAM) approach. 

We see that the circular velocity measurement of both modelling approaches are consistent for 
$\overline{v_{\phi}}/\sigma_R$ $\gtrsim 1.5$, corresponding to those radii in the Sb--Sbc galaxies where the disc luminosity is dominating over the bulge luminosity, 
i.e., where the thin solid curve in Fig.~\ref{fig:resRatio} approaches unity. 

Since some of our galaxies do not have well fitting models, we mark their data with a different symbol in order to check for possible biases. 
However, even after their removal, the general trends of the velocity and mass ratios are preserved.
The grey filled circles correspond to high-quality ADC and JAM model fits data, while the open pentagon and the open star correspond to low-quality ADC and JAM data, respectively. 
The black filled squares show the median of the distribution of the data per bin. 
The error bar corresponds to the uncertainties of the median value at the 25th and 75th percentiles of the distribution at each bin (see also Table \ref{tab:calc5}).
\begin{table}
\begin{center}
 \caption{Median values of the velocity and mass ratios with corresponding (25th and 75th percentile) uncertainties at each bin in Fig.\ref{fig:VSrel}: 
 (1) Ratio between the deprojected rotation and radial velocity dispersion; (2) Velocity ratio between JAM and ADC model and their 
 uncertainty at fixed $\overline{v_{\phi}}/\sigma_R$ ratio; (3) Ratio between the deprojected velocity and radial velocity dispersion; 
 (4) Mass ratio between JAM and ADC model and their uncertainty at fixed $\ensuremath{\sigma^{2}_R/\overline{v_\phi}^2}$ ratio;
 }
   \tabcolsep=1.2mm 
    \begin{tabular}{|*{2}{l}|*{2}{l}|}
    \hline
      \multicolumn{4}{|c|}{JAM-ADC conversion factors}      \\
      \hline
$\overline{v_{\phi}}/\sigma_R$ &  $\frac{v_{\mathrm{c,JAM}}}{v_{\mathrm{c,ADC}}}$  &
$\sigma^2_{R}/\overline{v_{\phi}}^2$ &  $\frac{M_{\mathrm{JAM}}}{M_{\mathrm{ADC}}}$\\
 (1)      &  (2)  &  (3)    & (4)      \\
\hline
0.1 & $1.550_{-0.089}^{+0.168}$ & 0.4 & $1.187_{-0.123}^{+0.055}$ \\
0.3 & $1.666_{-0.123}^{+0.147}$ & 0.8 & $1.425_{-0.224}^{+0.424}$ \\
0.5 & $1.593_{-0.146}^{+0.025}$ & 1.3 & $1.809_{-0.184}^{+0.227}$ \\
0.7 & $1.370_{-0.073}^{+0.157}$ & 2.4 & $1.959_{-0.243}^{+0.551}$ \\
0.9 & $1.345_{-0.089}^{+0.072}$ & 4.2 & $2.446_{-0.536}^{+0.162}$ \\
1.1 & $1.244_{-0.112}^{+0.184}$ & 8.3 & $2.606_{-0.414}^{+0.603}$ \\
1.3 & $1.116_{-0.031}^{+0.033}$ & 14.1 & $2.948_{-0.455}^{+0.404}$ \\
1.5 & $1.090_{-0.036}^{+0.026}$ & 24.8 & $2.648_{-0.504}^{+0.215}$ \\
1.7 & $1.004_{-0.044}^{+0.092}$ & 49.6 & $2.627_{-0.349}^{+0.414}$ \\
1.9 & $0.933_{-0.020}^{+0.037}$ & 85.5 & $2.448_{-0.347}^{+0.582}$ \\
2.0 & $0.953_{-0.007}^{+0.030}$ & 150.1 & $2.982_{-0.730}^{+0.719}$ \\
\hline
 \end{tabular}
  \label{tab:calc5}
\end{center}
\end{table}

Since we are probing the inner parts of these spiral galaxies it might well be the presence of bulges and/or thick stellar discs 
that causes a break-down of the thin-disc/epicycle approximation in ADC. As can be seen from Fig.~\ref{fig:Vc_a}, the stellar rotation 
(first column) is of the same order and often even lower than the stellar dispersion (second column). The small ordered-over-random motion 
values implies that the stars are far from dynamically cold. 

Detailed photometric studies indicate that most disc galaxies contain a thick disc \citep[e.g.][]{Dalcanton2002,Seth2005, Comeron2011}, 
and in low-mass galaxies with circular velocities $<$\,120\,\kms, like the Scd--Sd galaxies in this sample, thick disc stars can contribute 
nearly half the luminosity and dominate the stellar mass  (\citealt{Yoachim2006}). Moreover, recent hydrodynamical simulations that 
reproduce thick discs show that their typical scale lengths are around 3--5 kpc \citep[e.g.][]{Domenech-Moral2012}, i.e., twice range typically covered by our analysis. 
Additionally, various earlier studies have qualitatively indicated that the ADC approach might not be suitable in the case of stellar 
systems that are (locally) not dynamically cold \citep[e.g.,][]{Neistein1999, Bedregal2006,Williams2010}. We also find that the discrepancy 
between the two model is proportional to the error of the epicycle approximation in ADC (see \citealt{Vandervoort1975}), where 
$\ensuremath{\sigma^{2}_R/\overline{v_\phi}^2} \sim M_{\mathrm{JAM}}/M_{\mathrm{ADC}}$.

\cite{Davis2013} show that rotation curves of early-type galaxies based on the dynamically cold molecular gas are well traced by the $\vcjam$, 
and that the $H_{\beta}/\mathrm{OIII} $ rotation curves from their \sauron\ data exhibit noticeable asymmetric drift. 
They also find the correction of the stellar rotation curve using ADC approach is also consistent with $\vcjam$. 
However, they note that there is some indication that this becomes less true as velocity dispersion approaches the rotation speed, 
as we find in this study. It is not excluded that there are strong streaming motions in the centers of the \sauron\ galaxies due to presence 
of non-axisymmetric features, which prevent an accurate asymmetric-drft correction.

Nevertheless, some of the discrepancy might be arise from JAM overestimation of the M/L ratio. 
\cite{Lablanche2012} showed that the overestimation can be severe in face-on barred system. 
\cite{Davis2013} also discussed the possibility of an overestimation 
of the $M/L$ ratio due to disk averaging over various stellar populations. 
We do not exclude the latter since the region probed by our data is 
in the central part of the galaxies, where multiple stellar populations can be present.
The former case should not influence some of our measurements, since all of our galaxies have intermediate inclinations.  

In future studies, we will explore further this problem using large statistical sample of galaxies and morphologies.




\section{Summary}
\label{S:concl}

The rotation curves of spiral galaxies, traced by atomic gas, provide the most direct path to estimate the total galaxy masses outside the stellar disc. However, it is not straightforward to measure the rotation curve in the central parts of the galaxies, where the ionized and molecular gas dominate. Gas settles in the equatorial or polar plane and is thus less sensitive to the mass distribution perpendicular to it.  Gas is also dissipative and easily disturbed by perturbations in the plane from features like bars or spiral arms.  Thus, stars appear to be a better tracer as they are distributed in all three dimensions and, being collisionless, they are less sensitive to perturbations. However, stars are not cold tracers as they move in orbits that are neither circular nor confined to a single plane.  We therefore need to measure both velocity dispersions and line-of-sight velocities to recover the total mass distribution of galaxies.

In this paper we compare two different approaches for inferring dynamical masses of spiral galaxies: the commonly used asymmetric drift correction (ADC) and the axisymmetric Jeans equations. We used the stellar kinematics derived by integral field spectroscopy of a sample of 18 Sb -- Sd galaxies, observed 
with the \sauron\/ spectrograph. We obtained stellar mean velocity and velocity dispersion maps and derived the galaxies' circular velocity curves by fitting solutions to the Jeans equations. Using the same data we also derived the circular velocity curves via the ADC technique. We use Markov Chain Monte Carlo methods to determine credible values for model parameters and their associated uncertainties.

The ADC approach gives systematically lower values than JAM for the circular velocity curves, and hence the enclosed mass, when $\overline{v_{\phi}}/\sigma_R$ $\lesssim 1.5$. The velocity ratio $\vcjam/\vcadc$ for Sb--Sbc galaxies is larger in the inner parts ($R<R_e/5$) and decreases outwards.  However, for Scd--Sd galaxies the mass ratio stays roughly constant and in a few cases increases towards larger radii.

The complete Python codes, implementing MCMC analysis on both JAM and ADC models, described in this paper, can be downloaded from 
\texttt{https://github.com/Kalinova/Dyn\_models}. These codes provide a robust way to estimate the uncertainties in the derived mass distribution of galaxies through their circular velocity curves.


\section*{Acknowledgements}
We are extremely grateful to the anonymous referee for their valuable contribution towards the completeness of this paper.  His/her reports were constructive and thorough and dramatically improved the quality of the result. V.K. thanks Ronald L\"{a}sker for the fruitful discussions. 
We thank Michelle Cappellari for the helpful suggestions about the calculation of the uncertainties of the symmetrized $\Vrms$ field.
We acknowledge financial support to the DAGAL network from the People 
Programme (Marie Curie Actions) of the European Union's Seventh Framework 
Programme FP7/2007- 2013/ under REA grant agreement number PITN-GA-2011-289313. J.~F.-B. acknowledges support from grant
AYA2013-48226-C3-1-P from the Spanish Ministry of Economy and 
Competitiveness (MINECO). V.K, D.C. and E.R. are supported in part by a Discovery Grant from NSERC of Canada.
\footnotesize{
\bibliographystyle{mn2e_new}
\bibliography{manuscript_Kalinova}}
\bsp 
\label{lastpage}
\newpage
\appendix
\label{S:appendix}
%
\section{MGE Models}
\label{A:mgeTab} 
Here we provide a table with the parameters of our MGE models
for each galaxy, where $I_{0,j}$ is the central surface brightness, $\xi'_j$ is the the dispersion along the major $x'$-axis, and $q'_j$ is the flattening. Additionally, $\emph{ind}$ indicates the gaussians of the central, bulge or disc component of the galaxies with respective values of 0,1 and 2.

\begin{table*}
\begin{minipage}{170mm}
\begin{center}
\begin{tabular}{|c|cccc|cccc|cccc|} 
\hline
$j$ & $I_{0,j}$ & $\xi'_j$ & $q'_j$ & ind & $I_{0,j}$ & $\xi'_j$ & $q'_j$ & ind & $I_{0,j}$ & $\xi'_j$ & $q'_j$ & ind \\
& (L$_{\odot}$\,pc$^{-2})$ & (arcsec) & & & (L$_{\odot}$\,pc$^{-2})$ & (arcsec) & & & (L$_{\odot}$\,pc$^{-2})$ & (arcsec) & & \\ 
\hline
&\multicolumn{4}{c|}{NGC1042}&\multicolumn{4}{c|}{NGC2805}&\multicolumn{4}{c|}{NGC2964}\\
\hline
1 & 29277.46 & 0.23 & 1.000 & 0 & 12769.80 & 0.12 & 1.000 & 0 & 78000.01 & 0.20 & 1.000 & 0 \\
2 & 23.49 & 0.11 & 0.950 & 1 & 96.40 & 0.10 & 0.800 & 1 & 2617.84 & 0.10 & 0.950 & 1 \\
3 & 54.40 & 0.26 & 0.950 & 1 & 134.25 & 0.24 & 0.800 & 1 & 4856.21 & 0.20 & 0.950 & 1 \\
4 & 111.36 & 0.54 & 0.950 & 1 & 177.19 & 0.51 & 0.800 & 1 & 7703.26 & 0.35 & 0.950 & 1 \\
5 & 200.80 & 1.00 & 0.950 & 1 & 213.89 & 1.01 & 0.800 & 1 & 10286.09 & 0.53 & 0.950 & 1 \\
6 & 305.64 & 1.67 & 0.950 & 1 & 229.57 & 1.88 & 0.800 & 1 & 10381.19 & 0.74 & 0.950 & 1 \\
7 & 361.53 & 2.53 & 0.950 & 1 & 212.94 & 3.31 & 0.800 & 1 & 6767.80 & 0.98 & 0.950 & 1 \\
8 & 301.09 & 3.54 & 0.950 & 1 & 165.32 & 5.52 & 0.800 & 1 & 2380.00 & 1.23 & 0.950 & 1 \\
9 & 162.48 & 4.62 & 0.950 & 1 & 103.93 & 8.75 & 0.800 & 1 & 379.76 & 1.48 & 0.950 & 1 \\
10 & 51.72 & 5.76 & 0.950 & 1 & 51.22 & 13.28 & 0.800 & 1 & 24.00 & 1.73 & 0.950 & 1 \\
11 & 7.95 & 6.98 & 0.950 & 1 & 19.17 & 19.38 & 0.800 & 1 & 0.54 & 1.98 & 0.950 & 1 \\
12 & 0.15 & 7.97 & 0.950 & 1 & 5.24 & 27.40 & 0.800 & 1 & 0.00 & 2.24 & 0.950 & 1 \\
13 & 0.32 & 8.37 & 0.950 & 1 & 0.98 & 37.78 & 0.800 & 1 & 0.00 & 2.52 & 0.950 & 1 \\
14 & 0.00 & 9.78 & 0.950 & 1 & 0.12 & 51.15 & 0.800 & 1 & 0.00 & 17.76 & 0.950 & 1 \\
15 & 0.00 & 11.44 & 0.950 & 1 & 0.01 & 68.32 & 0.800 & 1 & 343.25 & 1.00 & 0.550 & 2 \\
16 & 0.00 & 23.70 & 0.950 & 1 & 0.00 & 90.34 & 0.800 & 1 & 847.42 & 4.13 & 0.550 & 2 \\
17 & 0.00 & 77.53 & 0.950 & 1 & 0.00 & 118.84 & 0.800 & 1 & 1253.73 & 10.52 & 0.550 & 2 \\
18 & 5.20 & 0.60 & 0.710 & 2 & 0.00 & 157.50 & 0.800 & 1 & 996.24 & 20.19 & 0.550 & 2 \\
19 & 14.74 & 2.50 & 0.710 & 2 & 0.00 & 560.32 & 0.800 & 1 & 374.58 & 32.24 & 0.550 & 2 \\
20 & 32.41 & 7.13 & 0.710 & 2 & 0.00 & 588.15 & 0.800 & 1 & 61.23 & 45.79 & 0.550 & 2 \\
21 & 57.25 & 16.36 & 0.710 & 2 & 2.97 & 0.62 & 0.760 & 2 & 4.04 & 60.39 & 0.550 & 2 \\
22 & 77.62 & 31.83 & 0.710 & 2 & 8.35 & 2.55 & 0.760 & 2 & 0.09 & 76.14 & 0.550 & 2 \\
23 & 73.91 & 54.04 & 0.710 & 2 & 18.29 & 7.25 & 0.760 & 2 & 0.00 & 93.57 & 0.550 & 2 \\
24 & 44.78 & 82.45 & 0.710 & 2 & 31.88 & 16.58 & 0.760 & 2 & 0.00 & 114.06 & 0.550 & 2 \\
25 & 15.40 & 116.63 & 0.710 & 2 & 41.75 & 32.02 & 0.760 & 2 & ... & ... & ... & ... \\
26 & 2.41 & 157.63 & 0.710 & 2 & 37.57 & 53.88 & 0.760 & 2 & ... & ... & ... & ... \\
27 & 0.10 & 210.10 & 0.710 & 2 & 21.02 & 81.47 & 0.760 & 2 & ... & ... & ... & ... \\
28 & ... & ... & ... & ... & 6.51 & 114.33 & 0.760 & 2 & ... & ... & ... & ... \\
29 & ... & ... & ... & ... & 0.90 & 153.37 & 0.760 & 2 & ... & ... & ... & ... \\
30 & ... & ... & ... & ... & 0.03 & 202.87 & 0.760 & 2 & ... & ... & ... & ... \\
\hline
\hline
$j$ & $I_{0,j}$ & $\xi'_j$ & $q'_j$ & ind & $I_{0,j}$ & $\xi'_j$ & $q'_j$ & ind & $I_{0,j}$ & $\xi'_j$ & $q'_j$ & ind \\
& (L$_{\odot}$\,pc$^{-2})$ & (arcsec) & & & (L$_{\odot}$\,pc$^{-2})$ & (arcsec) & & & (L$_{\odot}$\,pc$^{-2})$ & (arcsec) & & \\ 
\hline
&\multicolumn{4}{c|}{NGC3346}&\multicolumn{4}{c|}{NGC3423}&\multicolumn{4}{c|}{NGC3949}\\
\hline
1 & 8592.16 & 0.12 & 1.000 & 0 & 13792.68 & 0.12 & 1.000 & 0 & 33147.01 & 0.12 & 1.000 & 0 \\
2 & 5.52 & 0.10 & 0.750 & 1 & 123.67 & 0.10 & 0.870 & 1 & 247.94 & 0.10 & 0.700 & 1 \\
3 & 12.93 & 0.21 & 0.750 & 1 & 180.26 & 0.23 & 0.870 & 1 & 415.67 & 0.22 & 0.700 & 1 \\
4 & 23.81 & 0.36 & 0.750 & 1 & 250.35 & 0.47 & 0.870 & 1 & 648.53 & 0.45 & 0.700 & 1 \\
5 & 52.10 & 0.59 & 0.750 & 1 & 322.26 & 0.89 & 0.870 & 1 & 915.74 & 0.81 & 0.700 & 1 \\
6 & 127.03 & 0.95 & 0.750 & 1 & 372.27 & 1.58 & 0.870 & 1 & 1122.66 & 1.37 & 0.700 & 1 \\
7 & 257.57 & 1.47 & 0.750 & 1 & 374.59 & 2.64 & 0.870 & 1 & 1141.20 & 2.17 & 0.700 & 1 \\
8 & 358.46 & 2.04 & 0.750 & 1 & 317.69 & 4.20 & 0.870 & 1 & 911.23 & 3.25 & 0.700 & 1 \\
9 & 266.26 & 2.59 & 0.750 & 1 & 219.51 & 6.36 & 0.870 & 1 & 534.22 & 4.61 & 0.700 & 1 \\
10 & 81.56 & 3.06 & 0.750 & 1 & 119.93 & 9.18 & 0.870 & 1 & 209.63 & 6.30 & 0.700 & 1 \\
11 & 7.86 & 3.49 & 0.750 & 1 & 50.95 & 12.73 & 0.870 & 1 & 48.61 & 8.33 & 0.700 & 1 \\
12 & 0.17 & 3.88 & 0.750 & 1 & 16.66 & 17.08 & 0.870 & 1 & 5.69 & 10.75 & 0.700 & 1 \\
13 & 0.00 & 4.25 & 0.750 & 1 & 3.89 & 22.51 & 0.870 & 1 & 0.28 & 13.60 & 0.700 & 1 \\
14 & 0.00 & 30.87 & 0.750 & 1 & 0.55 & 29.44 & 0.870 & 1 & 0.00 & 16.93 & 0.700 & 1 \\
15 & 20.03 & 1.12 & 0.840 & 2 & 0.04 & 38.25 & 0.870 & 1 & 0.00 & 20.87 & 0.700 & 1 \\
16 & 54.82 & 4.86 & 0.840 & 2 & 0.00 & 49.35 & 0.870 & 1 & 0.00 & 25.54 & 0.700 & 1 \\
17 & 102.72 & 13.39 & 0.840 & 2 & 0.00 & 63.23 & 0.870 & 1 & 0.00 & 31.02 & 0.700 & 1 \\
18 & 125.36 & 28.07 & 0.840 & 2 & 0.00 & 80.66 & 0.870 & 1 & 0.00 & 190.20 & 0.700 & 1 \\
19 & 88.40 & 48.50 & 0.840 & 2 & 0.00 & 103.25 & 0.870 & 1 & 255.39 & 1.02 & 0.640 & 2 \\
20 & 32.26 & 73.22 & 0.840 & 2 & 0.00 & 588.15 & 0.870 & 1 & 619.84 & 4.19 & 0.640 & 2 \\
21 & 5.54 & 101.15 & 0.840 & 2 & 20.07 & 1.09 & 0.770 & 2 & 886.68 & 10.58 & 0.640 & 2 \\
22 & 0.38 & 132.44 & 0.840 & 2 & 55.32 & 4.74 & 0.770 & 2 & 663.27 & 20.12 & 0.640 & 2 \\
23 & 0.01 & 169.72 & 0.840 & 2 & 105.74 & 13.13 & 0.770 & 2 & 227.58 & 31.90 & 0.640 & 2 \\
24 & 0.00 & 588.15 & 0.840 & 2 & 134.74 & 27.78 & 0.770 & 2 & 32.95 & 45.07 & 0.640 & 2 \\
25 & ... & ... & ... & ... & 101.83 & 48.48 & 0.770 & 2 & 1.87 & 59.22 & 0.640 & 2 \\
26 & ... & ... & ... & ... & 40.74 & 73.85 & 0.770 & 2 & 0.04 & 74.44 & 0.640 & 2 \\
27 & ... & ... & ... & ... & 7.85 & 102.70 & 0.770 & 2 & 0.00 & 91.29 & 0.640 & 2 \\
28 & ... & ... & ... & ... & 0.62 & 135.13 & 0.770 & 2 & 0.00 & 111.07 & 0.640 & 2 \\
29 & ... & ... & ... & ... & 0.01 & 173.88 & 0.770 & 2 & ... & ... & ... & ... \\
30 & ... & ... & ... & ... & 0.00 & 588.15 & 0.770 & 2 & ... & ... & ... & ... \\
\hline
\label{Tab:MGE_a}
\end{tabular}
\end{center}
\end{minipage}
\end{table*}

\begin{table*}
\begin{minipage}{170mm}
\begin{center}
\begin{tabular}{|c|cccc|cccc|cccc|}
\hline
$j$ & $I_{0,j}$ & $\xi'_j$ & $q'_j$ & ind & $I_{0,j}$ & $\xi'_j$ & $q'_j$ & ind & $I_{0,j}$ & $\xi'_j$ & $q'_j$ & ind \\
& (L$_{\odot}$\,pc$^{-2})$ & (arcsec) & & & (L$_{\odot}$\,pc$^{-2})$ & (arcsec) & & & (L$_{\odot}$\,pc$^{-2})$ & (arcsec) & & \\  
\hline
\hline
&\multicolumn{4}{c|}{NGC4030}&\multicolumn{4}{c|}{NGC4102}&\multicolumn{4}{c|}{NGC4254}\\
\hline
1 & 48113.13 & 0.13 & 1.000 & 0 & 0.00 & 1.52 & 1.000 & 0 & 55276.68 & 0.21 & 1.000 & 0 \\
2 & 6683.20 & 0.10 & 0.850 & 1 & 780479.16 & 0.10 & 0.800 & 1 & 2360.27 & 0.10 & 0.900 & 1 \\
3 & 6306.38 & 0.23 & 0.850 & 1 & 430239.29 & 0.23 & 0.800 & 1 & 2650.37 & 0.23 & 0.900 & 1 \\
4 & 5769.43 & 0.44 & 0.850 & 1 & 225993.57 & 0.44 & 0.800 & 1 & 2884.28 & 0.48 & 0.900 & 1 \\
5 & 5280.87 & 0.77 & 0.850 & 1 & 105795.77 & 0.79 & 0.800 & 1 & 2932.75 & 0.93 & 0.900 & 1 \\
6 & 4689.74 & 1.30 & 0.850 & 1 & 41165.86 & 1.37 & 0.800 & 1 & 2689.19 & 1.71 & 0.900 & 1 \\
7 & 3905.92 & 2.13 & 0.850 & 1 & 12914.85 & 2.25 & 0.800 & 1 & 2177.31 & 2.97 & 0.900 & 1 \\
8 & 2987.56 & 3.39 & 0.850 & 1 & 3209.39 & 3.58 & 0.800 & 1 & 1527.83 & 4.96 & 0.900 & 1 \\
9 & 2069.12 & 5.26 & 0.850 & 1 & 623.15 & 5.52 & 0.800 & 1 & 912.72 & 7.94 & 0.900 & 1 \\
10 & 1282.12 & 7.99 & 0.850 & 1 & 93.51 & 8.24 & 0.800 & 1 & 456.20 & 12.23 & 0.900 & 1 \\
11 & 702.74 & 11.89 & 0.850 & 1 & 10.74 & 12.00 & 0.800 & 1 & 187.78 & 18.20 & 0.900 & 1 \\
12 & 336.43 & 17.36 & 0.850 & 1 & 0.93 & 17.08 & 0.800 & 1 & 63.06 & 26.20 & 0.900 & 1 \\
13 & 138.18 & 24.94 & 0.850 & 1 & 0.06 & 23.90 & 0.800 & 1 & 17.31 & 36.60 & 0.900 & 1 \\
14 & 47.39 & 35.41 & 0.850 & 1 & 0.00 & 33.05 & 0.800 & 1 & 3.86 & 50.03 & 0.900 & 1 \\
15 & 13.07 & 49.82 & 0.850 & 1 & 0.00 & 45.35 & 0.800 & 1 & 0.64 & 67.87 & 0.900 & 1 \\
16 & 2.76 & 69.65 & 0.850 & 1 & 0.00 & 61.94 & 0.800 & 1 & 0.07 & 92.37 & 0.900 & 1 \\
17 & 0.42 & 97.03 & 0.850 & 1 & 0.00 & 84.36 & 0.800 & 1 & 0.00 & 127.13 & 0.900 & 1 \\
18 & 0.04 & 135.26 & 0.850 & 1 & 0.00 & 115.03 & 0.800 & 1 & 0.00 & 180.77 & 0.900 & 1 \\
19 & 0.00 & 190.51 & 0.850 & 1 & 0.00 & 159.32 & 0.800 & 1 & 0.00 & 588.15 & 0.900 & 1 \\
20 & 0.00 & 279.53 & 0.850 & 1 & 0.00 & 317.40 & 0.800 & 1 & 41.78 & 0.67 & 0.730 & 2 \\
21 & 94.59 & 0.79 & 0.760 & 2 & 417.54 & 0.99 & 0.555 & 2 & 116.76 & 2.77 & 0.730 & 2 \\
22 & 257.16 & 3.32 & 0.760 & 2 & 1034.48 & 4.11 & 0.555 & 2 & 247.55 & 7.80 & 0.730 & 2 \\
23 & 490.15 & 9.09 & 0.760 & 2 & 1546.11 & 10.50 & 0.555 & 2 & 402.28 & 17.44 & 0.730 & 2 \\
24 & 622.90 & 19.14 & 0.760 & 2 & 1252.87 & 20.20 & 0.555 & 2 & 465.56 & 32.71 & 0.730 & 2 \\
25 & 471.99 & 33.32 & 0.760 & 2 & 484.60 & 32.33 & 0.555 & 2 & 345.32 & 53.37 & 0.730 & 2 \\
26 & 190.98 & 50.64 & 0.760 & 2 & 82.09 & 46.01 & 0.555 & 2 & 148.27 & 78.49 & 0.730 & 2 \\
27 & 37.79 & 70.24 & 0.760 & 2 & 5.66 & 60.77 & 0.555 & 2 & 32.87 & 107.57 & 0.730 & 2 \\
28 & 3.26 & 91.86 & 0.760 & 2 & 0.14 & 76.67 & 0.555 & 2 & 3.02 & 141.30 & 0.730 & 2 \\
29 & 0.10 & 116.04 & 0.760 & 2 & 0.00 & 94.30 & 0.555 & 2 & 0.07 & 183.07 & 0.730 & 2 \\
30 & 0.00 & 144.89 & 0.760 & 2 & 0.00 & 115.01 & 0.555 & 2 & ... & ... & ... & ... \\
\hline
\hline
$j$ & $I_{0,j}$ & $\xi'_j$ & $q'_j$ & ind & $I_{0,j}$ & $\xi'_j$ & $q'_j$ & ind & $I_{0,j}$ & $\xi'_j$ & $q'_j$ & ind \\
& (L$_{\odot}$\,pc$^{-2})$ & (arcsec) & & & (L$_{\odot}$\,pc$^{-2})$ & (arcsec) & & & (L$_{\odot}$\,pc$^{-2})$ & (arcsec) & & \\ 
\hline
\hline
&\multicolumn{4}{c|}{NGC4487}&\multicolumn{4}{c|}{NGC4775}&\multicolumn{4}{c|}{NGC0488}\\
\hline
1 & 29433.24 & 0.17 & 1.000 & 0 & 21385.18 & 0.12 & 1.000 & 0 & 24161.14 & 0.38 & 1.000 & 0 \\
2 & 128.29 & 0.11 & 0.600 & 1 & 85.94 & 0.12 & 0.830 & 1 & 16980.54 & 0.10 & 0.900 & 1 \\
3 & 205.54 & 0.29 & 0.600 & 1 & 104.35 & 0.28 & 0.830 & 1 & 14945.13 & 0.23 & 0.900 & 1 \\
4 & 289.40 & 0.67 & 0.600 & 1 & 70.83 & 0.43 & 0.830 & 1 & 12930.46 & 0.46 & 0.900 & 1 \\
5 & 350.12 & 1.40 & 0.600 & 1 & 137.73 & 0.65 & 0.830 & 1 & 10797.95 & 0.84 & 0.900 & 1 \\
6 & 338.21 & 2.63 & 0.600 & 1 & 156.92 & 0.98 & 0.830 & 1 & 8245.04 & 1.48 & 0.900 & 1 \\
7 & 256.84 & 4.48 & 0.600 & 1 & 206.29 & 1.49 & 0.830 & 1 & 5595.02 & 2.53 & 0.900 & 1 \\
8 & 154.95 & 6.99 & 0.600 & 1 & 231.29 & 2.26 & 0.830 & 1 & 3318.18 & 4.17 & 0.900 & 1 \\
9 & 79.42 & 10.36 & 0.600 & 1 & 231.40 & 3.42 & 0.830 & 1 & 1694.88 & 6.66 & 0.900 & 1 \\
10 & 27.00 & 14.96 & 0.600 & 1 & 176.99 & 5.18 & 0.830 & 1 & 734.89 & 10.33 & 0.900 & 1 \\
11 & 6.26 & 20.87 & 0.600 & 1 & 86.68 & 7.85 & 0.830 & 1 & 266.76 & 15.61 & 0.900 & 1 \\
12 & 0.82 & 28.64 & 0.600 & 1 & 18.69 & 11.90 & 0.830 & 1 & 79.95 & 23.02 & 0.900 & 1 \\
13 & 0.05 & 38.42 & 0.600 & 1 & 0.96 & 18.03 & 0.830 & 1 & 19.44 & 33.25 & 0.900 & 1 \\
14 & 0.00 & 50.53 & 0.600 & 1 & 0.00 & 27.32 & 0.830 & 1 & 3.72 & 47.28 & 0.900 & 1 \\
15 & 0.00 & 65.51 & 0.600 & 1 & 0.00 & 41.39 & 0.830 & 1 & 0.53 & 66.51 & 0.900 & 1 \\
16 & 0.00 & 84.25 & 0.600 & 1 & 0.00 & 62.71 & 0.830 & 1 & 0.05 & 93.11 & 0.900 & 1 \\
17 & 0.00 & 108.71 & 0.600 & 1 & 0.00 & 330.56 & 0.830 & 1 & 0.00 & 130.94 & 0.900 & 1 \\
18 & 0.00 & 588.15 & 0.600 & 1 & 50.98 & 0.92 & 0.865 & 2 & 0.00 & 189.04 & 0.900 & 1 \\
19 & 21.17 & 1.08 & 0.630 & 2 & 131.09 & 3.83 & 0.865 & 2 & 0.00 & 307.69 & 0.900 & 1 \\
20 & 58.35 & 4.71 & 0.630 & 2 & 216.63 & 10.04 & 0.865 & 2 & 0.00 & 323.27 & 0.900 & 1 \\
21 & 111.87 & 13.06 & 0.630 & 2 & 209.75 & 19.96 & 0.865 & 2 & 30.42 & 0.69 & 0.770 & 2 \\
22 & 143.33 & 27.67 & 0.630 & 2 & 104.72 & 32.85 & 0.865 & 2 & 84.00 & 2.87 & 0.770 & 2 \\
23 & 109.17 & 48.36 & 0.630 & 2 & 24.46 & 47.73 & 0.865 & 2 & 176.60 & 8.01 & 0.770 & 2 \\
24 & 44.17 & 73.72 & 0.630 & 2 & 2.48 & 64.00 & 0.865 & 2 & 287.49 & 17.80 & 0.770 & 2 \\
25 & 8.66 & 102.54 & 0.630 & 2 & 0.10 & 81.62 & 0.865 & 2 & 340.68 & 33.29 & 0.770 & 2 \\
26 & 0.71 & 134.83 & 0.630 & 2 & 0.00 & 101.14 & 0.865 & 2 & 267.03 & 54.34 & 0.770 & 2 \\
27 & 0.02 & 173.29 & 0.630 & 2 & 0.00 & 124.14 & 0.865 & 2 & 125.56 & 80.14 & 0.770 & 2 \\
28 & 0.00 & 588.15 & 0.630 & 2 & ... & ... & ... & ... & 31.53 & 110.27 & 0.770 & 2 \\
29 & ... & ... & ... & ... & ... & ... & ... & ... & 3.38 & 145.59 & 0.770 & 2 \\
30 & ... & ... & ... & ... & ... & ... & ... & ... & 0.09 & 189.76 & 0.770 & 2 \\
\hline
\label{Tab:MGE_b}
\end{tabular}
\end{center}
\end{minipage}
\end{table*}

\begin{table*}
\begin{minipage}{170mm}
\begin{center}
\begin{tabular}{|c|cccc|cccc|cccc|}
\hline
$j$ & $I_{0,j}$ & $\xi'_j$ & $q'_j$ & ind & $I_{0,j}$ & $\xi'_j$ & $q'_j$ & ind & $I_{0,j}$ & $\xi'_j$ & $q'_j$ & ind \\
& (L$_{\odot}$\,pc$^{-2})$ & (arcsec) & & & (L$_{\odot}$\,pc$^{-2})$ & (arcsec) & & & (L$_{\odot}$\,pc$^{-2})$ & (arcsec) & & \\ 
\hline
\hline
&\multicolumn{4}{c|}{NGC5585}&\multicolumn{4}{c|}{NGC5668}&\multicolumn{4}{c|}{NGC5678}\\
\hline
1 & 5628.53 & 0.27 & 1.000 & 0 & 10359.16 & 0.12 & 1.000 & 0 & 34754.46 & 0.32 & 1.000 & 0 \\
2 & 4.28 & 0.12 & 0.800 & 1 & 86.95 & 0.10 & 0.800 & 1 & 25905.05 & 0.10 & 0.800 & 1 \\
3 & 8.11 & 0.28 & 0.800 & 1 & 126.05 & 0.24 & 0.800 & 1 & 17168.58 & 0.22 & 0.800 & 1 \\
4 & 1.55 & 0.43 & 0.800 & 1 & 173.70 & 0.51 & 0.800 & 1 & 12031.87 & 0.42 & 0.800 & 1 \\
5 & 18.27 & 0.65 & 0.800 & 1 & 220.00 & 1.01 & 0.800 & 1 & 8529.89 & 0.76 & 0.800 & 1 \\
6 & 5.84 & 0.98 & 0.800 & 1 & 249.90 & 1.89 & 0.800 & 1 & 4679.15 & 1.39 & 0.800 & 1 \\
7 & 38.14 & 1.49 & 0.800 & 1 & 247.84 & 3.34 & 0.800 & 1 & 1949.35 & 2.47 & 0.800 & 1 \\
8 & 23.61 & 2.26 & 0.800 & 1 & 207.79 & 5.58 & 0.800 & 1 & 629.79 & 4.20 & 0.800 & 1 \\
9 & 72.02 & 3.42 & 0.800 & 1 & 142.25 & 8.86 & 0.800 & 1 & 156.65 & 6.86 & 0.800 & 1 \\
10 & 70.47 & 5.18 & 0.800 & 1 & 76.86 & 13.45 & 0.800 & 1 & 29.61 & 10.81 & 0.800 & 1 \\
11 & 112.16 & 7.85 & 0.800 & 1 & 31.67 & 19.61 & 0.800 & 1 & 4.21 & 16.44 & 0.800 & 1 \\
12 & 95.48 & 11.90 & 0.800 & 1 & 9.54 & 27.69 & 0.800 & 1 & 0.45 & 24.24 & 0.800 & 1 \\
13 & 45.38 & 18.03 & 0.800 & 1 & 1.97 & 38.11 & 0.800 & 1 & 0.04 & 34.80 & 0.800 & 1 \\
14 & 3.29 & 27.32 & 0.800 & 1 & 0.25 & 51.47 & 0.800 & 1 & 0.00 & 48.82 & 0.800 & 1 \\
15 & 0.00 & 41.39 & 0.800 & 1 & 0.02 & 68.50 & 0.800 & 1 & 0.00 & 67.20 & 0.800 & 1 \\
16 & 0.00 & 62.71 & 0.800 & 1 & 0.00 & 90.18 & 0.800 & 1 & 0.00 & 90.83 & 0.800 & 1 \\
17 & 0.00 & 330.56 & 0.800 & 1 & 0.00 & 117.92 & 0.800 & 1 & 0.00 & 121.04 & 0.800 & 1 \\
18 & 3.22 & 0.59 & 0.640 & 2 & 0.00 & 154.81 & 0.800 & 1 & 0.00 & 164.61 & 0.800 & 1 \\
19 & 9.10 & 2.43 & 0.640 & 2 & 0.00 & 588.15 & 0.800 & 1 & 0.00 & 309.03 & 0.800 & 1 \\
20 & 20.23 & 6.96 & 0.640 & 2 & 10.14 & 0.75 & 0.845 & 2 & 0.00 & 588.15 & 0.800 & 1 \\
21 & 36.28 & 16.06 & 0.640 & 2 & 27.94 & 3.13 & 0.845 & 2 & 168.11 & 0.89 & 0.525 & 2 \\
22 & 50.54 & 31.49 & 0.640 & 2 & 55.57 & 8.68 & 0.845 & 2 & 439.46 & 3.72 & 0.525 & 2 \\
23 & 50.21 & 53.94 & 0.640 & 2 & 77.06 & 18.68 & 0.845 & 2 & 749.06 & 9.87 & 0.525 & 2 \\
24 & 32.24 & 82.97 & 0.640 & 2 & 67.35 & 33.27 & 0.845 & 2 & 769.26 & 19.83 & 0.525 & 2 \\
25 & 11.93 & 118.18 & 0.640 & 2 & 33.11 & 51.61 & 0.845 & 2 & 418.84 & 32.99 & 0.525 & 2 \\
26 & 2.05 & 160.73 & 0.640 & 2 & 8.34 & 72.73 & 0.845 & 2 & 109.35 & 48.32 & 0.525 & 2 \\
27 & 0.09 & 215.52 & 0.640 & 2 & 0.96 & 96.26 & 0.845 & 2 & 12.68 & 65.17 & 0.525 & 2 \\
28 & ... & ... & ... & ... & 0.04 & 122.79 & 0.845 & 2 & 0.57 & 83.47 & 0.525 & 2 \\
29 & ... & ... & ... & ... & 0.00 & 154.70 & 0.845 & 2 & 0.01 & 103.77 & 0.525 & 2 \\
30 & ... & ... & ... & ... & ... & ... & ... & ... & 0.00 & 127.71 & 0.525 & 2 \\
\hline
\hline
$j$ & $I_{0,j}$ & $\xi'_j$ & $q'_j$ & ind & $I_{0,j}$ & $\xi'_j$ & $q'_j$ & ind & $I_{0,j}$ & $\xi'_j$ & $q'_j$ & ind \\
& (L$_{\odot}$\,pc$^{-2})$ & (arcsec) & & & (L$_{\odot}$\,pc$^{-2})$ & (arcsec) & & & (L$_{\odot}$\,pc$^{-2})$ & (arcsec) & & \\  
\hline
\hline
&\multicolumn{4}{c|}{NGC0628}&\multicolumn{4}{c|}{NGC0772}&\multicolumn{4}{c|}{NGC0864}\\
\hline
1 & 32854.43 & 0.16 & 1.000 & 0 & 96530.87 & 0.32 & 1.000 & 0 & 834.18 & 1.90 & 1.000 & 0 \\
2 & 125.78 & 0.10 & 0.950 & 1 & 11343.60 & 0.10 & 0.830 & 1 & 26446.20 & 0.10 & 0.950 & 1 \\
3 & 225.98 & 0.24 & 0.950 & 1 & 8973.63 & 0.23 & 0.830 & 1 & 12341.46 & 0.22 & 0.950 & 1 \\
4 & 374.32 & 0.52 & 0.950 & 1 & 7132.09 & 0.46 & 0.830 & 1 & 6151.01 & 0.41 & 0.950 & 1 \\
5 & 564.49 & 1.04 & 0.950 & 1 & 5671.02 & 0.85 & 0.830 & 1 & 3110.85 & 0.72 & 0.950 & 1 \\
6 & 758.45 & 1.91 & 0.950 & 1 & 4271.96 & 1.50 & 0.830 & 1 & 1406.19 & 1.21 & 0.950 & 1 \\
7 & 876.39 & 3.27 & 0.950 & 1 & 2958.77 & 2.58 & 0.830 & 1 & 526.61 & 2.02 & 0.950 & 1 \\
8 & 831.86 & 5.26 & 0.950 & 1 & 1857.68 & 4.33 & 0.830 & 1 & 160.13 & 3.31 & 0.950 & 1 \\
9 & 616.02 & 7.97 & 0.950 & 1 & 1046.19 & 7.08 & 0.830 & 1 & 38.78 & 5.33 & 0.950 & 1 \\
10 & 336.16 & 11.49 & 0.950 & 1 & 523.09 & 11.31 & 0.830 & 1 & 7.30 & 8.43 & 0.950 & 1 \\
11 & 125.58 & 15.89 & 0.950 & 1 & 229.91 & 17.66 & 0.830 & 1 & 1.05 & 13.08 & 0.950 & 1 \\
12 & 28.95 & 21.29 & 0.950 & 1 & 88.03 & 27.01 & 0.830 & 1 & 0.11 & 19.87 & 0.950 & 1 \\
13 & 3.59 & 27.84 & 0.950 & 1 & 29.12 & 40.50 & 0.830 & 1 & 0.01 & 29.56 & 0.950 & 1 \\
14 & 0.20 & 35.71 & 0.950 & 1 & 8.24 & 59.73 & 0.830 & 1 & 0.00 & 43.04 & 0.950 & 1 \\
15 & 0.00 & 45.12 & 0.950 & 1 & 1.95 & 87.17 & 0.830 & 1 & 0.00 & 61.32 & 0.950 & 1 \\
16 & 0.00 & 56.40 & 0.950 & 1 & 0.37 & 127.32 & 0.830 & 1 & 0.00 & 85.89 & 0.950 & 1 \\
17 & 0.00 & 70.00 & 0.950 & 1 & 0.05 & 190.42 & 0.830 & 1 & 0.00 & 120.15 & 0.950 & 1 \\
18 & 0.00 & 86.58 & 0.950 & 1 & 0.00 & 307.88 & 0.830 & 1 & 0.00 & 173.33 & 0.950 & 1 \\
19 & 0.00 & 489.20 & 0.950 & 1 & 0.00 & 587.78 & 0.830 & 1 & 0.00 & 303.66 & 0.950 & 1 \\
20 & 9.91 & 0.57 & 0.810 & 2 & 20.98 & 0.65 & 0.660 & 2 & 0.00 & 588.15 & 0.950 & 1 \\
21 & 28.13 & 2.32 & 0.810 & 2 & 58.45 & 2.67 & 0.660 & 2 & 49.80 & 0.77 & 0.680 & 2 \\
22 & 63.36 & 6.67 & 0.810 & 2 & 125.53 & 7.53 & 0.660 & 2 & 137.00 & 3.23 & 0.680 & 2 \\
23 & 117.06 & 15.56 & 0.810 & 2 & 212.58 & 16.98 & 0.660 & 2 & 265.67 & 8.90 & 0.680 & 2 \\
24 & 173.22 & 30.99 & 0.810 & 2 & 268.28 & 32.34 & 0.660 & 2 & 352.74 & 18.92 & 0.680 & 2 \\
25 & 190.87 & 54.10 & 0.810 & 2 & 228.94 & 53.78 & 0.660 & 2 & 287.21 & 33.30 & 0.680 & 2 \\
26 & 143.02 & 84.87 & 0.810 & 2 & 119.18 & 80.57 & 0.660 & 2 & 128.14 & 51.11 & 0.680 & 2 \\
27 & 65.22 & 123.15 & 0.810 & 2 & 33.60 & 112.23 & 0.660 & 2 & 28.60 & 71.43 & 0.680 & 2 \\
28 & 14.54 & 170.51 & 0.810 & 2 & 4.11 & 149.61 & 0.660 & 2 & 2.85 & 93.96 & 0.680 & 2 \\
29 & 0.89 & 232.94 & 0.810 & 2 & 0.12 & 196.62 & 0.660 & 2 & 0.10 & 119.25 & 0.680 & 2 \\
30 & ... & ... & ... & ... & ... & ... & ... & ... & 0.00 & 149.53 & 0.680 & 2 \\
\hline
\label{Tab:MGE_c}
\end{tabular}
\end{center}
\end{minipage}
\end{table*}
\section{JAM-MCMC burn-in chains, posterior chains and corner plots}
\label{B:chains} 

\begin{figure*}
{\includegraphics[width=1.0\textwidth]{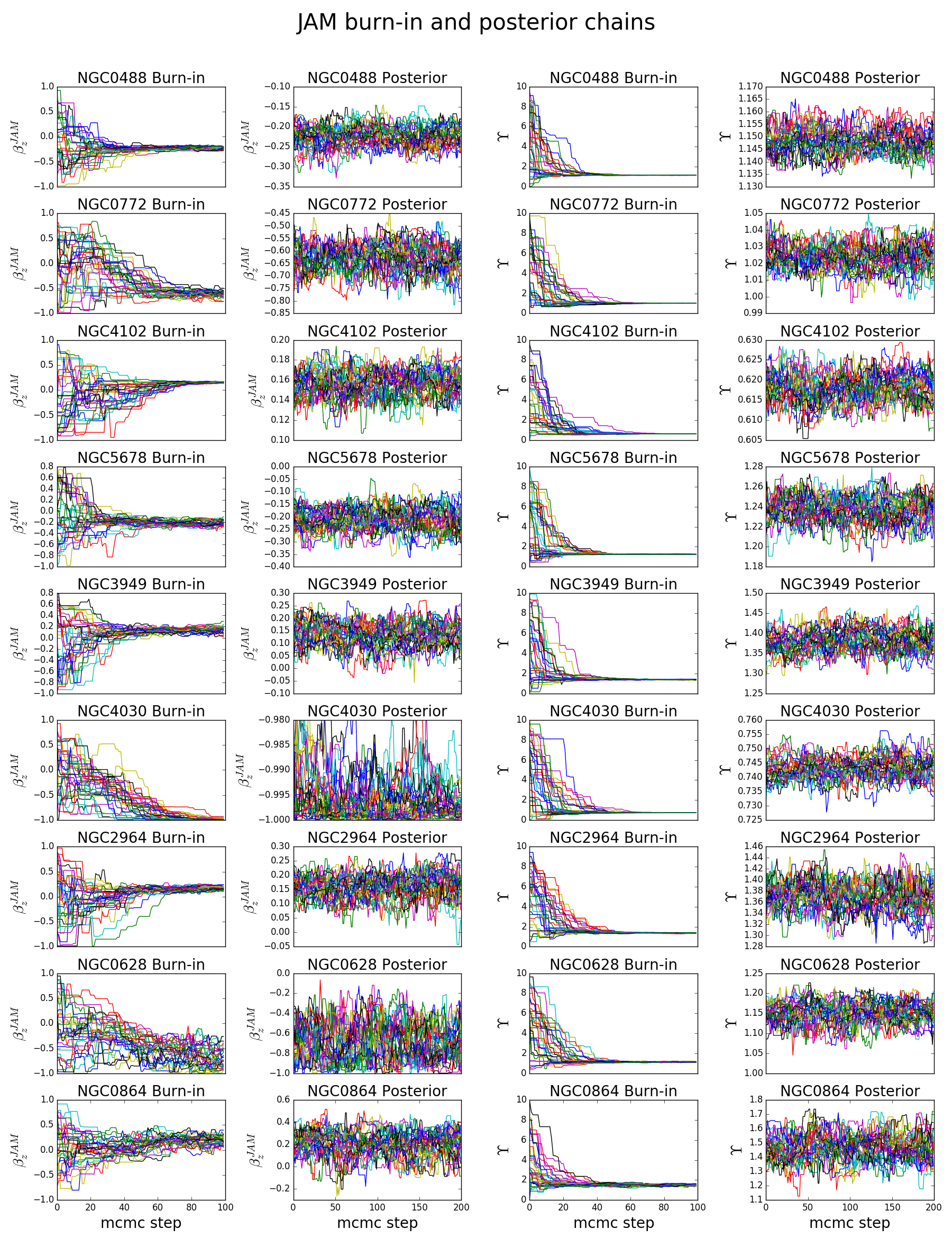}
}
\caption{Burn-in and posterior chain plots of the velocity anisotropy ($\beta^{\mathrm{JAM}}_z$) and mass-to-light ratio ($\Upsilon$) in JAM-MCMC method.}
\label{fig:Ca_jam}
\end{figure*}

\addtocounter{figure}{-1}
\addtocounter{subfigure}{1}

\begin{figure*}
{\includegraphics[width=1.0\textwidth]{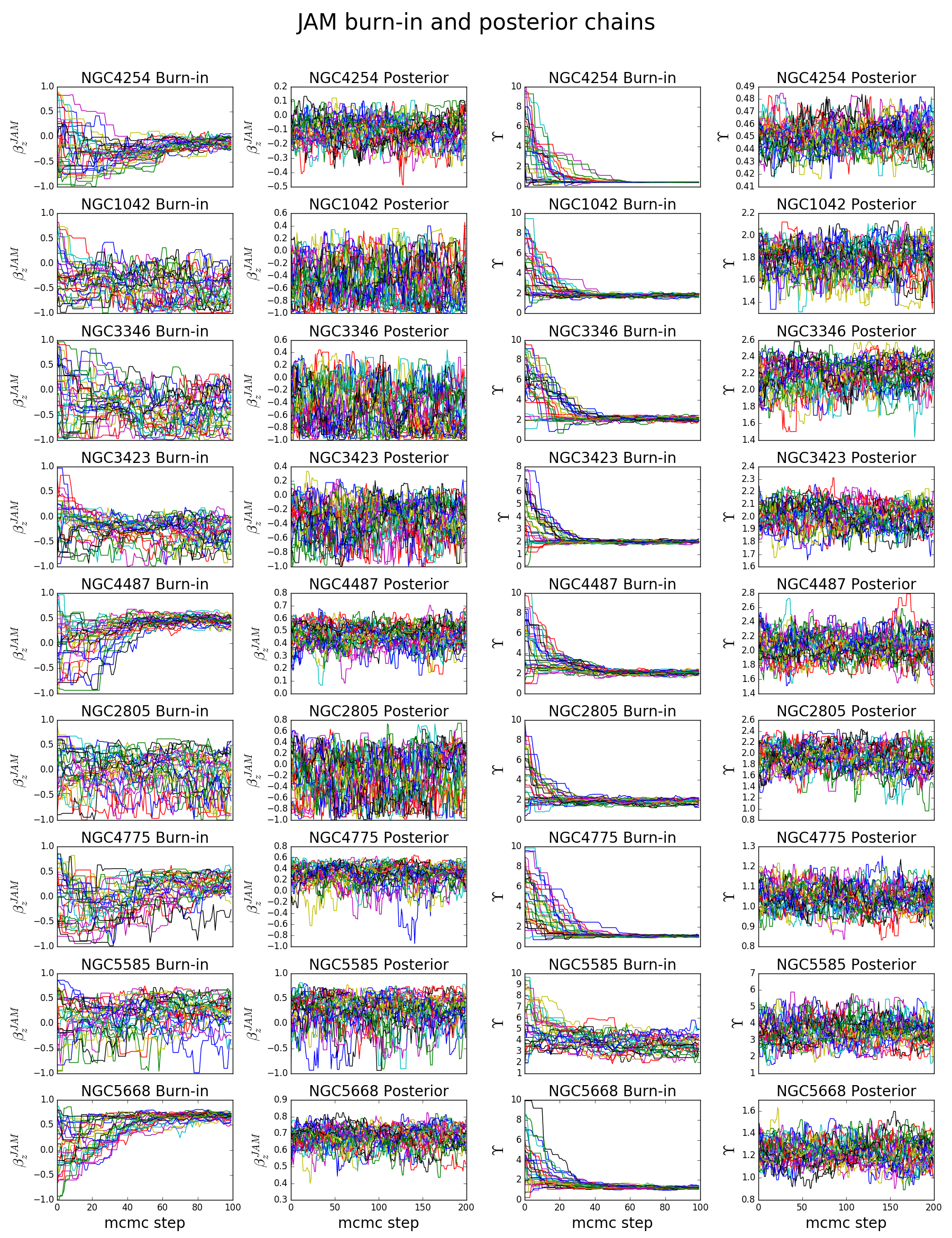}
}
\caption{{\it -- continuation}}
\label{fig:Cb_jam}
\end{figure*}

\begin{figure*}
{\includegraphics[width=0.4\textwidth]{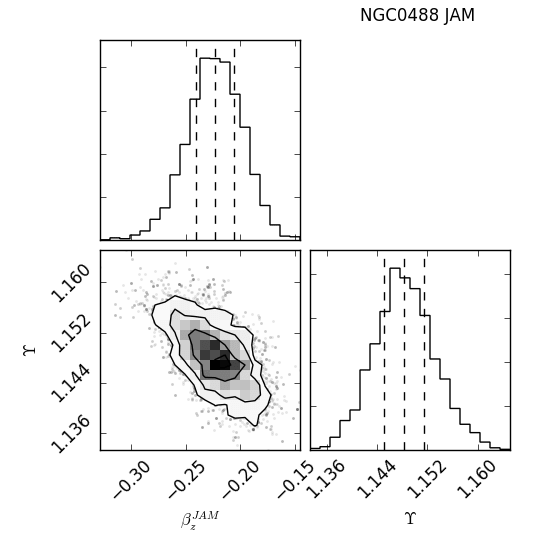}
\includegraphics[width=0.4\textwidth]{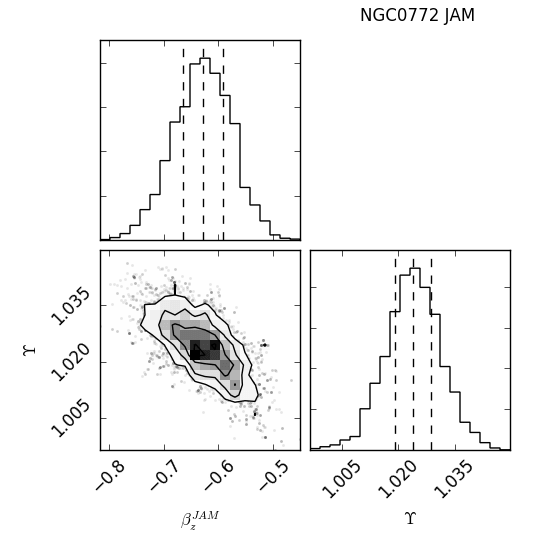}
\includegraphics[width=0.4\textwidth]{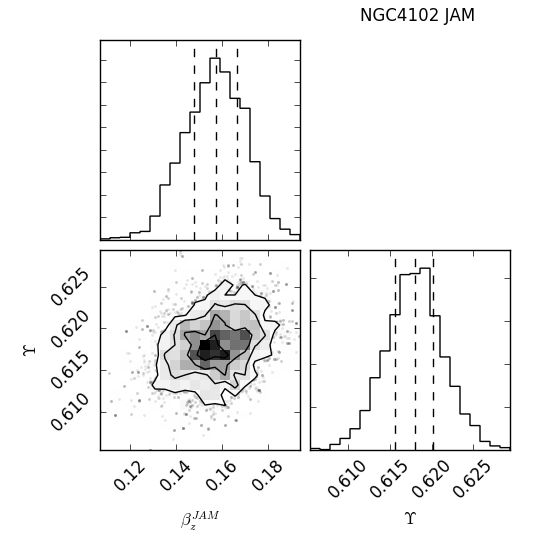}
\includegraphics[width=0.4\textwidth]{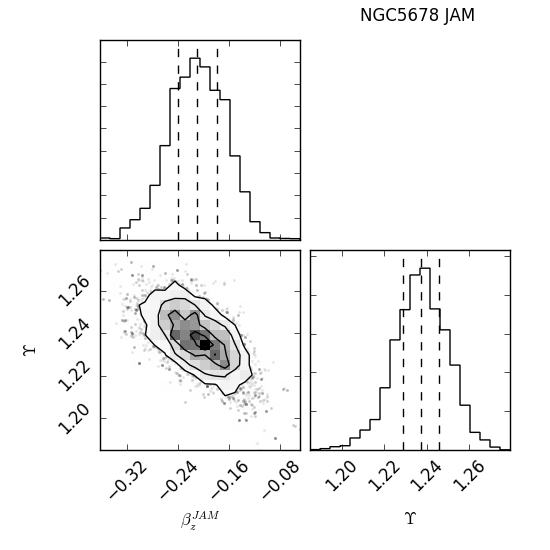}
\includegraphics[width=0.4\textwidth]{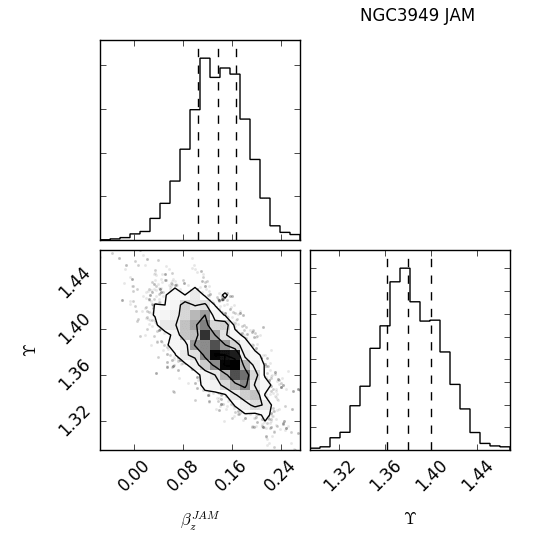}
\includegraphics[width=0.4\textwidth]{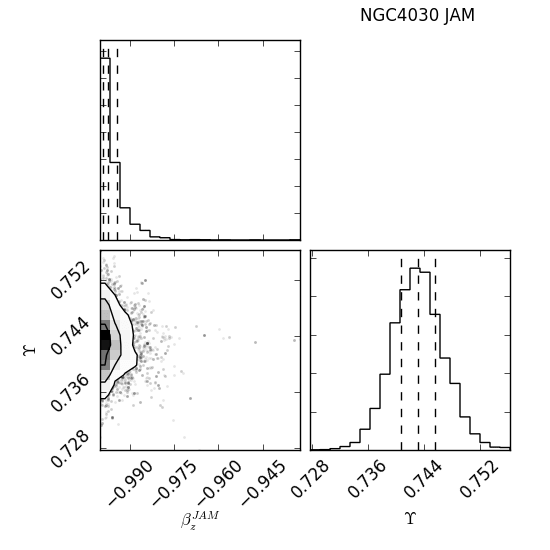}
}
\caption{Corner plot that shows the covariance between the model parameters of the 18 galaxies from the JAM-MCMC method.
 Overplotted are the median value of the distributions related to the fitted parameters, together with their uncertainties at 25 and 75 percentiles.
}
\label{fig:Tr_JAM_a}
\end{figure*}

\addtocounter{figure}{-1}
\addtocounter{subfigure}{1}

\begin{figure*}
{\includegraphics[width=0.4\textwidth]{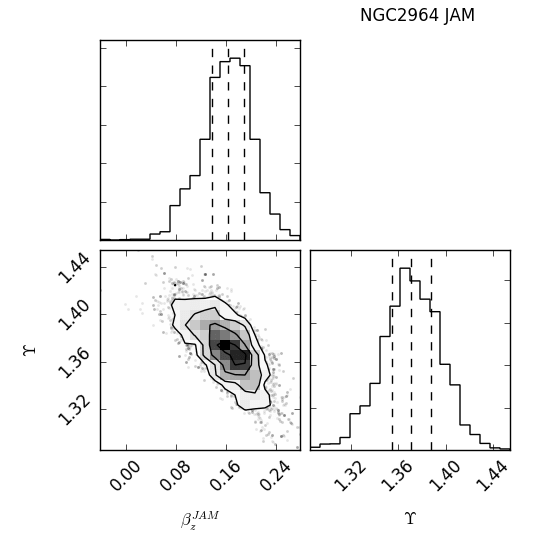}
\includegraphics[width=0.4\textwidth]{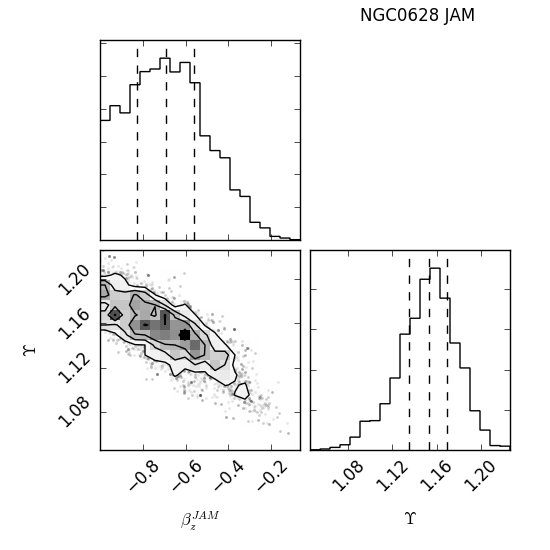}
\includegraphics[width=0.4\textwidth]{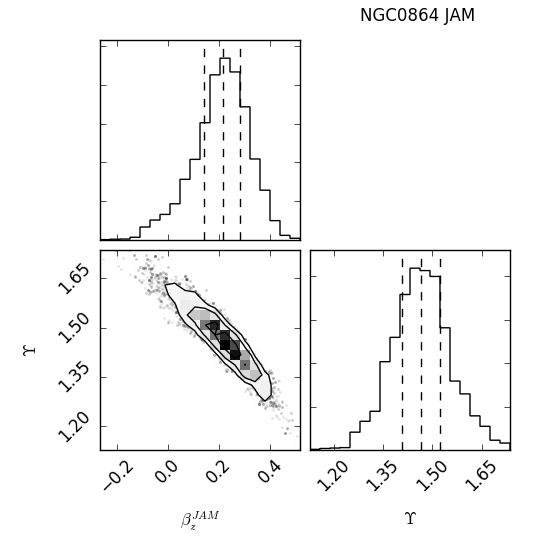}
\includegraphics[width=0.4\textwidth]{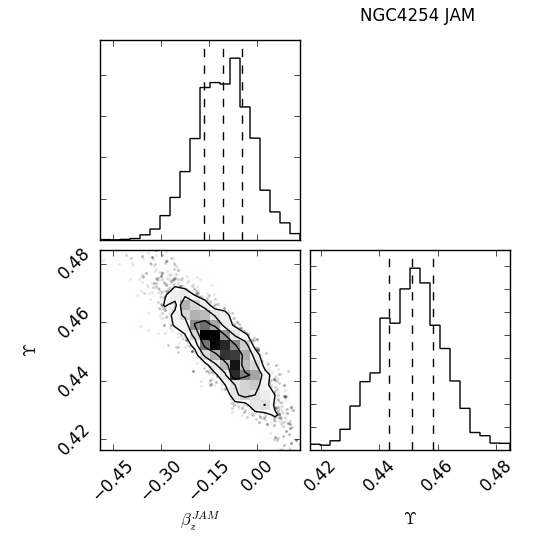}
\includegraphics[width=0.4\textwidth]{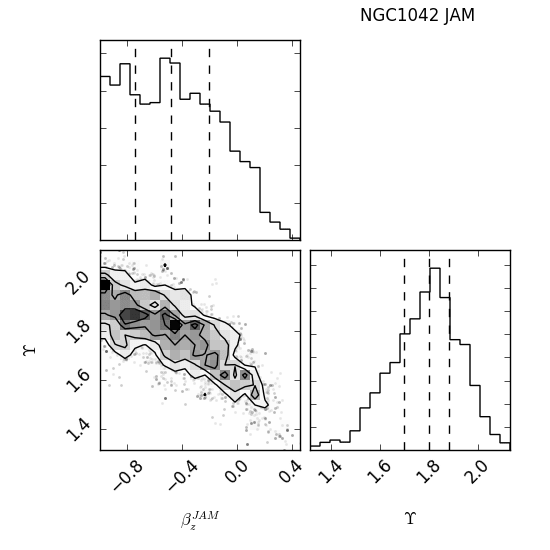}
\includegraphics[width=0.4\textwidth]{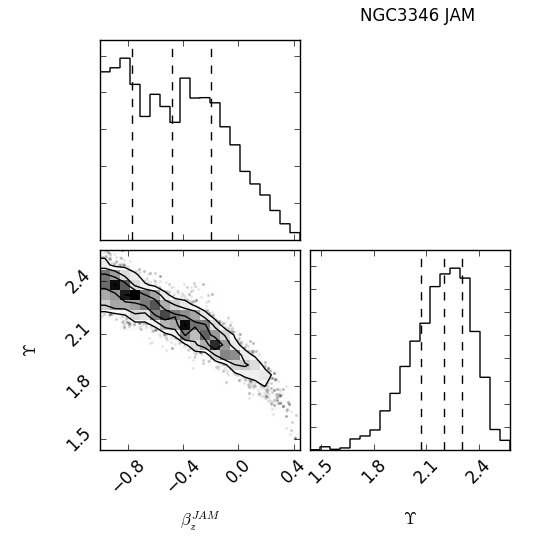}
}
\caption{{\it -- continuation}}
\label{fig:Tr_JAM_b}
\end{figure*}

\addtocounter{figure}{-1}
\addtocounter{subfigure}{1}

\begin{figure*}
{\includegraphics[width=0.4\textwidth]{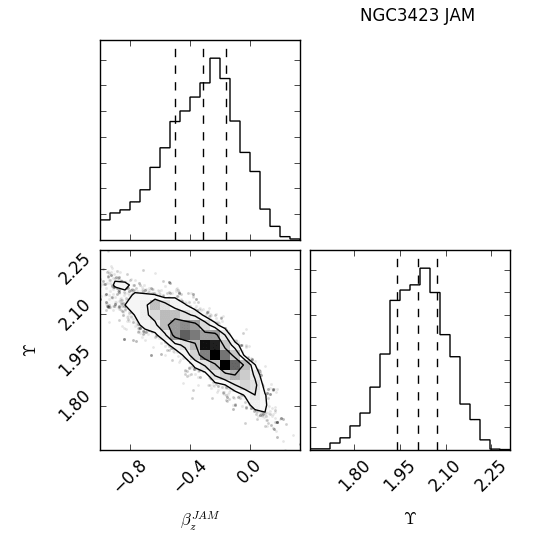}
\includegraphics[width=0.4\textwidth]{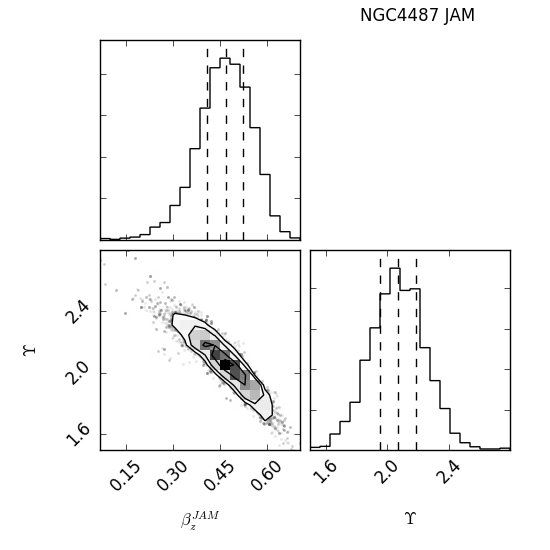}
\includegraphics[width=0.4\textwidth]{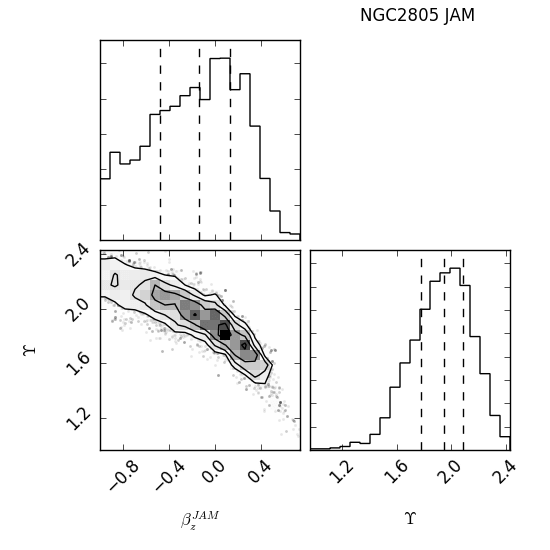}
\includegraphics[width=0.4\textwidth]{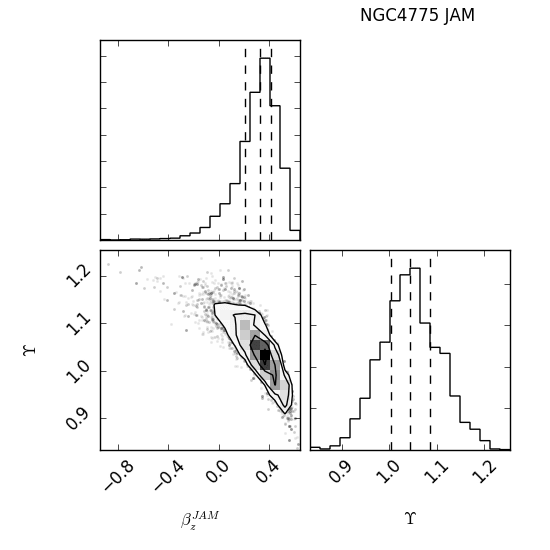}
\includegraphics[width=0.4\textwidth]{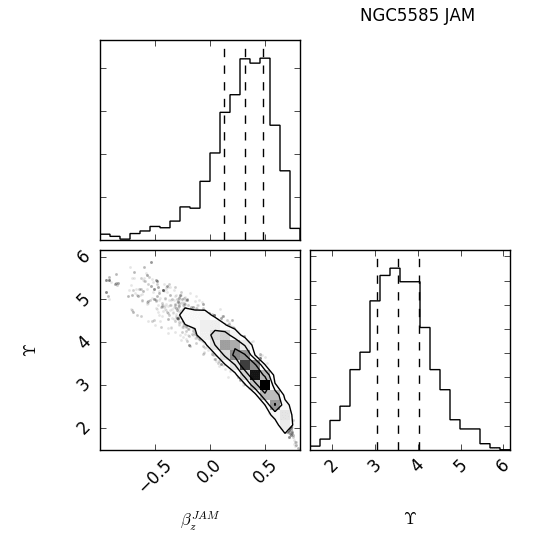}
\includegraphics[width=0.4\textwidth]{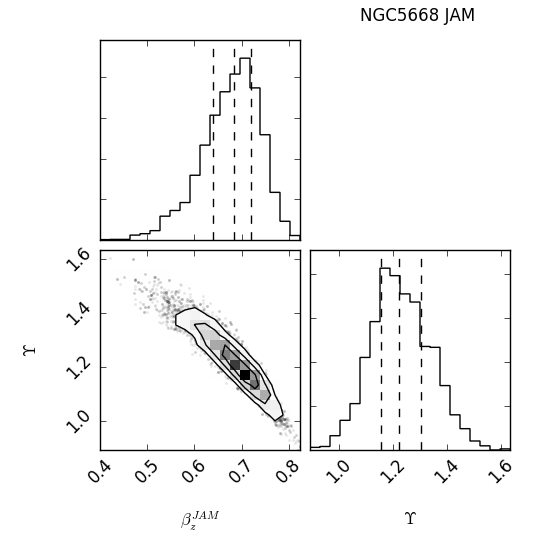}
}
\caption{{\it -- continuation}}
\label{fig:Tr_JAM_c}
\end{figure*}

\section{ADC-MCMC burn-in chains, posterior chains and corner plots}
\label{C:chains} 
\begin{figure*}
{\includegraphics[width=0.8\textwidth]{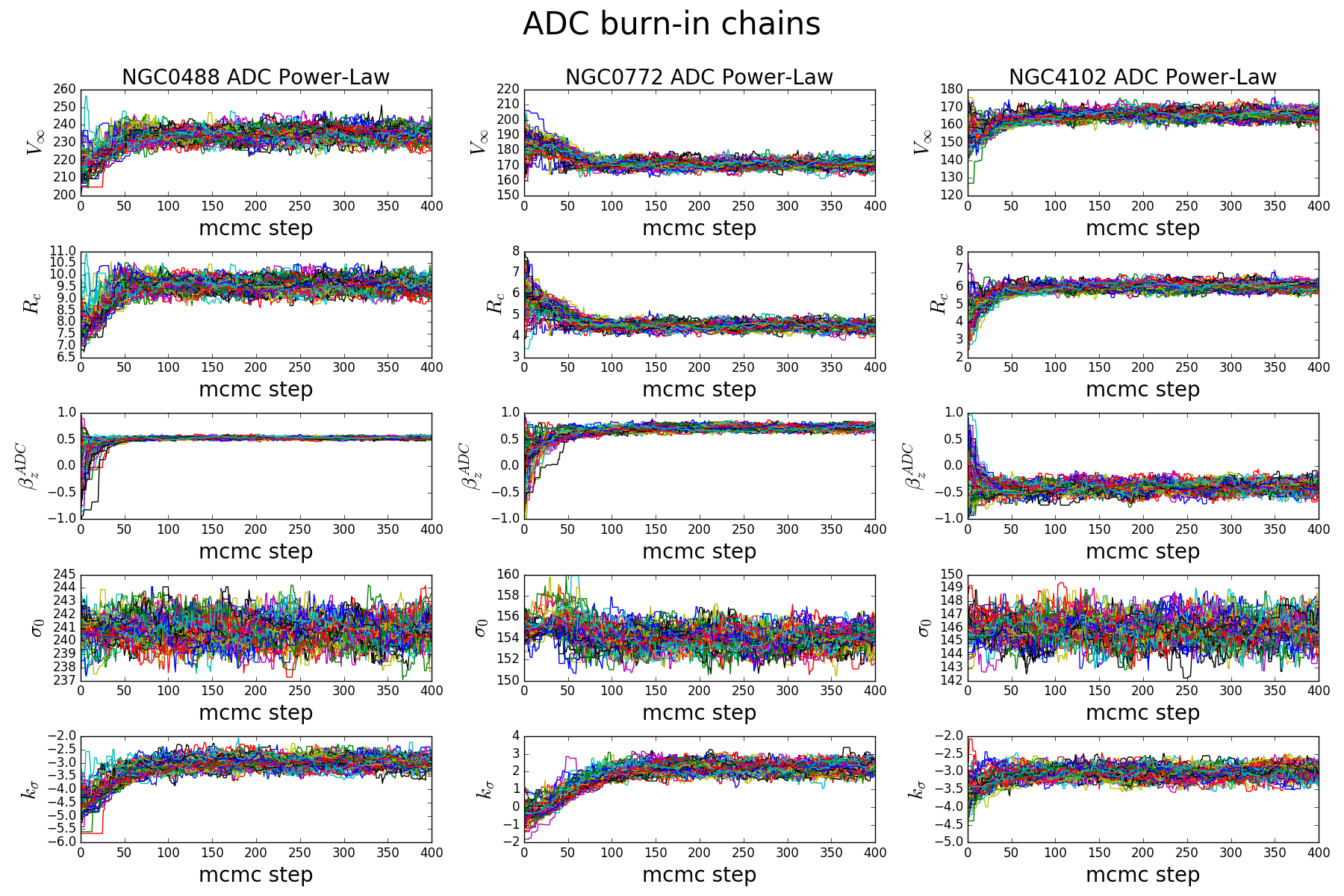}
\includegraphics[width=0.8\textwidth]{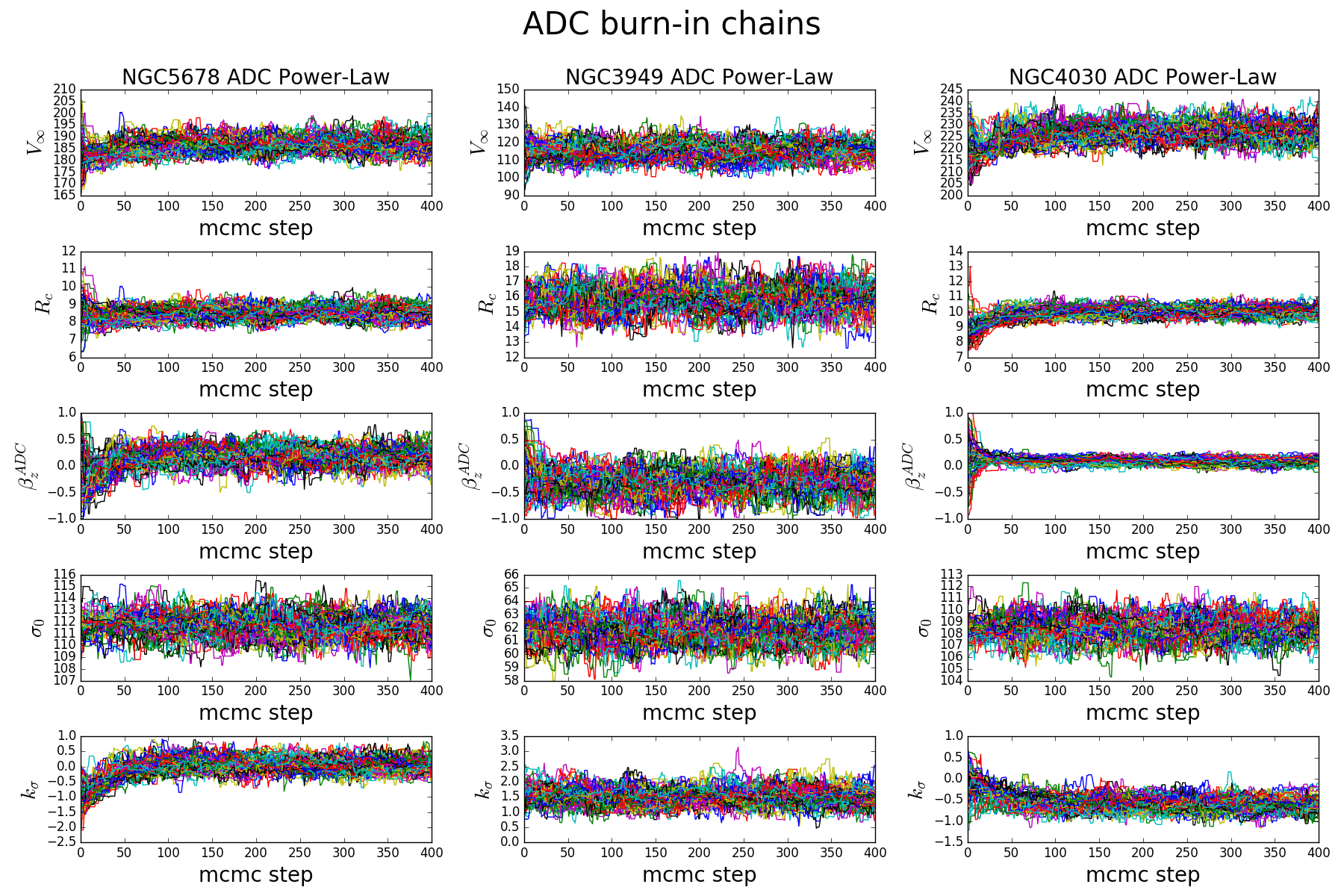}
}
\caption{Burn-in chain plots of model parameters for the 18 galaxies from ADC-MCMC method.}
\label{fig:Ca_pw}
\end{figure*}

\addtocounter{figure}{-1}
\addtocounter{subfigure}{1}

\begin{figure*}
{\includegraphics[width=0.8\textwidth]{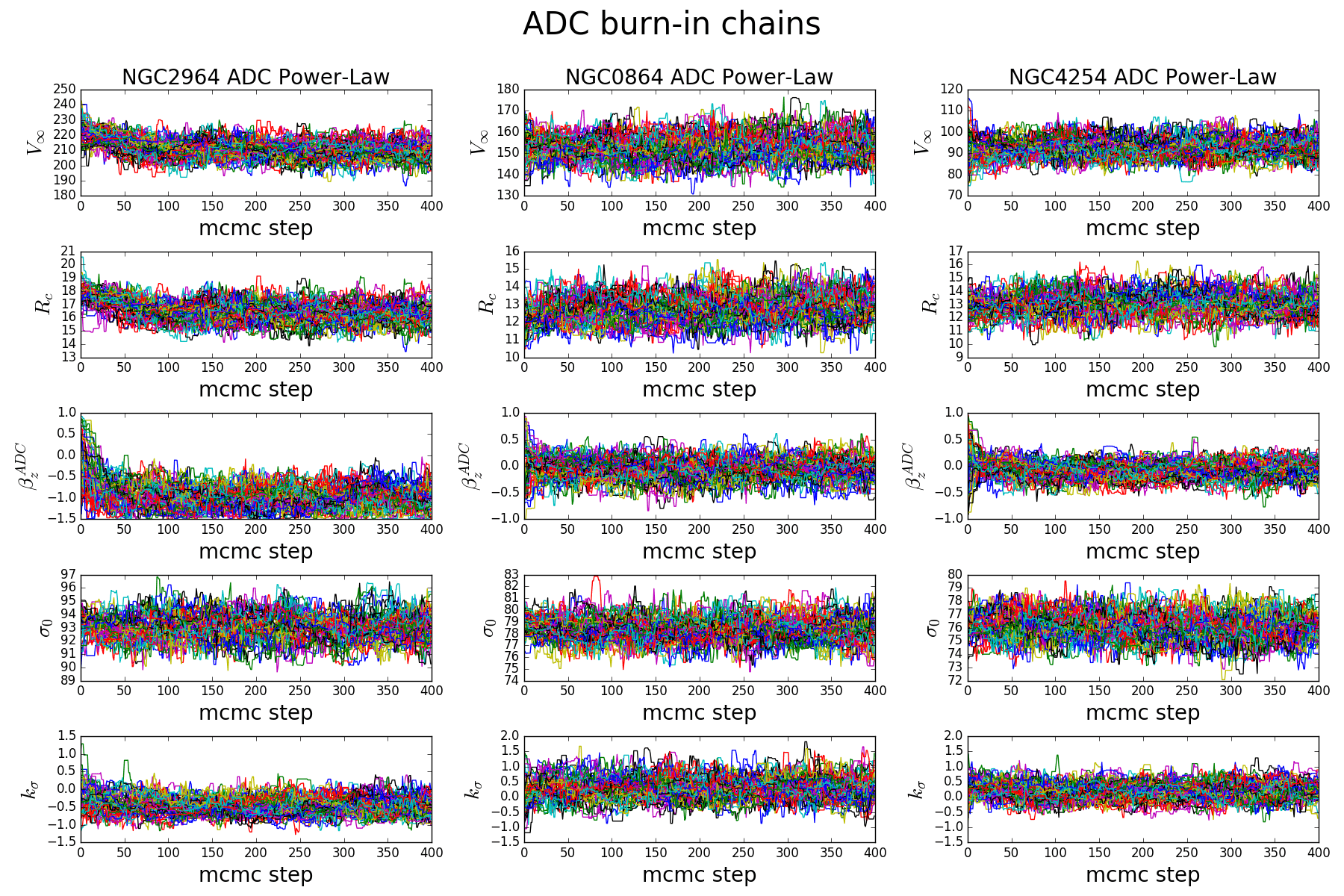}
\includegraphics[width=0.8\textwidth]{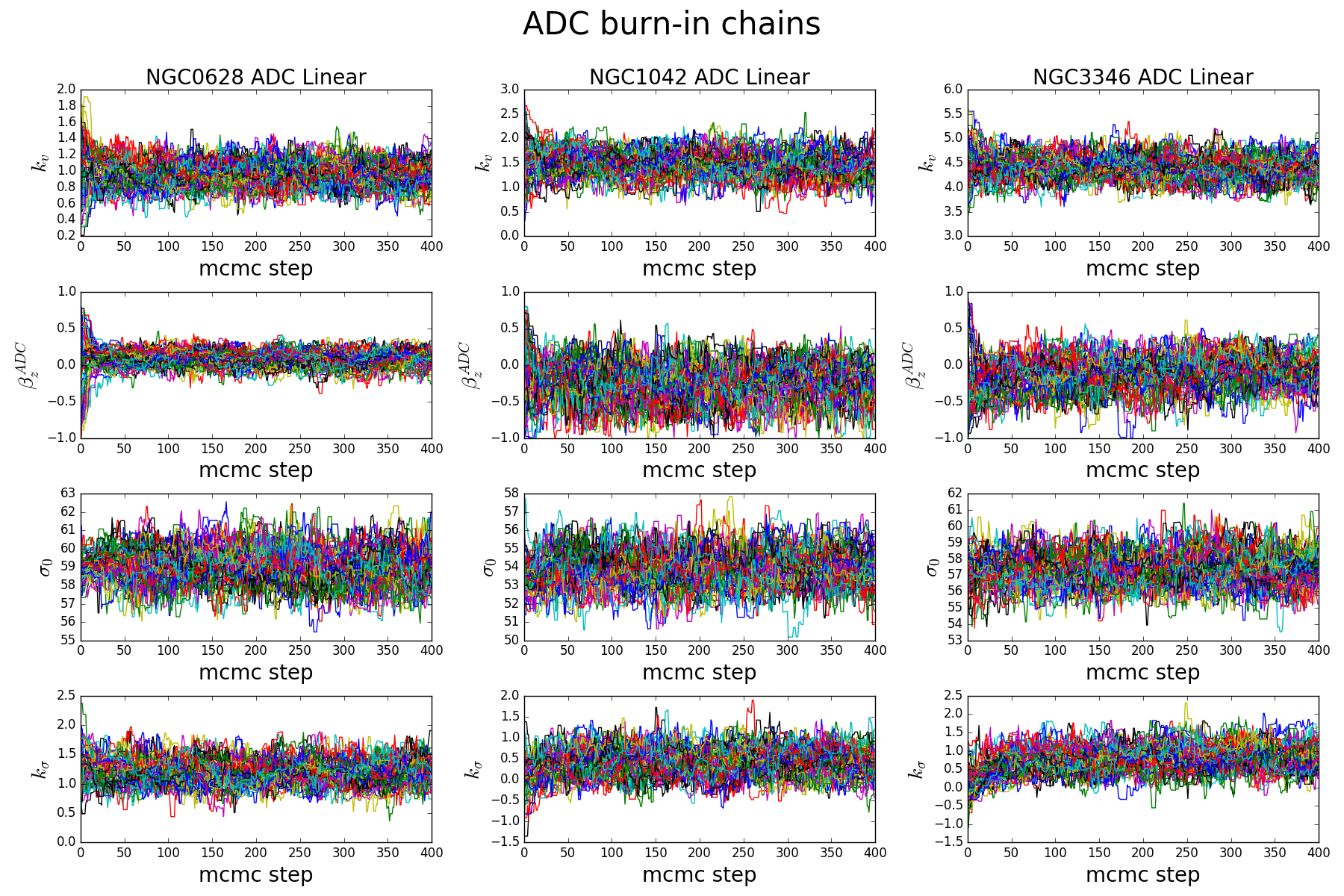}
}
\caption{{\it -- continuation}}
\label{fig:Cc_pw}
\end{figure*}

\addtocounter{figure}{-1}
\addtocounter{subfigure}{1}

\begin{figure*}
{\includegraphics[width=0.8\textwidth]{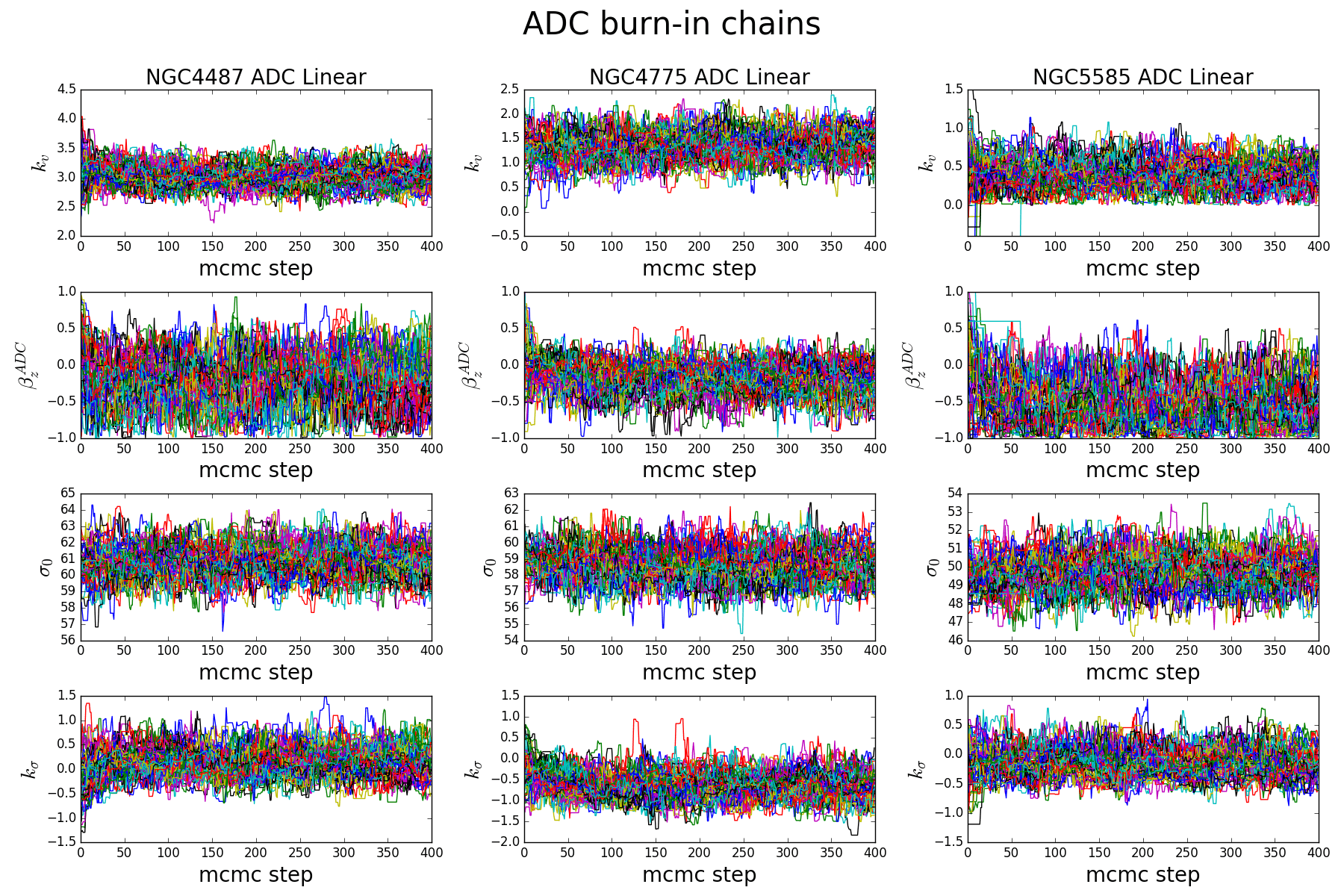}
}
\caption{{\it -- continuation}}
\label{fig:Cb_lin}
\end{figure*}


\begin{figure*}
{\includegraphics[width=0.8\textwidth]{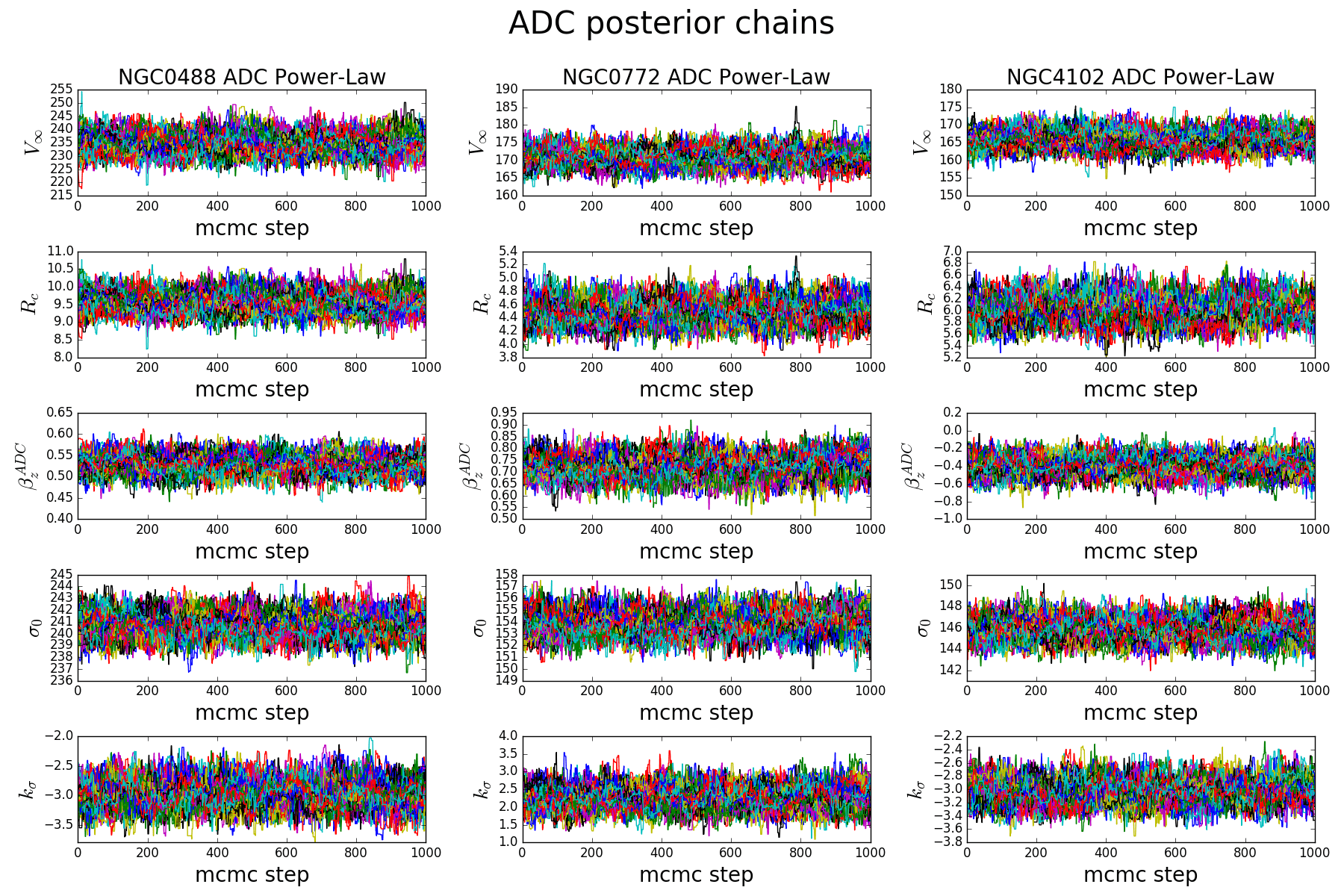}
\includegraphics[width=0.8\textwidth]{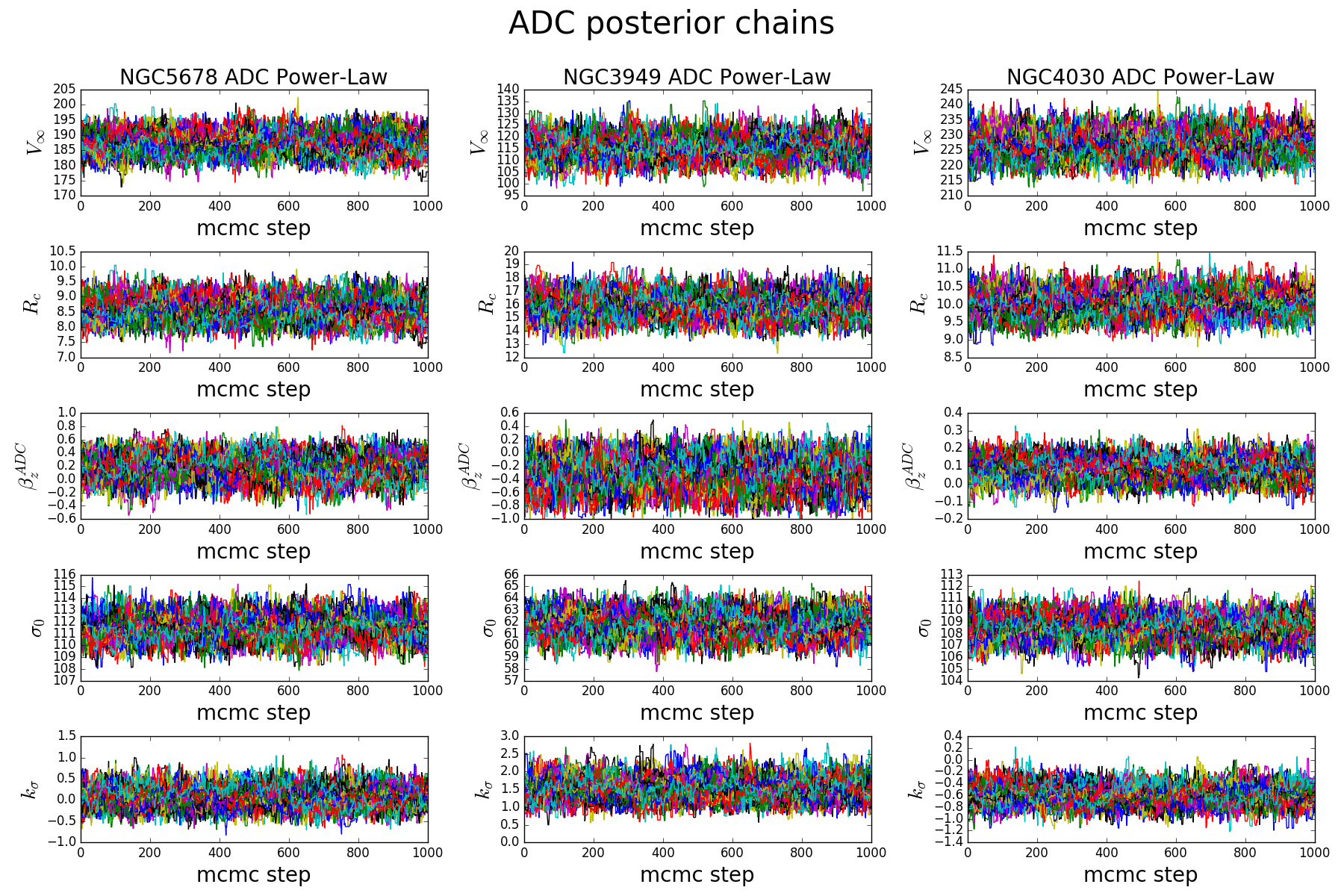}
}
\caption{Posterior chain plots of model parameters for the 18 galaxies from ADC-MCMC method.}
\label{fig:Ca_pw}
\end{figure*}

\addtocounter{figure}{-1}
\addtocounter{subfigure}{1}

\begin{figure*}
{\includegraphics[width=0.8\textwidth]{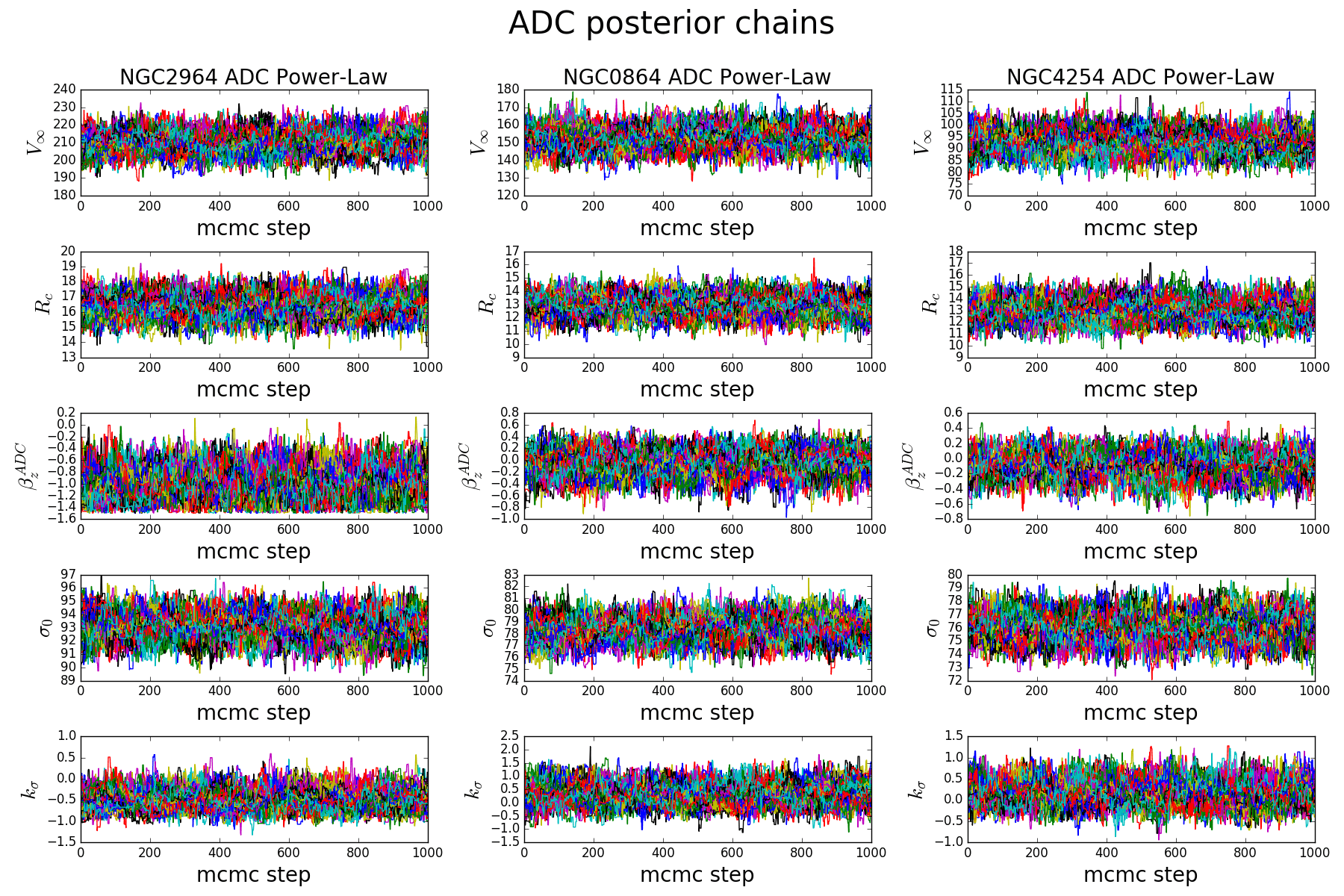}
\includegraphics[width=0.8\textwidth]{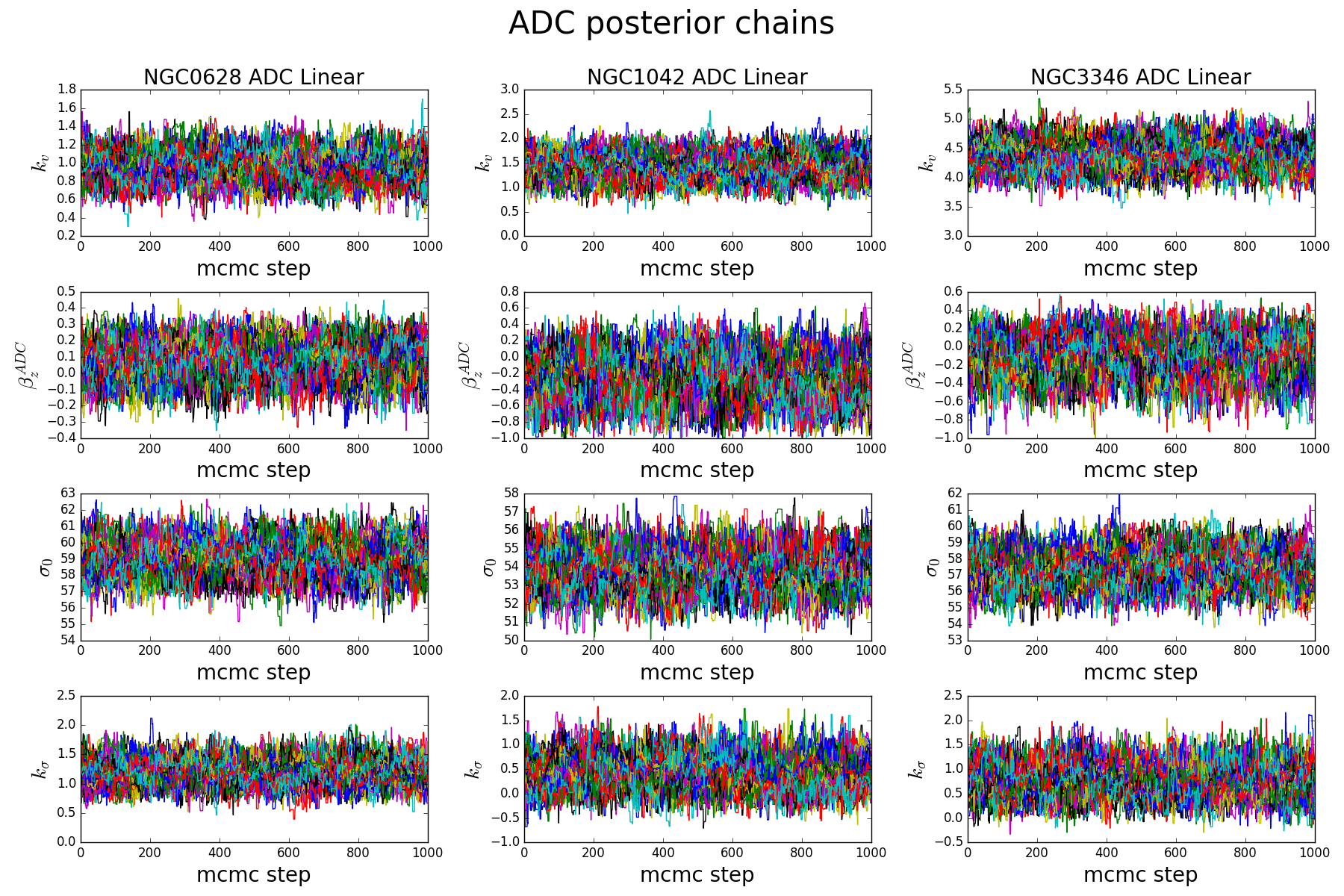}
}
\caption{{\it -- continuation}}
\label{fig:Cc_pw}
\end{figure*}

\addtocounter{figure}{-1}
\addtocounter{subfigure}{1}

\begin{figure*}
{\includegraphics[width=0.8\textwidth]{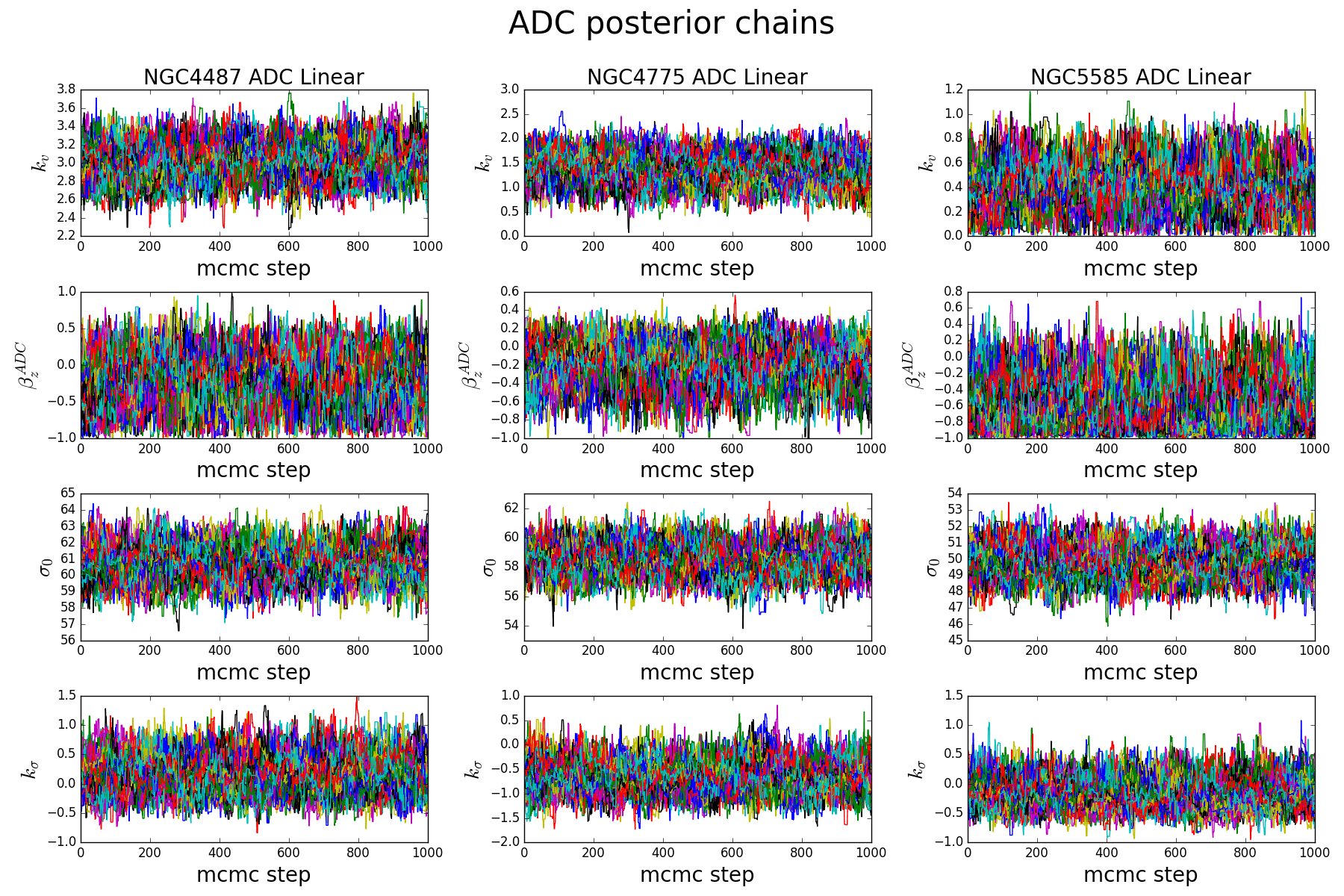}
}
\caption{{\it -- continuation}}
\label{fig:Cb_lin}
\end{figure*}
\begin{figure*}
{\includegraphics[width=0.65\textwidth]{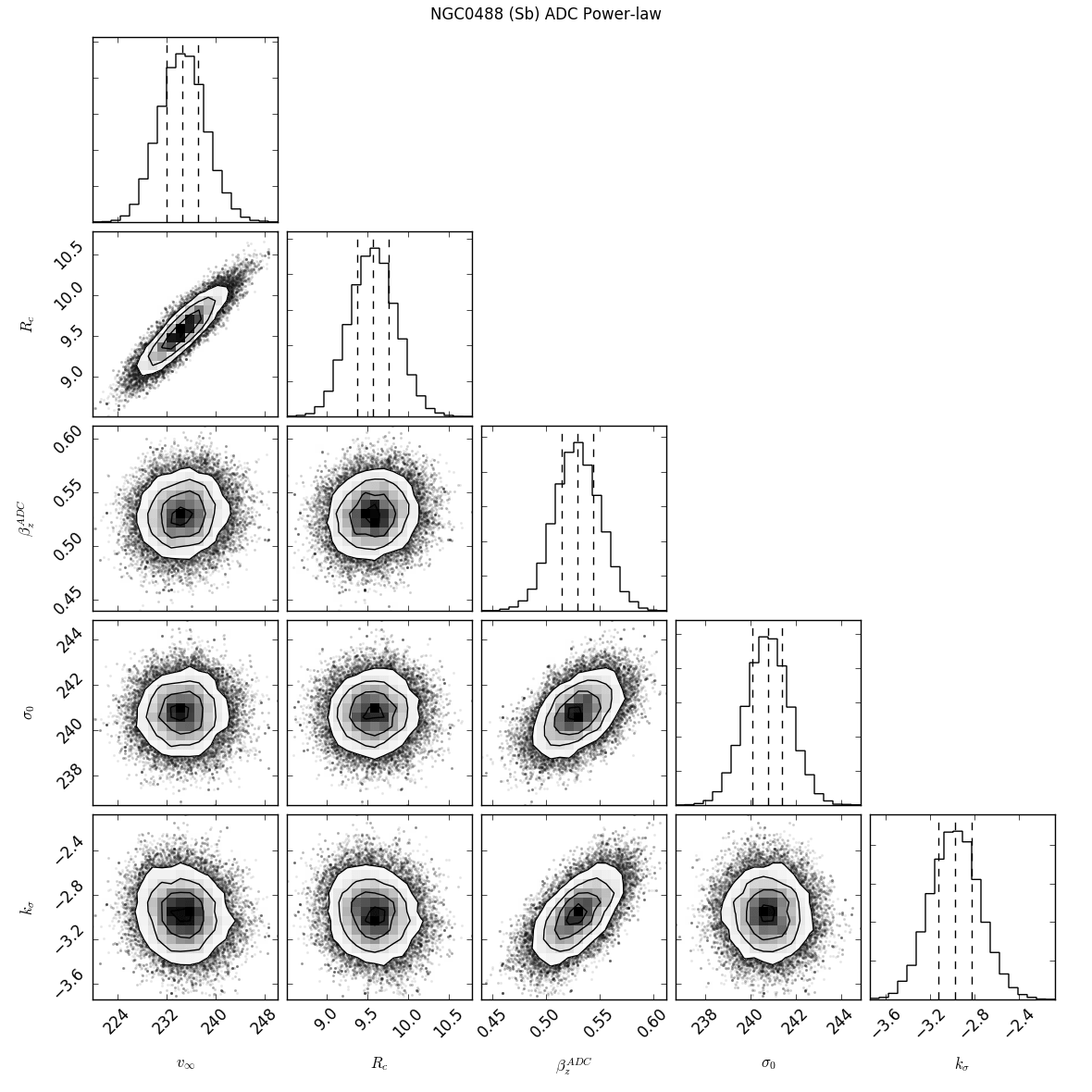}
\includegraphics[width=0.65\textwidth]{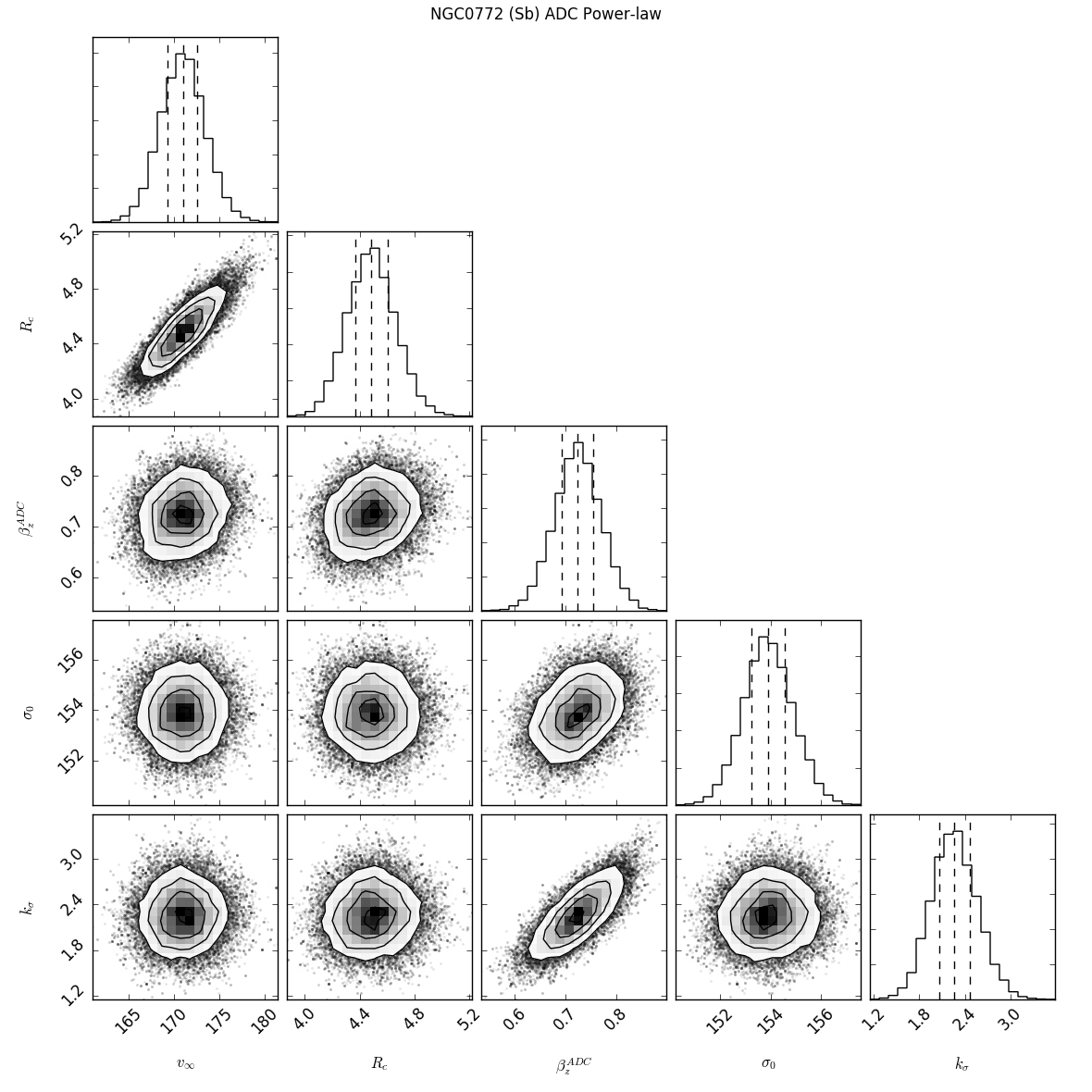}}
\caption{Corner plot shows the covariance between the model parameters of the 18 galaxies from ADC-MCMC method.
 Overplotted are the median value of the distributions related to the fitted parameters, together with their uncertainties at 25 and 75 percentiles.
}
\label{fig:TrA1}
\end{figure*}

\addtocounter{figure}{-1}
\addtocounter{subfigure}{1}

\begin{figure*}
{\includegraphics[width=0.65\textwidth]{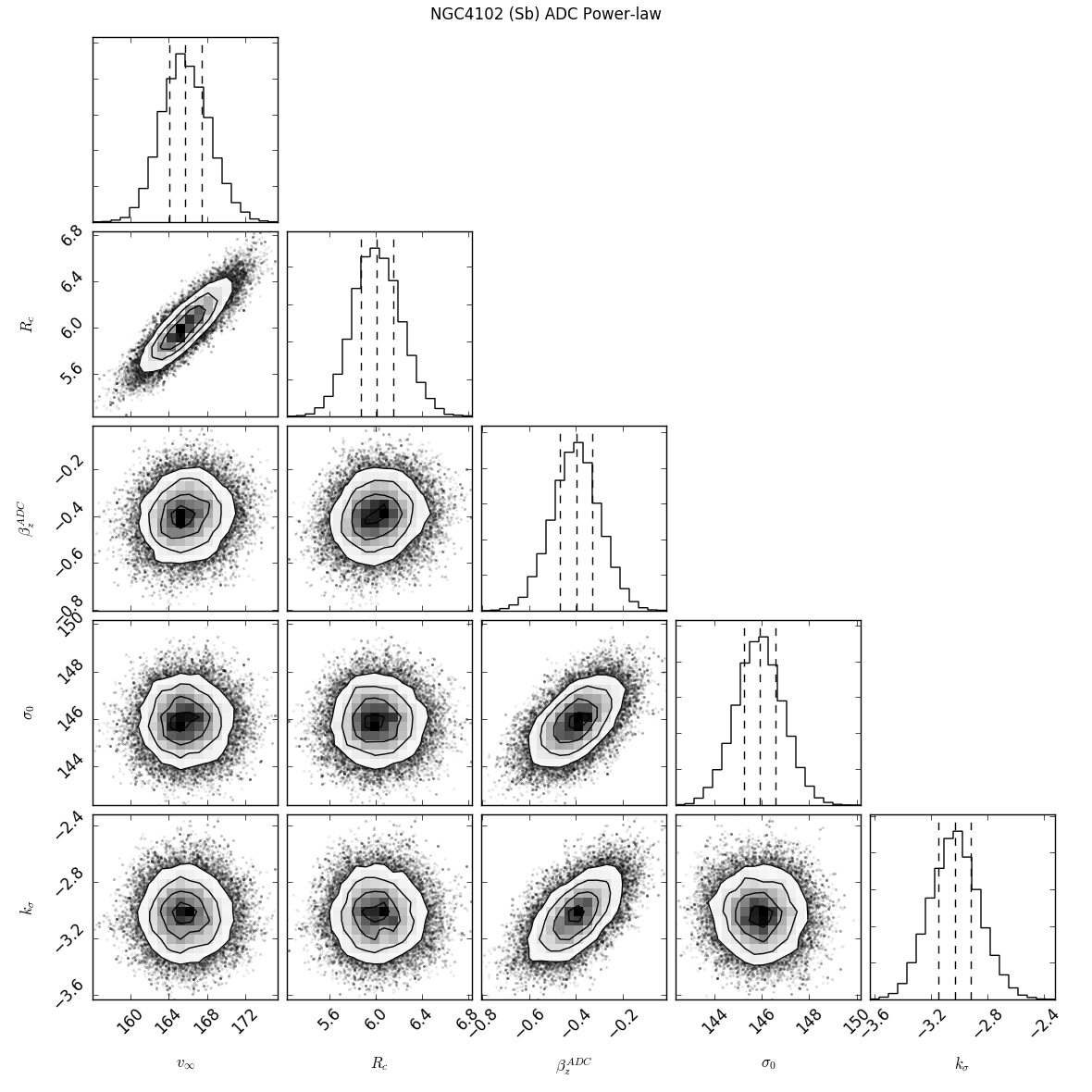}
\includegraphics[width=0.65\textwidth]{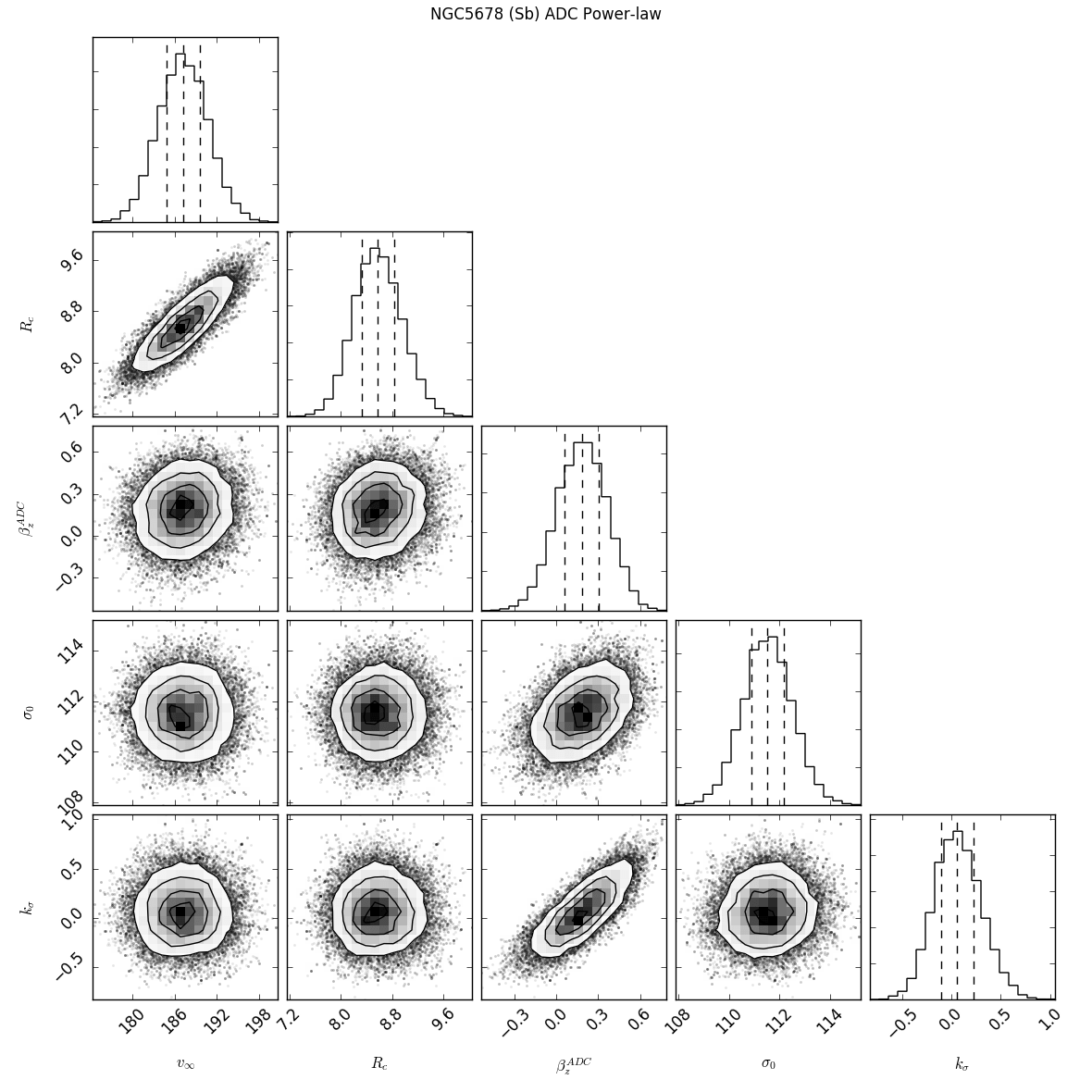}}
\caption{{\it -- continuation}}
\label{fig:TrA2}
\end{figure*}

\addtocounter{figure}{-1}
\addtocounter{subfigure}{1}

\begin{figure*}
{\includegraphics[width=0.65\textwidth]{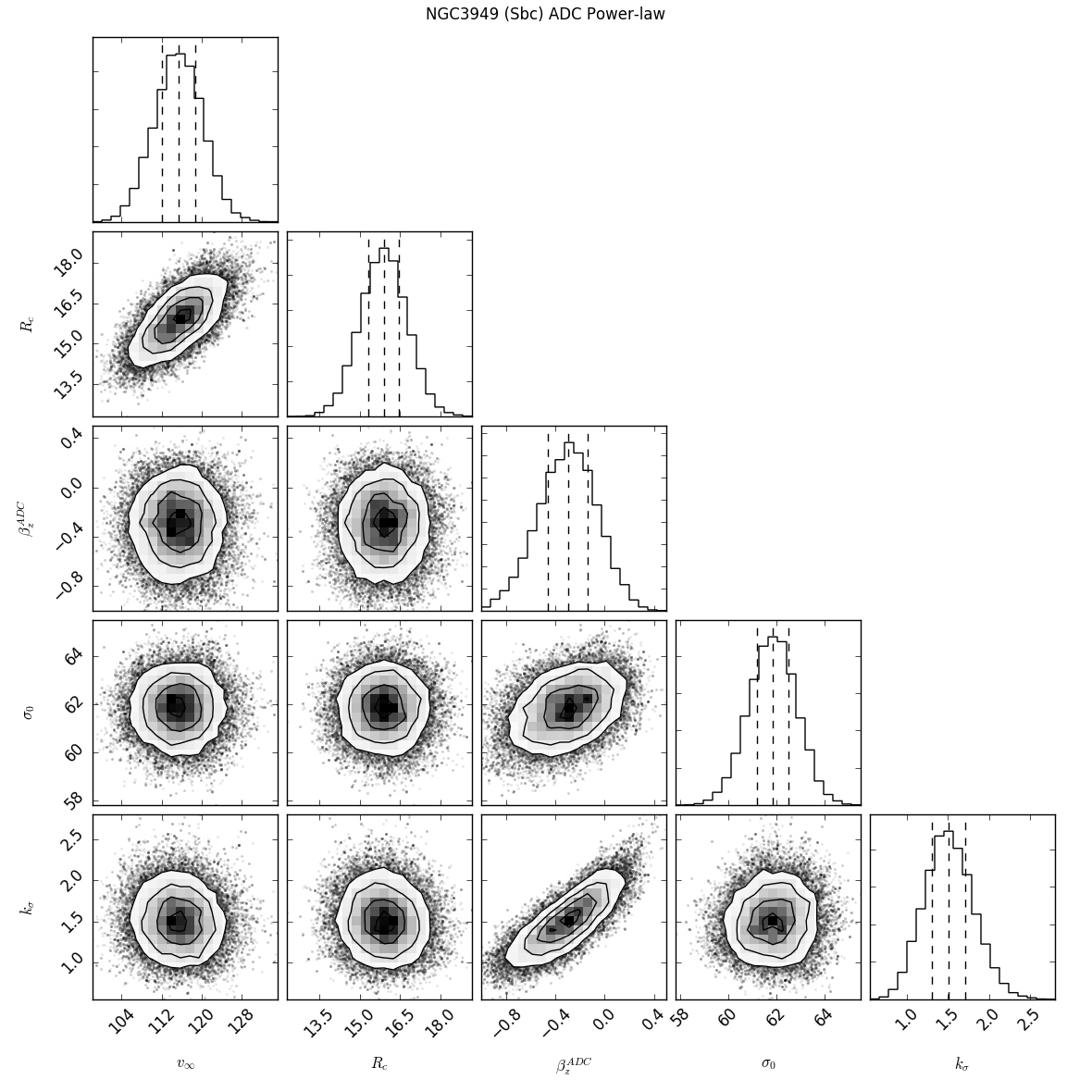}
\includegraphics[width=0.65\textwidth]{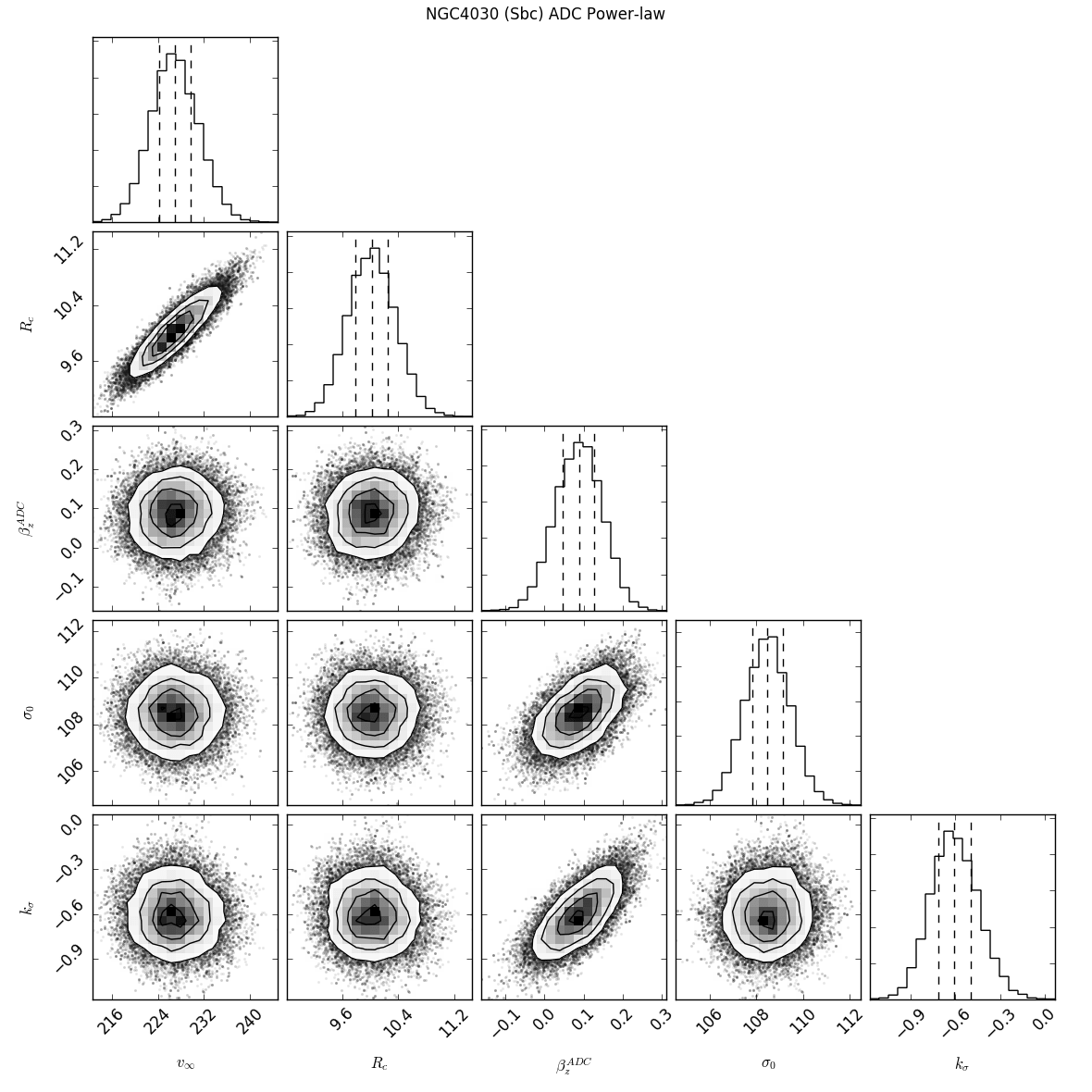}}
\caption{{\it -- continuation}}
\label{fig:TrA3}
\end{figure*}

\clearpage
\addtocounter{figure}{-1}
\addtocounter{subfigure}{1}

\begin{figure*}
{\includegraphics[width=0.65\textwidth]{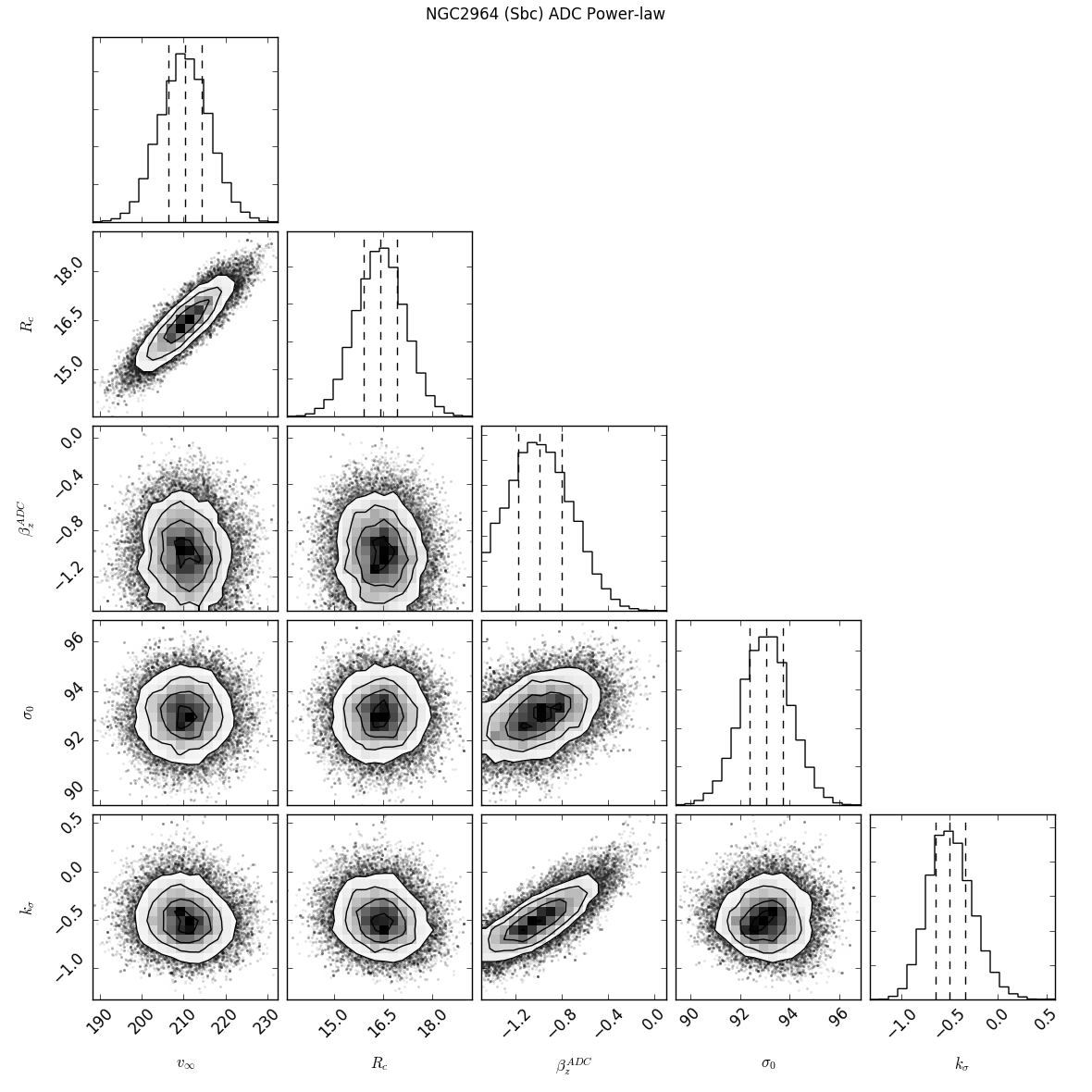}
\includegraphics[width=0.65\textwidth]{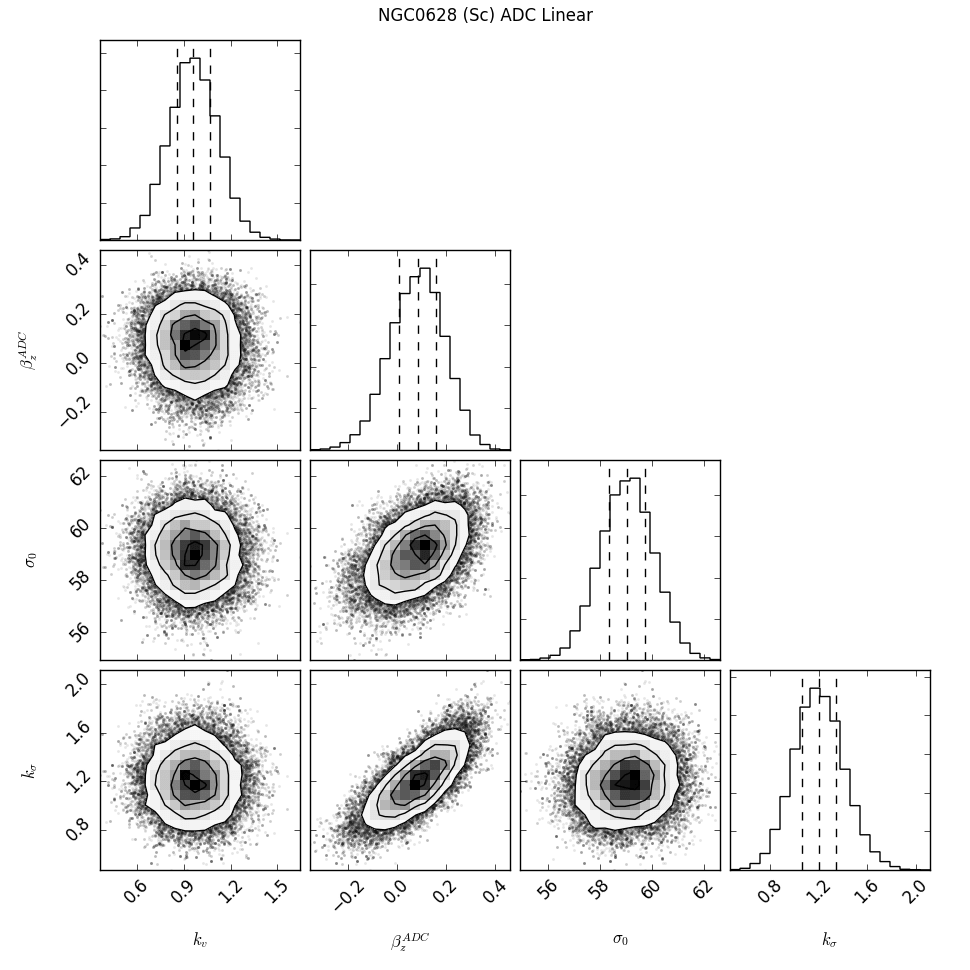}}
\caption{{\it -- continuation}}
\label{fig:TrA4}
\end{figure*}

\addtocounter{figure}{-1}
\addtocounter{subfigure}{1}

\begin{figure*}
{\includegraphics[width=0.65\textwidth]{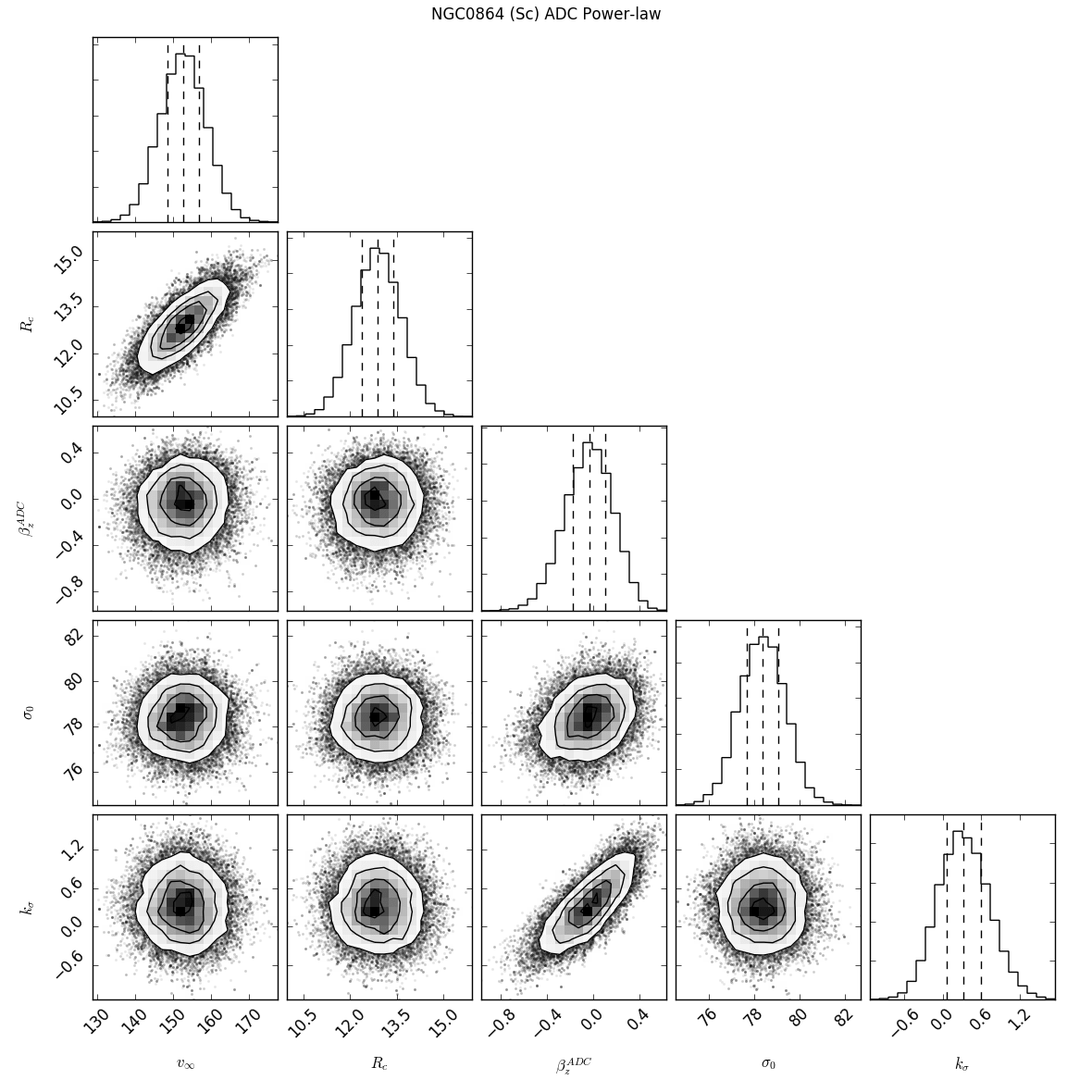}
\includegraphics[width=0.65\textwidth]{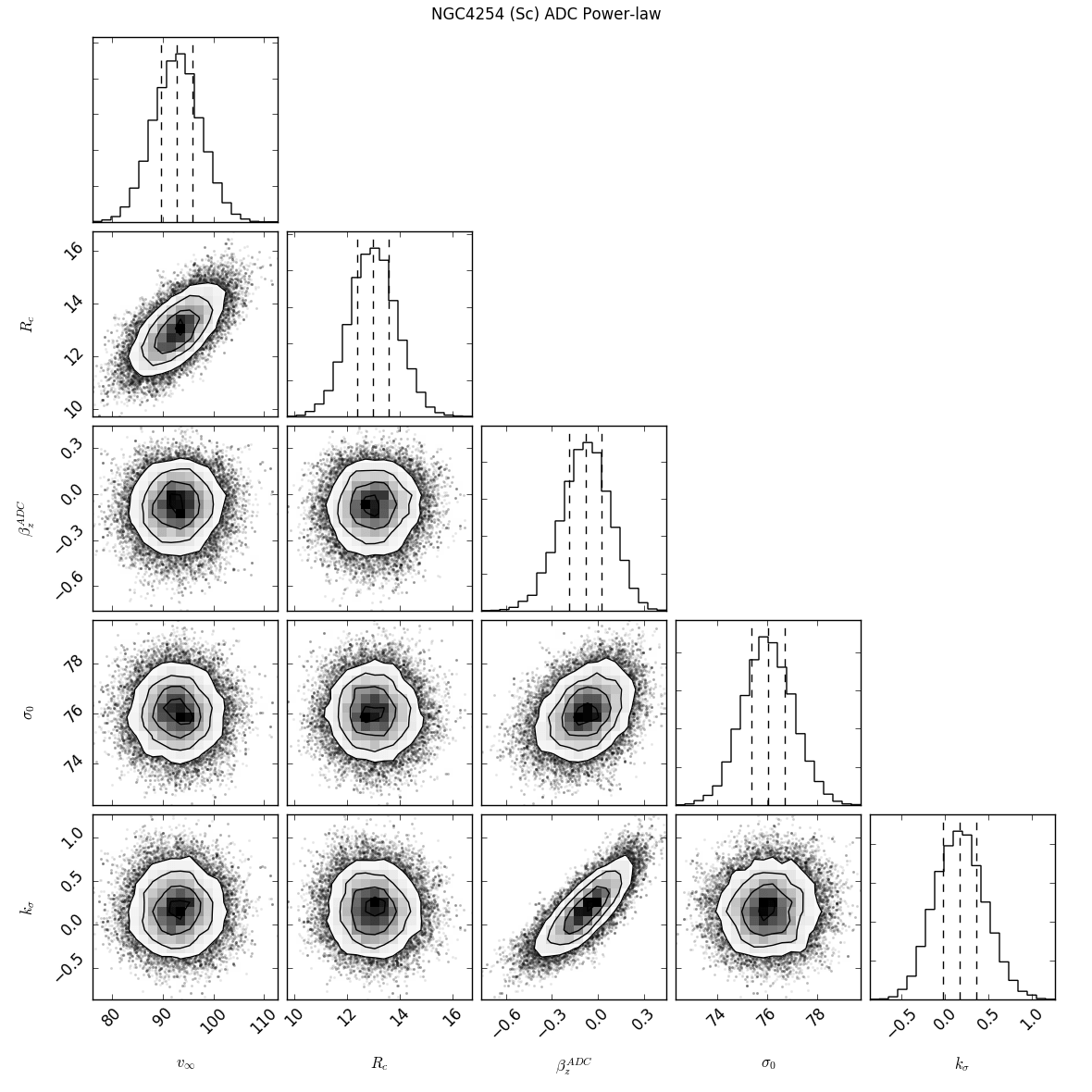}}
\caption{{\it -- continuation}}
\label{fig:TrA4}
\end{figure*}

\addtocounter{figure}{-1}
\addtocounter{subfigure}{1}

\begin{figure*}
{\includegraphics[width=0.65\textwidth]{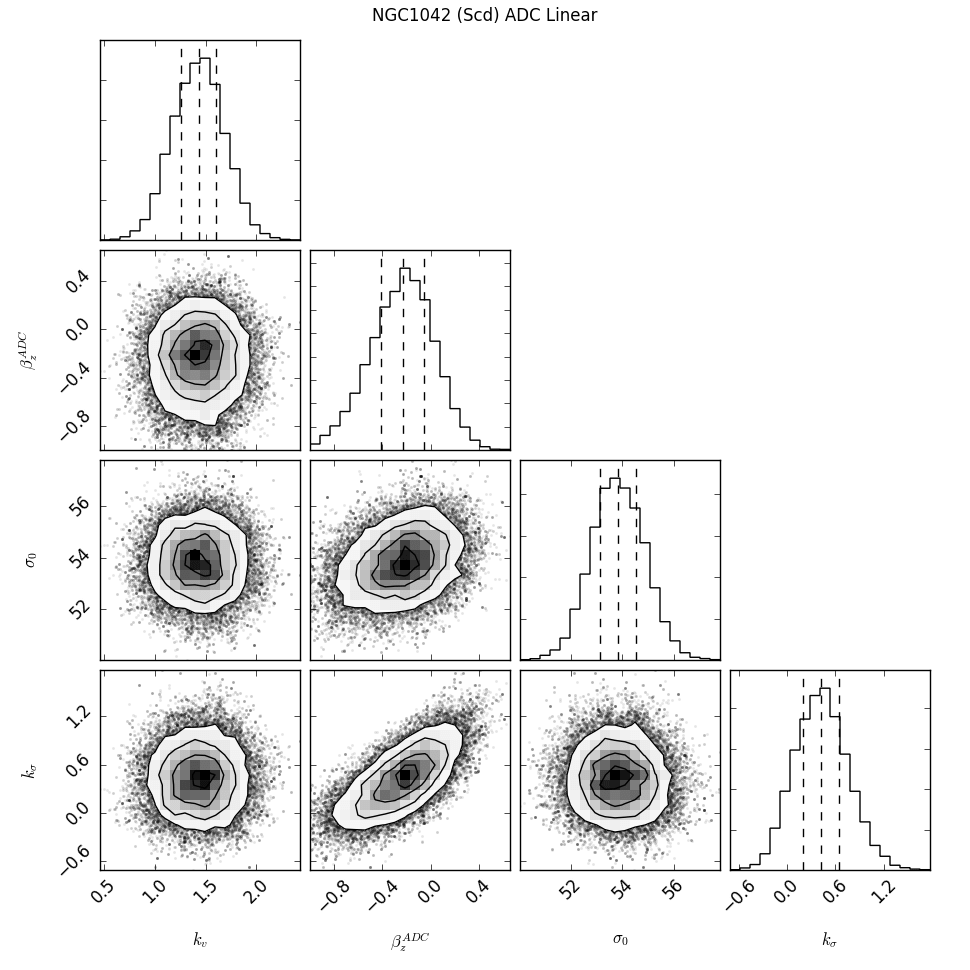}
\includegraphics[width=0.65\textwidth]{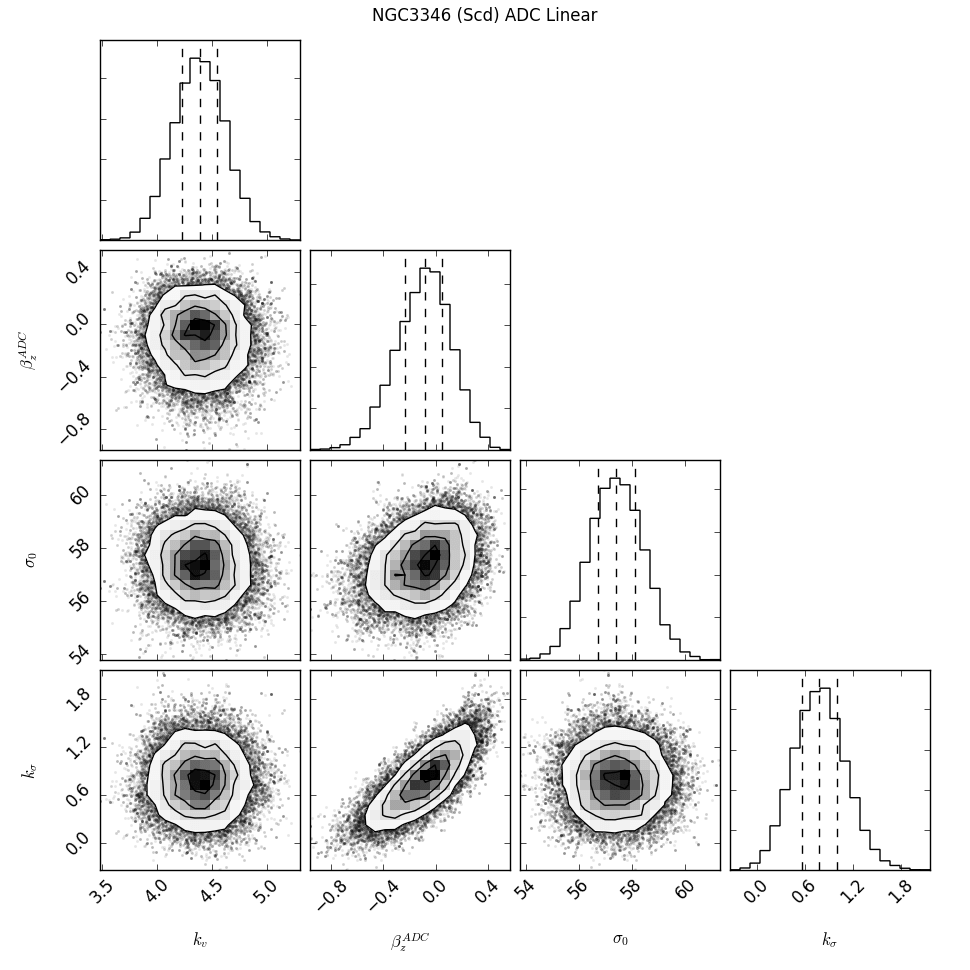}}
\caption{{\it -- continuation}}
\label{fig:TrA4}
\end{figure*}

\begin{figure*}
{\includegraphics[width=0.65\textwidth]{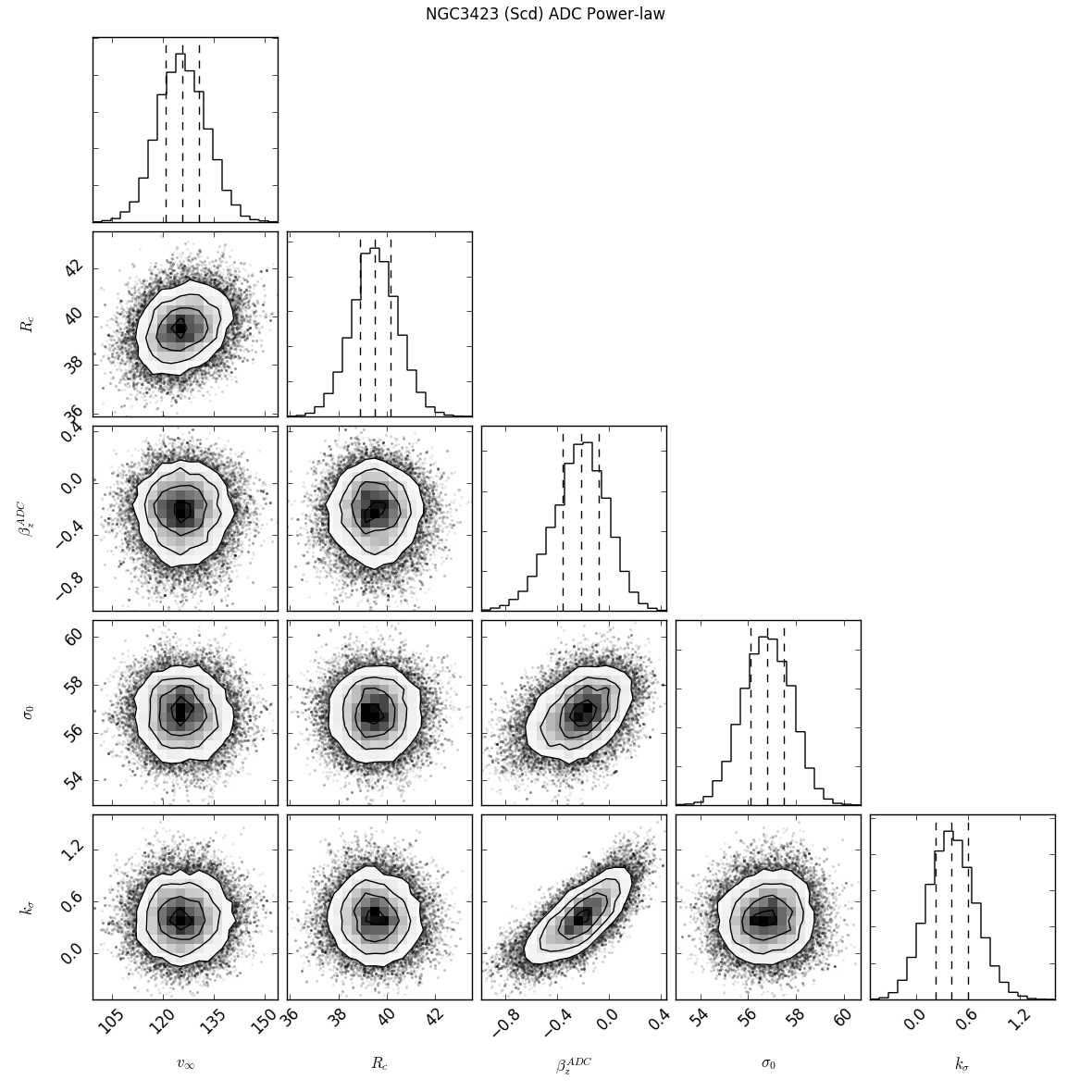}
\includegraphics[width=0.65\textwidth]{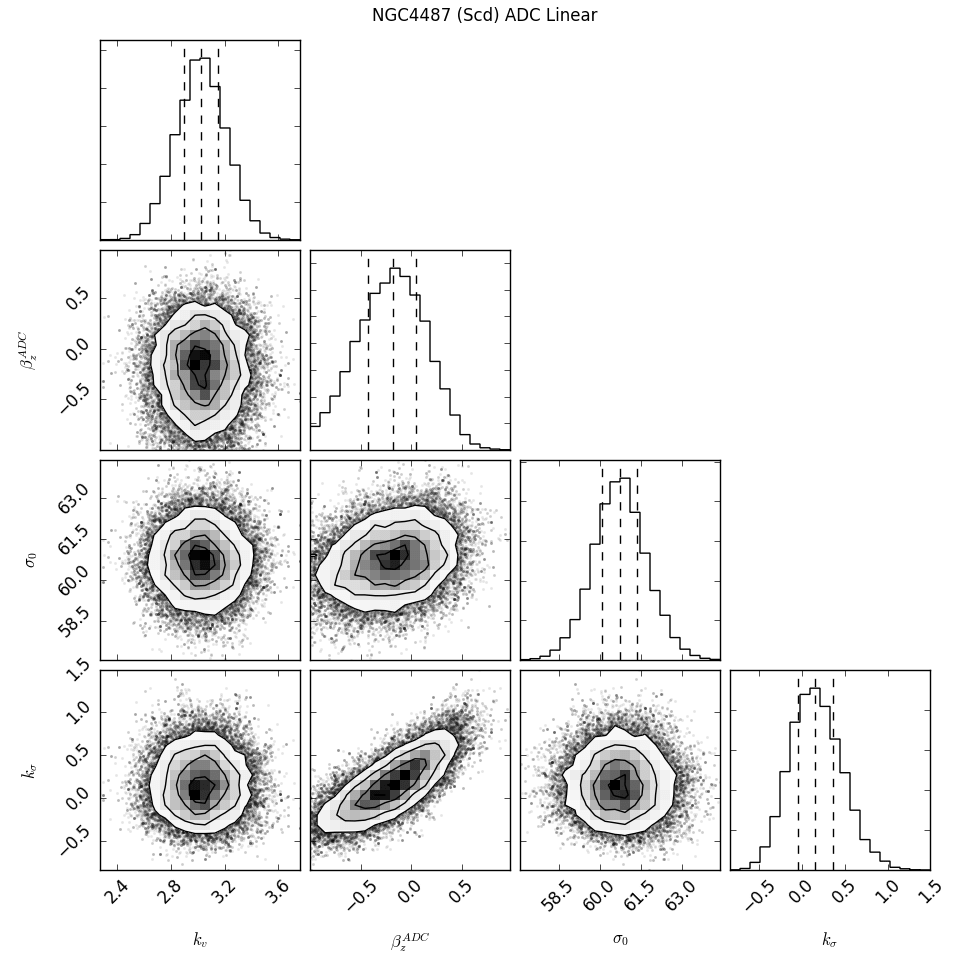}}
\caption{ {\it -- continuation}}
\label{fig:TrA5}
\end{figure*}

\addtocounter{figure}{-1}
\addtocounter{subfigure}{1}

\begin{figure*}
{\includegraphics[width=0.65\textwidth]{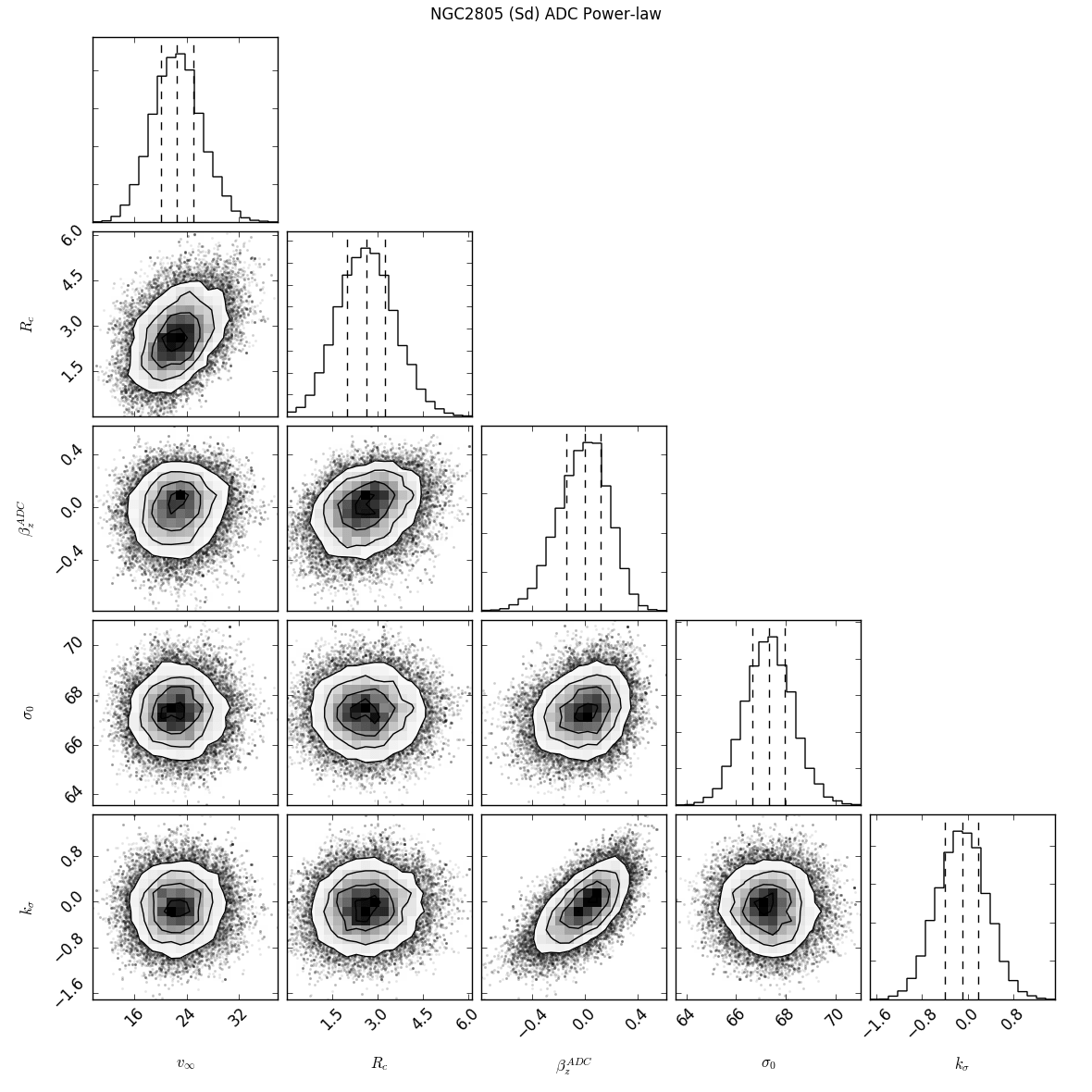}
\includegraphics[width=0.65\textwidth]{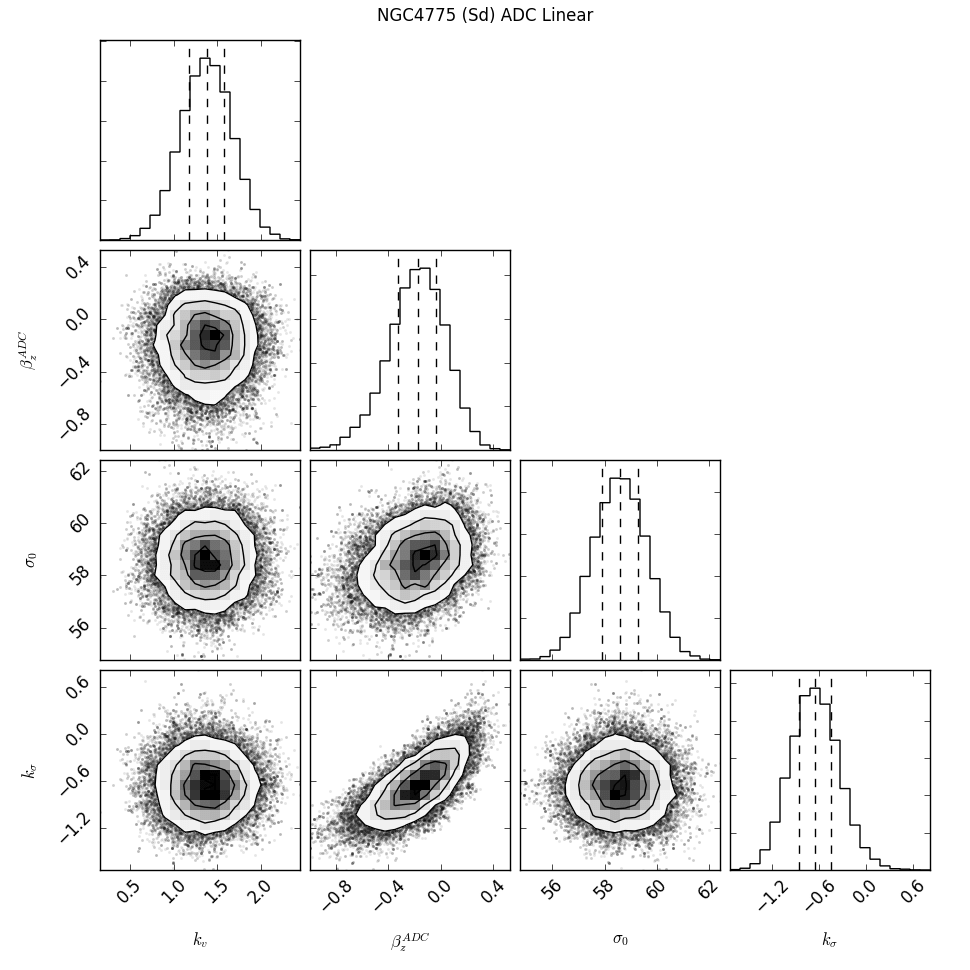}}
\caption{ {\it -- continuation}}
\label{fig:TrA5}
\end{figure*}

\addtocounter{figure}{-1}
\addtocounter{subfigure}{1}

\begin{figure*}
{\includegraphics[width=0.65\textwidth]{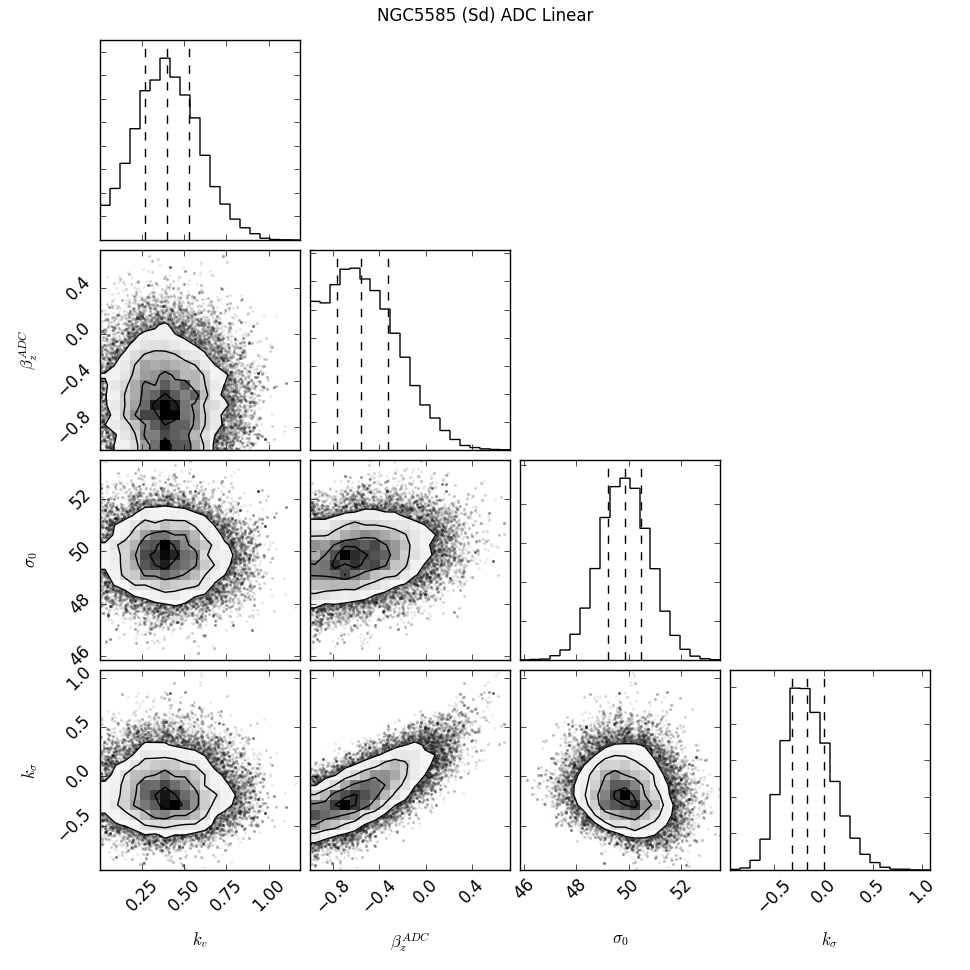}
\includegraphics[width=0.65\textwidth]{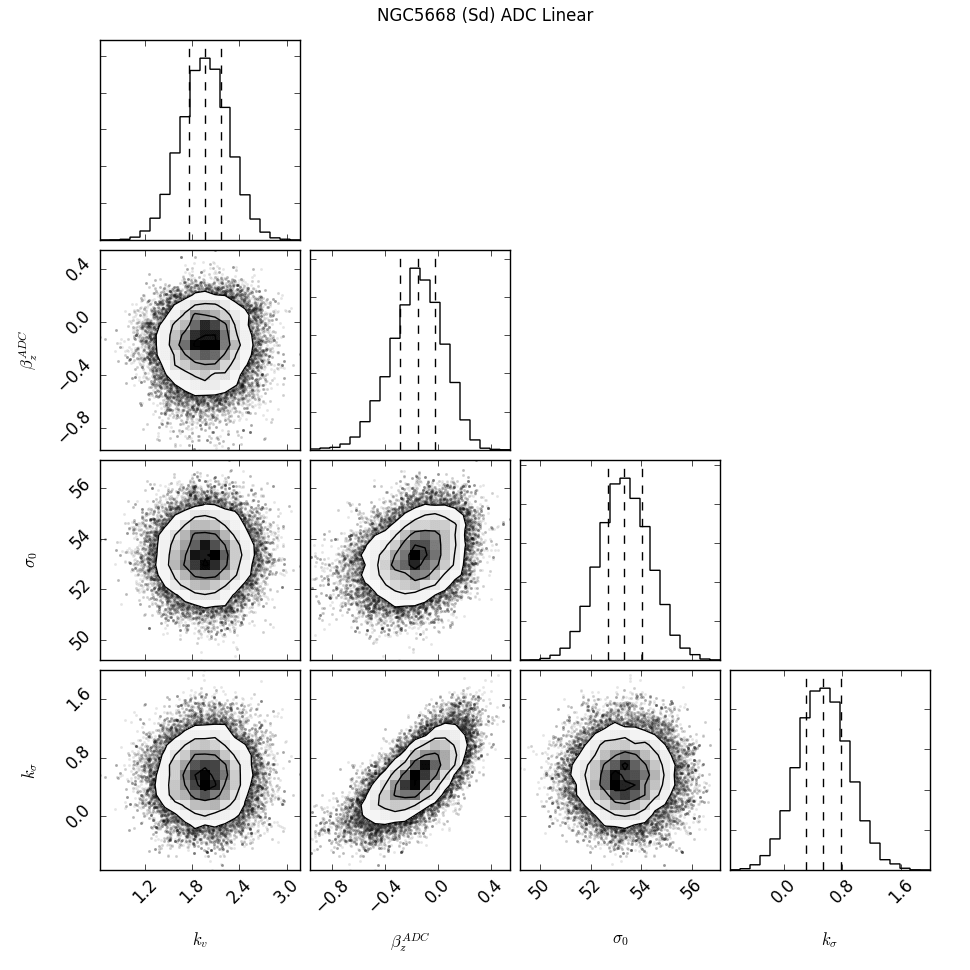}}
\caption{ {\it -- continuation}}
\label{fig:TrA5}
\end{figure*}

\end{document}